\documentclass[12pt]{JHEP3}
\usepackage[dvips]{graphicx}
\usepackage{epsfig}
\usepackage{amsmath}
\usepackage{bm}
\usepackage{times}
\newcommand{\ens}{\epsilon_{ns}}
\newcommand{\es}{\epsilon_{s}}
\newcommand{\vlus}{V_L^{us}}
\newcommand{\vlud}{V_L^{ud}}
\newcommand{\vrus}{V_R^{us}}

\newcommand{\veff}{\mathcal{V}_{\mathit{eff}}}
\newcommand{\aeff}{\mathcal{A}_{\mathit{eff}}}
\newcommand{\vfus}{\mathcal{V}^{us}_\mathit{eff}}
\newcommand{\vfud}{\mathcal{V}^{ud}_\mathit{eff}}
\newcommand{\afus}{\mathcal{A}^{us}_\mathit{eff}}
\newcommand{\afud}{\mathcal{A}^{ud}_\mathit{eff}}
\newcommand{\atau}{\alpha_s(m_\tau)}
\newcommand{\dpns}{\Delta^{+,kl}_{ud}}
\newcommand{\asgg}{\langle\frac{\alpha_s}{\pi} \,GG\rangle}

\newcommand{\gf}{1-\xi^2 \rho_L}
\newcommand{\sweff}{\tilde{s}^2}

\def\roughly#1{\mathrel{\raise.3ex\hbox{$#1$\kern-.75em%
\lower1ex\hbox{$\sim$}}}}
\def\lsim{\roughly<}
\def\gsim{\roughly>}

\title{Tests of non-standard electroweak couplings of right-handed quarks}
\author{V\'eronique Bernard \\
Universit\'{e} Louis Pasteur, Laboratoire de Physique Th\'{e}orique, \\
3-5 rue de l'Universit\'{e}, 67084 Strasbourg, France\\
E-mail: \email{bernard@lpt6.u-strasbg.fr}}
\author{Micaela Oertel\\LUTH, CNRS, Observatoire de Paris,
  Universit\'e Paris Diderot\\ 5 place
  Jules Janssen, 92195 Meudon, France\\ E-mail: 
\email{micaela.oertel@obspm.fr}}
\author{Emilie Passemar\footnote{Present Address: Institute for theoretical physics,
University of Bern, Sidlerstr. 5, CH-3012 Bern, Switzerland}\ and Jan Stern\\ 
Groupe de Physique Th\'{e}orique, IPN,
           CNRS, Universit\'{e} de Paris Sud-XI, 91406 Orsay, France\\
           E-mail: \email{passemar@ipno.in2p3.fr, stern@ipno.in2p3.fr}}

\abstract{The standard model can be interpreted as the
leading order of a Low-Energy Effective Theory (LEET) invariant under
a higher non linearly realized symmetry $S_{nat}\supset SU(2)_W \times
U(1)_Y$ equipped with a systematic power counting. Within the minimal
version of this ``not quite decoupling'' LEET, the dominant
non-standard effect appears at next-to-leading order (NLO) and is a
modification of the couplings of fermions to $W$ and $Z$. In
particular, the coupling of right-handed quarks to $Z$ is modified and
a direct coupling of right-handed quarks to $W$ emerges.  Charged
right-handed lepton currents are forbidden by an additional discrete
symmetry in the lepton sector originally designed to suppress Dirac
neutrino masses. A complete NLO analysis of experimental constraints
on these modified couplings is presented. Concerning couplings of
light quarks, the interface of the electroweak tests with QCD aspects is
discussed in detail.}


\begin{document}

\section{Introduction}

In parallel to the direct searches of New Physics aiming at production
of new heavy particles at LHC and other colliders it is important to
further develop precision low-energy searches of small modifications
of Electroweak (EW) couplings of known light particles that could not be
explained within the Standard Model (SM). A combination of these two
complementary approaches could help to correctly interpret
forthcoming experimental results. Indirect low-energy tests may be
based on a particular model of a high-energy completion of the SM and
predict its observable low-energy signatures. This traditional ``top -
down'' approach has the advantage of being well defined from the onset
and the disadvantage of representing just one possibility among many
others. This latter point becomes more relevant to the extent that a
larger variety of high energy Models become theoretically
conceivable. Indeed, during the last years many Models have appeared
(and disappeared - both without any deeper experimental motivation)
ranging from SUSY standard models with variable degree of
minimality (see for example~\cite{susy}), passing through variants of
Technicolor~\cite{technicolor} and even including non renormalizable
models such as Little Higgs~\cite{LH} and extradimensional models with or
without Higgs particle(s)~\cite{ED}.

This landscape of open possibilities calls for
an alternative ``bottom - up'' approach in which one starts with the
experimentally established features of the SM viewed as the leading order
of a non-decoupling Low Energy Effective Theory (LEET) and one asks
what the higher orders may be without assuming any
specific high energy completion of the LEET. This model independence is
conceivable provided one respects two requirements:
\begin{itemize} 

\item[i)] The LEET is formulated as a consistent Quantum Field Theory
renormalized and unitarized order by order in momentum expansion
following the well elaborated example of Chiral Perturbation
Theory~\cite{W79, GL84,GL85,Ecker+}. The lack of renormalizability in
the traditional sense (i.e, order by order in powers of coupling
constant(s)) does not mean the lack of consistency but a limitation of
predictivity at all scales.

\item[ii)] The second requirement concerns ``naturality'': At each
order the LEET should contain all operators that are allowed by its
symmetries. This principle stating that everything that is not
forbidden by a symmetry is allowed and should be effectively there,
partially restores predictivity. First, the actual (non linear)
symmetries of the LEET can be essentially inferred asking that no
non-standard operators appear at Leading Order (LO). Furthermore, the
power counting of the LEET provides a natural classification of
effects beyond the SM according to their importance at low
energies. In this respect, our results appear as non trivial and not
quite expected: The first non-standard effects arise already at
next-to-leading order (NLO) and resume to non-standard universal
couplings of fermions to $W$ and $Z$. They mainly concern right-handed
quarks.  The intensively studied oblique corrections, universality breaking
effects and flavor changing neutral currents (FCNC) only appear at
NNLO together with loops and are much more suppressed.
\end{itemize}

A detailed analysis of the NLO of our LEET and its confrontation with
experiment is the main subject of this paper.  The paper is organized
as follows: Section~\ref{LEET} reviews, comments and completes the
theoretical framework of the ``minimal not-quite decoupling LEET''
developed some time ago~\cite{HS04a, HS04b, HS06}. In
Section~\ref{NLO}, the physical content of the NLO is described and
the notation is settled. Section~\ref{Zanalysis} is devoted to the
analysis of couplings to $Z$. A full NLO fit to the standard EW
precision $Z$-pole and atomic parity violation observables is
performed and discussed. In Section~\ref{Wanalysis} the main
prediction of the LEET - the occurrence of couplings of right-handed
quarks to $W$ at NLO is discussed in the light of recent experimental
tests involving light quarks. The latter include $K_{\mu3}$ decays, as
well as inclusive tests using hadronic tau decays, deep inelastic
scattering (DIS) of (anti-) neutrinos, and the leptonic branching
fraction of $W$. These results are further discussed in
Sections~\ref{sec:discussion} and \ref{sec:otp}. Section~\ref{sec:summary}
concludes the paper.
\section{Minimal not quite decoupling EW effective theory}
\label{LEET}
First, we present a concise overview of effective theory framework
suitable for a bottom-up analysis of possible extensions of the SM
. No new material is involved in this Section that would not be
already contained in Refs.~\cite{HS04a, HS04b, HS06} (see also
Ref.~\cite{JS06}).

The theoretical framework is intended to encompass a large class of
(renormalizable) models extending to energies much larger than the
scale at which the effective description operates (typically $E\ll
\Lambda_W \sim 3$ TeV). These models remain so far unspecified except
for a common symmetry pattern and the common light particle
content. The latter then form the basis of the common Low Energy
Effective Theory (LEET).  The LEET is required to define, at least in
principle \footnote{i.e. regardless to global questions of
convergence}, a consistent Quantum Theory (characterized, in
particular, by a finite, unitary, analytic and crossing symmetric
S-matrix) through a well defined low-energy expansion in powers of
momenta and gauge couplings. The crucial ingredient of this expansion
is {\bf the infrared power counting} which allows one to classify bare
vertices (operators) as well as loops according to their importance in
the low-energy limit. This in turn allows one to define a systematic
``order by order renormalization'' as formulated and experienced in
Chiral Perturbation Theory~\cite{W79, GL84,GL85,Ecker+}. The UV
behavior and the requirement of ''renormalizability at all scales''
are not essential here, they concern more particularly the models
susceptible to provide the high energy completion of the
LEET. Similarly, the mass (or UV) dimension of an operator is not
necessarily the sole indication of its relevance at low energies:
Instead of representing the effective Lagrangian as the familiar
(decoupling) expansion
\begin{equation}
\mathcal{L}_{\mathit{eff}} = \mathcal{L}_{ren} + \sum_{D>4 }\frac{\mathcal{O}_D}{\Lambda^{D-4}}
\label{decoupexpansion}
\end{equation}
which adds to the {\bf renormalizable} SM Lagrangian (irrelevant)           
operators with
increasing mass dimension $D$ suppressed by inverse powers of an unspecified
scale  $\Lambda$, it is more convenient to organize the low-energy
expansion as
\begin{equation}
\mathcal{L}_{\mathit{eff}} = \sum_{d \ge 2} \mathcal{L}_{d}
\label{leffexpansion}
\end{equation}
where $d$ denotes the chiral (or IR) dimension indicating the low-momentum
behavior
\begin{equation}
\mathcal{L}_{d}  =  \mathcal{O}([p/\Lambda_W]^d) .
\end{equation}
Each term in the expansion, Eq.~(\ref{leffexpansion}), contains a
finite number of operators with the same infrared dimension $d$.
Similarly, the order by order renormalization makes use at each order
$d$ of a finite number of new counterterms of dimension $d$. This makes
appear new low-energy constants which are not fixed by the LEET
itself. They reflect the missing information hidden in the high-energy
completion of the LEET by a so far unspecified (renormalizable) Model.
Even in the absence of this information, the LEET framework allows one
to identify the most important effects beyond the SM and to provide an
efficient parametrization of their experimental signature at low
energies. The present paper illustrates how this statement works in
practice and how it compares with experiment.

Eqs.~(\ref{decoupexpansion}) and (\ref{leffexpansion}) are two
complementary representations of the same effective Lagrangian and
they are not necessarily in contradiction with each other. Their
comparison calls for few comments:
\begin{itemize}
\item[i)] Except for counting derivatives, {\bf the chiral (IR)
    dimension $d$ and the mass (UV) dimension $D$ do not coincide}.
  For theories involving gauge fields, chiral fermions and Goldstone
  bosons, the chiral dimension of a local operator is given by
  \cite{W79,GL84,Wu94,NS99}
\begin{equation}
d = n_{\delta} + n_g  +  n_f / 2
\label{irdimension}
\end{equation}
where $n_{\delta}$, $n_g$, and $n_f$ stand for the number of derivatives,
number of gauge coupling constants and number of fermion fields, 
respectively 
\footnote{This concerns canonically normalized fields. For instance, a gauge
field and GB fields have $d=0$, whereas the gauge connection has
$d=1$ so that this counting respects gauge invariance.}.

\item[ii)] {\bf The importance of loops in the low-energy expansion}
  is given by a generalization of the Weinberg power counting formula
  originally established for Goldstone bosons~\cite{W79} and
  subsequently extended to include gauge fields and chiral
  fermions~\cite{Wu94,NS99,Urech}. The infrared dimension of a
  connected Feynman diagram made up from vertices of
  $\mathcal{L}_{\mathit{eff}}$ labelled $v=1{\ldots} V$ and containing $L$
  loops should read
\begin{equation}
d = 2 + 2 L + \sum_{v} (d_v - 2).
\label{powercounting}
\end{equation}
This provides a close link between momentum and loop expansions and it
guarantees the renormalizability order by order, provided all vertices
satisfy $d_v \ge 2$. This last condition means that the interaction
must be suppressed in the low-energy limit as a consequence of the
symmetry enjoyed by the effective theory.  This may be viewed as a
generalization of Adler's theorem stating that the interaction of GBs
vanishes at $E=0$ due to chiral symmetry.

\item[iii)] The validity of the power counting formula,
  Eq.~(\ref{powercounting}),  finally
  requires that all particles contained in the LEET should be {\bf
    naturally light as a consequence of a symmetry}. Indeed, in order
  to guarantee the appropriate scaling of all propagators in the low -
  energy limit, the masses should scale as
\begin{equation}
\mathrm{mass} = \mathcal{O}(p^n),\quad n \ge 1~.
\label{masscondition}
\end{equation}
How this can happen is best illustrated by the example of the mass of
a gauge field arising from the Higgs mechanism. Due to gauge symmetry
such a mass takes the generic form $M_W =\frac{1}{2} g F_W$, where
$\Lambda_W = 4 \pi F_W $ is an {\bf intrinsic scale of the LEET}. The
power counting, Eq.~(\ref{irdimension}), then implies $M_W =
\mathcal{O}(p)$.  Fermion masses protected by a chiral symmetry will
be discussed shortly.

Let us stress, however, that no example of a low-energy symmetry except SUSY is
known that would protect masses of scalar particles which are not
Goldstone bosons. In this way the well known
difficulty to construct a non SUSY {\bf renormalizable model} with
naturally light Higgs particles reappears within the LEET framework.

\item[iv)] It is worth stressing the difference between the starting
  points of the expansions, Eq.~(\ref{decoupexpansion}) and
  (\ref{leffexpansion}), $\mathcal{L}_{\mathit{ren}}$ and
  $\mathcal{L}_2$, respectively.  In the decoupling case (cf
  Eq.~(\ref{decoupexpansion})), the scale $\Lambda$ is not fixed by the
  low-energy dynamics. The effective theory,
  Eq.~(\ref{decoupexpansion}), should be internally consistent for an
  arbitrarily large $\Lambda$ including in the decoupling limit
  $\Lambda \to \infty$. In this limit one should recover the full SM
  Lagrangian {\bf including the Higgs sector as dictated by
  renormalizability.}  In the alternative case, see
  Eq.~(\ref{leffexpansion}), the scale $\Lambda_W$ is a fixed
  characteristic of the theory, cf.
\begin{equation}
\Lambda_W = 4 \pi F_W \sim 3 ~\mathrm{TeV}
\end{equation}
and there is no point in considering the limit of large $\Lambda_W$. 
$\mathcal{L}_2$ is then the collection of all $d=2$ terms compatible
with the symmetries of the LEET. In particular, whether a light Higgs
particle should be included is no more dictated by the requirement of
renormalizability but rather by  symmetry considerations          
 and last but not least by experiment.
        
\item[v)] The argument of dimensional suppression of operators with $D
  > 4$ can very well coexist with the infrared power counting.
  Operators of higher mass dimension $D$ and a lower chiral dimension
  $d < D$ may still be dimensionally suppressed by $\Lambda^{4-D}$,
  where \mbox{$\Lambda\gg\Lambda_W$} is a scale exterior to the LEET. The
  best example is provided by the four fermion operators without
  derivatives and no insertion of gauge coupling $g$ which have $d =
  2$ and $D = 6$. Even if one does not include these terms into
  $\mathcal{L}_2$, they will appear at the tree level proportional to
  the inverse squared of the LEET scale $F_W$. It is an assumption
  that extra four fermion operators with $d=2$ that are not generated
  within the LEET will be suppressed by a scale $\Lambda \gg \Lambda_W
  = 4 \pi F_W$ and can be disregarded. A similar reasoning can be
  developed for magnetic moment type operators with $D=5$.
\end{itemize}
\subsection{Symmetry and particle content}
We do not know which new particles exist at scales much larger than
$\Lambda_W$ and which (local) symmetries beyond $SU(2)_W \times
U(1)_Y$ govern their interaction. If the energy decreases, particles
above $\Lambda_W$ will gradually decouple, meaning that the
corresponding heavy degrees of freedom can be integrated out. This
does, however, not imply that only symmetries acting linearly on light
degrees of freedom will be relevant in the LEET. Heavy particles
decouple whereas symmetries associated with them can reappear in the
LEET and can become non-linearly realized. Such symmetries usually do
not show up in the light particle spectrum but they can restrict the
form of the effective interactions of light particles.  In electroweak
LEETs, this possibility was so far not enough exploited, despite the
fact that it is realized and well understood in QCD below the chiral
scale $ \Lambda_{ch}= 4 \pi F_{\pi}$.

The minimal version of the LEET just contains all observed particles:
$W, Z$, photon and three generations of doublets of quarks and leptons,
including right-handed neutrinos.  They transform in the standard
linear way under the EW group $SU(2)_W\times U(1)_Y$. Following the
remark developed above, the latter may be embedded into a larger
symmetry group
\begin{equation}
S_{\mathit{nat}} \supset S_{\mathit{ew}} = SU(2)_W \times U(1)_Y ,
\end{equation}
such that $S_{\mathit{nat}}/S_{\mathit{ew}}$ is non-linearly realized. $S_{\mathit{nat}}$ and its
low-energy representation will be specified shortly. In addition, the
theory must contain three real Goldstone bosons collected into an
$SU(2)$ matrix $\Sigma(x)$ transforming under
$S_{\mathit{ew}}$ as
\begin{equation}
\Sigma(x) \longrightarrow G_L(x) \Sigma(x) G_R^{-1} (x)~,
\end{equation}
where $G_L$ represents the weak (left) isospin and the action of $U(1)_Y$ is 
represented by the right multiplication by a $SU(2)$ matrix $G_R$ satisfying
\begin{equation}
G_R(x) \tau_3 G_R^{-1}(x) = \tau_3.
\label{custodial}
\end{equation}
Such $G_R$ may be viewed as the right isospin pointing in the third
direction.  It is convenient to organize all right-handed fermions
into right isospin doublets (presuming the existence of right-handed
neutrinos). This is known to be strictly equivalent to the usual SM
assignment, provided the spectrum of hypercharges of right-handed
fermions satisfies \mbox{$Y/2 = T_{R}^3 + (B-L)/2$}.
 
$\Sigma$ represents the three GBs contained in a complex doublet of
Higgs fields that are needed to give masses to $W$ and $Z$ via the GB
kinetic term~\footnote{We use the notation $\langle A\rangle =
\mathrm{Tr}[A]$.}
\begin{equation}
\mathcal{L}_{\mathrm{mass}} = \frac{1}{4} F^2_W \langle D_{\mu}\Sigma^{\dagger}
D^{\mu}\Sigma\rangle~. 
\label{GBcin}
\end{equation}
Notice that $\mathcal{L}_{\mathrm{mass}}$ has chiral dimension $d=2$
as well as kinetic terms of $S_{\mathit{ew}}$ gauge fields and the
usual gauge invariant fermion action.

Hence, among the $SU(2)_W \times U(1)_Y$ invariants of the leading IR
dimension $d=2$ one finds all the Higgsless vertices of the SM. The
converse is, however, not true: As pointed out in Ref.~\cite{HS04a},
there are several ``unwanted'' $S_{\mathit{ew}}$ invariant operators with the
leading chiral dimension $d=2$ that are absent in the SM and which are
not observed. In the decoupling effective theory (cf.
Eq.~(\ref{decoupexpansion})) such operators do not appear at the
leading order because they carry the mass (UV) dimension $D>4$ and
they are not renormalizable. In the not quite decoupling alternative
the lack of renormalizability at low energies should be compensated by
a higher symmetry $S_{\mathit{nat}} \supset S_{\mathit{ew}}$. The primary role of
$S_{\mathit{nat}}$ is to forbid all ``unwanted operators'' that appear at the
leading order $d=2$. In the ``bottom - up'' approach to the LEET it
should be possible to infer the symmetry $S_{\mathit{nat}}$ from the known SM
interaction vertices below the scale $\Lambda_{W}$, before one
identifies heavy states associated with a probable (linear)
manifestation of $S_{\mathit{nat}}$ in the spectrum of states above
$\Lambda_W$.
\subsection{Bottom-up reconstruction of the symmetry $S_{\mathit{nat}}$}
Inspecting the list of $d=2$ ``unwanted'' operators~\cite{HS04a,HS06},
the symmetry $S_{\mathit{nat}}$ can be inferred in two steps: The first step
involves the well known custodial
symmetry~\cite{Sikivie80,Longhitano80} protecting the standard model
relation between gauge bosons mixing and masses, $\rho = 1$. It
concerns operators that are invariant under $S_{\mathit{ew}}$ thanks
to the constraints, Eq.~(\ref{custodial}) reducing the right isospin
group $G_R$ to its $U(1)_{T_3}$ subgroup. A typical example is the
$\mathcal{O}(p^2)$ operator
\begin{equation}
\mathcal{O}_T = \langle  \tau_3 \Sigma^{\dagger} D_{\mu} \Sigma 
\rangle^2
\label{Tparam}
\end{equation}
which directly affects the GB kinetic term, Eq.~(\ref{GBcin}), inducing
a potentially large modification of the SM gauge boson mixing.
Unwanted operators of the type (\ref{Tparam}) are eliminated by the familiar
left-right extension~\cite{LRmodels,Mohapatra83}
\begin{equation}
S_{\mathit{ew}} \to S_{\mathit{elem}} = SU(2)_{G_L} \times SU(2)_{G_R} \times
U(1)^{B-L}_{G_B} \subset S_{\mathit{nat}}
\label{selem}
\end{equation}
which is achieved by relaxing the condition (\ref{custodial}) and
allowing for a general $G_R \in SU(2)_R$. The same extension is
operated for (RH) fermion doublets.  The latter then transform under
$S_{\mathit{elem}}$ (\ref{selem}) as
\begin{equation}
\psi_L \in [ 1/2 , 0 ; B-L]    ,  \psi_R \in [0 , 1/2 ; B-L]
\end{equation}
where
\begin{equation}
\psi_{L/R} = \frac{1}{2} ( 1 \mp \gamma_5) \psi , 
\end{equation}
and $\psi$ denotes a generic fermion doublet.

This first step is neither surprising nor new: as already mentioned,
it is reminiscent of the custodial symmetry and of L-R extensions of
the SM~\cite{LRmodels,Mohapatra83}. The difference concerns the non
linear realization of the right isospin in the LEET that does not
necessarily require the existence of a light gauge particle $W_R$
below the scale $\Lambda_W$. Before developing this point, it is worth
stressing that the symmetry $S_{\mathit{elem}}$ (\ref{selem}) does not
eliminate all unwanted $d=2$ operators and that further extension of
$S_{\mathit{nat}}$ beyond (\ref{selem}) is necessary~\cite{HS04a,HS06}.

Among the remaining $d=2$ unwanted operators invariant under
$S_{\mathit{elem}}$ (\ref{selem}) there is
\begin{equation}
\mathcal{O}_S = \langle G_{L,\mu\nu} \Sigma G_R^{\mu\nu} \Sigma^{\dagger}\rangle~,
\label{os}
\end{equation}
where $G_{L,\mu\nu}$ and $G_{R,\mu\nu}$ are the (canonically normalized)
field strengths of $SU(2)_{G_L} \times SU(2)_{G_R}$. (They both carry the chiral
dimension $d=1$.) This operator represents an unsuppressed contribution to the
parameter $S$. Then we turn to non-standard $d=2$ operators involving
fermions:
\begin{equation}
\mathcal{O}_L = \bar\psi_L \gamma^{\mu} \Sigma D_{\mu} \Sigma^ {\dagger}\psi_L~,
\end{equation}
\begin{equation}
\mathcal{O}_R = \bar\psi_R \gamma^{\mu} \Sigma^{\dagger} D_{\mu} \Sigma \psi_R~.
\end{equation}
These two operators represent potentially large (tree level) modifications
of SM couplings of  fermions to electroweak gauge bosons.
They need to be suppressed, too. Finally, let us mention the unsuppressed
Yukawa coupling
\begin{equation}
\mathcal{O}_{\mathit{Yukawa}} = \bar\psi_{L} \Sigma \psi_R
\label{yukawa}
\end{equation}
which is invariant under $S_{\mathit{elem}}$ and carries the chiral
dimension $d=1$.  Such an operator would naturally generate fermion
masses of the order $m_f \sim \Lambda_W$ contradicting the condition,
Eq.~(\ref{masscondition}), for a fermion to belong to the LEET.
                        
The problem with unsuppressed operators (\ref{os} - \ref{yukawa})
concerns the origin and quantum numbers of GBs $\Sigma(x) \in SU(2)$,
which were tacitly assumed to arise from the spontaneous breaking of
the symmetry $S_{\mathit{elem}}$, i.e., to transform under the latter as
the representation $[1/2 , 1/2 ; 0 ]$. The alternative to this
oversimplified scenario takes example in QCD and in Technicolor models
without necessarily adopting all their (model dependent) consequences:
The GBs that represent active agents of the Higgs mechanism need not
be considered as ``elementary'' but rather as bound states of some
so far unspecified Strong Dynamics (SD) operating at scales above
$\Lambda_W $ and involving new degrees of freedom.  The GBs of the
spontaneously broken chiral symmetry of the SD
\begin{equation}
S_{\mathit{comp}} = SU(2)_{\Gamma_L} \times SU(2)_{\Gamma_R}
\label{scomp}
\end{equation}
then appear as the only manifestation of the SD much below the scale
$\Lambda_W$.  Such GBs now transform as
\begin{equation}
\Sigma(x) \to \Gamma_L(x) \Sigma(x) \left[\Gamma_R(x)\right ]^{-1}
\end{equation}
where $\Gamma_{L/R}$ denote elements of $S_{\mathit{comp}}$
(\ref{scomp}). Accordingly, in the GB kinetic term, Eq.~(\ref{GBcin}),
the covariant derivatives now involves the corresponding connections
\footnote{The composite gauge sector is entirely described by
  the connections $\Gamma_{L/R,\mu}$ of chiral dimension $d=1$.
  Unless one specifies the corresponding gauge field ($d=0$) and/or
  gauge coupling ($d=1$) there is no way to write a corresponding Yang-Mills
  action with $d=2$. The square of the curvature $\Gamma_{\mu\nu} =
  \partial_{\mu} \Gamma_{\nu} - \partial_{\nu} \Gamma_{\mu} - i \left[
    \Gamma_{\mu} , \Gamma_{\nu} \right ]$ has $d=4$.}
\begin{equation}
D_{\mu} \Sigma = \partial_{\mu} \Sigma - i \,\Gamma_{L,\mu}\, \Sigma +
i\, \Sigma\, \Gamma_{R,\mu}~.
\end{equation}
On the other hand, the description of the ``elementary sector''
with the gauge group (\ref{selem}) and chiral fermion doublets transforming as
\begin{equation}
 \psi_{L/R} \to G_{L/R}\exp\left[-i \frac{B-L}{2} \alpha \right] \psi_{L/R}
\label{trafofermions}
\end{equation}
remains as before. Consequently, the ``unwanted operators'' such as
(\ref{os}-\ref{yukawa}) are no more invariant and are suppressed.
 
The above picture combining elementary and composite sectors should be
merely viewed as a possible physical motivation of the extension of
the symmetry $S_{\mathit{nat}}$ from $S_{\mathit{elem}}$ to $S_{\mathit{elem}}
\times S_{\mathit{comp}}$.  This step is necessary within the class of
LEETs considered here and it is not tied to any particular model. The
result may be summarized as
\begin{equation}
S_{\mathit{nat}} = \left [ SU(2)_{G_L} \times SU(2)_{G_R} \times U(1)^{B-L}_{G_B} \right ]_{\mathit{elem}}
\times \left [ SU(2)_{\Gamma_L} \times SU(2)_{\Gamma_R} \right ]_{\mathit{comp}}.
\label{snat}
\end{equation}
\FIGURE[t]{
\includegraphics*[scale=0.35,angle=-90,bb=125 230 432 620]{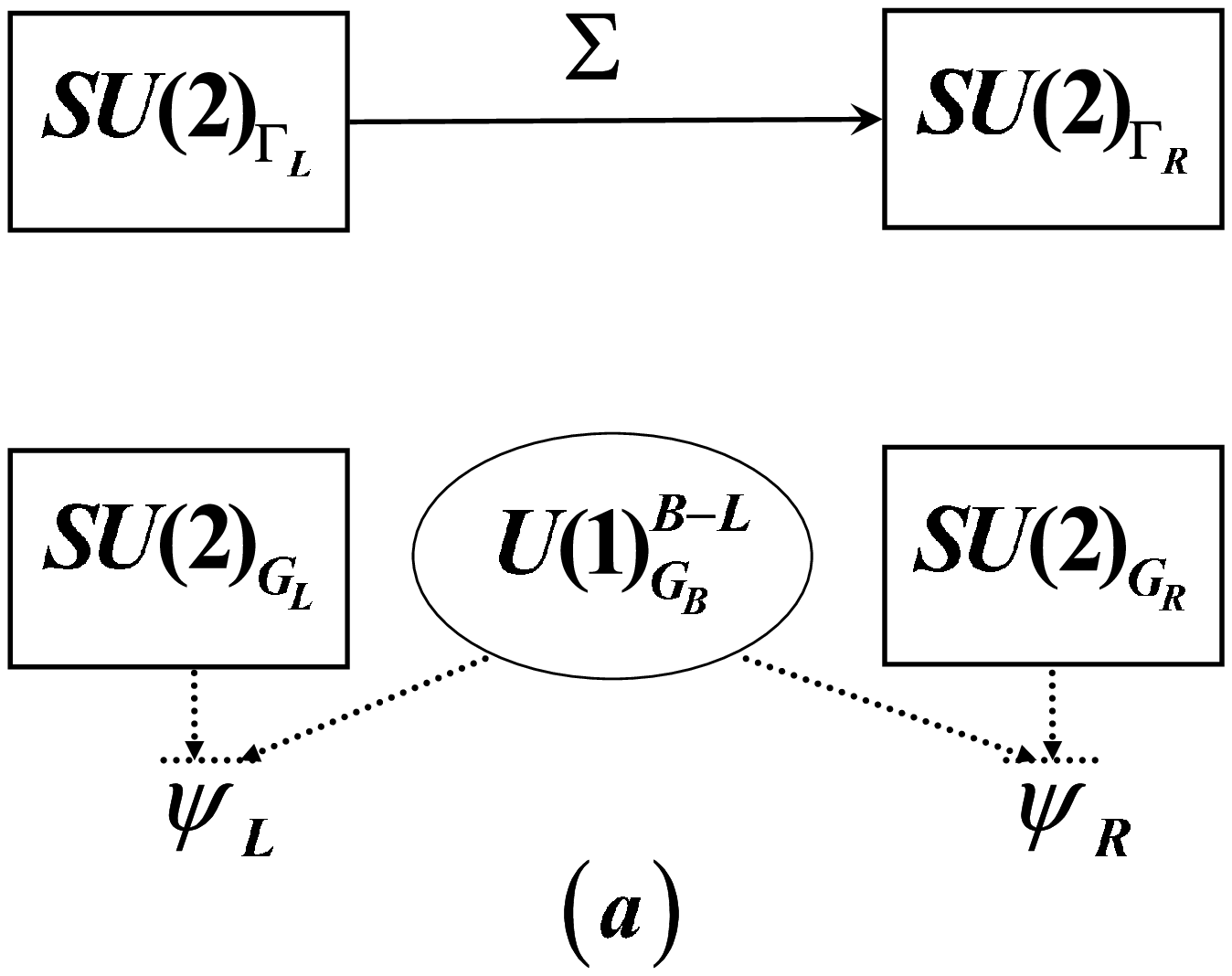}
\hspace{1cm}
\includegraphics*[scale=0.35,angle=-90,bb=100 120 421 710]{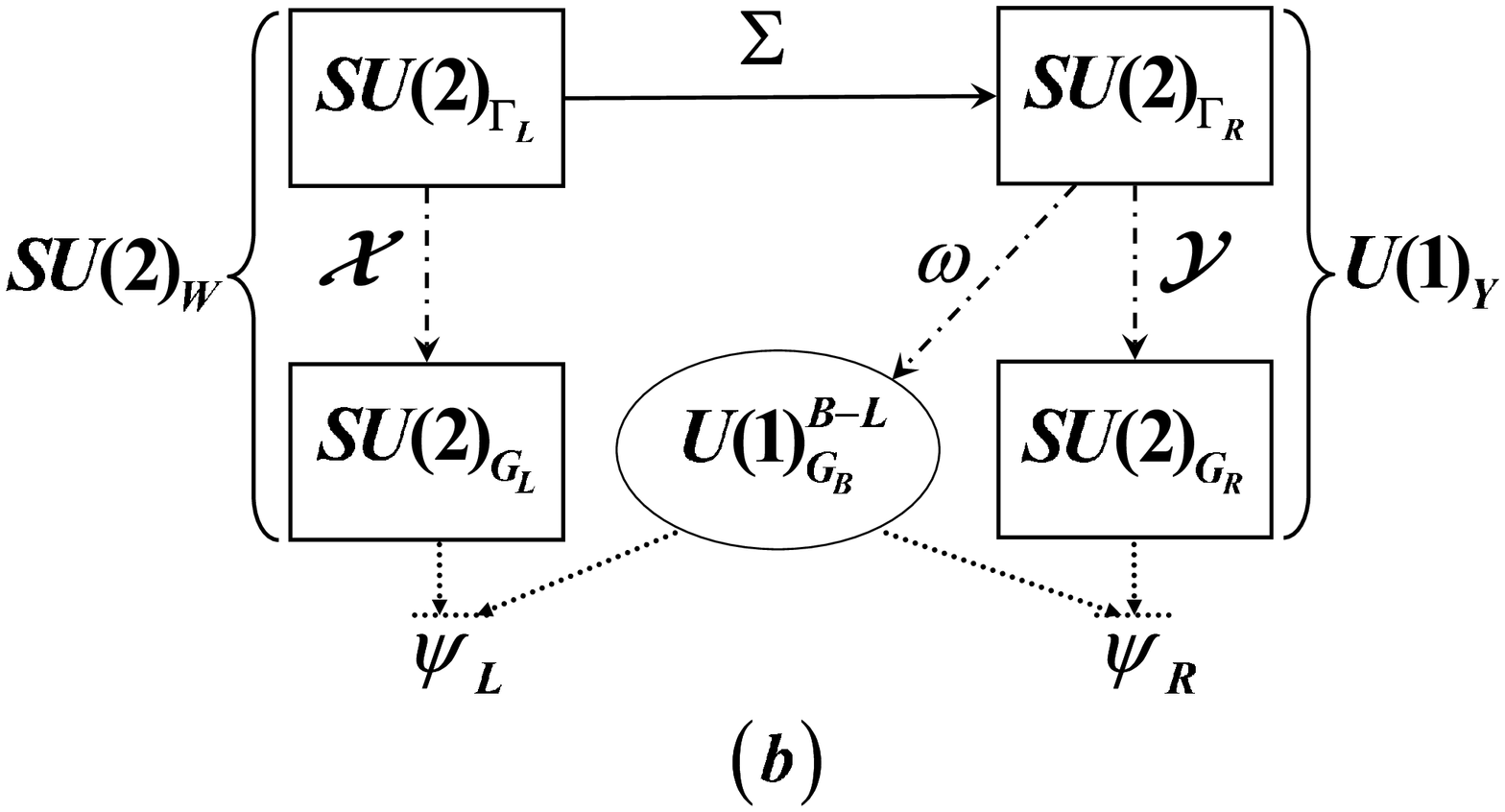}
\caption{\it Moose diagram showing the structure of the LEET without
  spurions (left) and with spurions (right) connecting the elementary
  and the composite sector.}
\label{figmoose}
}
The corresponding transformation properties of GBs and elementary
fermions are represented in Fig.~\ref{figmoose}a
using the ``Moose notation''~\cite{Moose}.  At the leading order
$d=2$, the most general Lagrangian (see the comment v) in
Section~\ref{LEET} about four fermion interactions) invariant under
{\bf the linear action} of $S_{\mathit{nat}}$, Eq.~(\ref{snat}), reads
\begin{eqnarray}
  \mathcal{L} \left( p^2 \right)  &=& \frac{F_W^2}{4}  \left\langle D_{\mu}
  \Sigma^{\dag} D^{\mu} \Sigma \right\rangle + i\, \overline{\psi_L}
  \gamma^{\mu} D_{\mu} \psi_L + i\, \overline{\psi_R} \gamma^{\mu} D_{\mu}
  \psi_R \nonumber \\
  && -  \frac{1}{2}  \left\langle G_{L \mu \nu} G_L^{\mu \nu} + G_{R \mu \nu}
  G^{\mu \nu}_R \right\rangle - \frac{1}{4} G_{B \mu \nu} G_B^{\mu \nu} . 
  \label{lag}
\end{eqnarray}
It contains more gauge fields than actually observed at low energies. Most
of them remain massless. The only hint of a mass term arises from the GB
kinetic term. In the physical gauge, $\Sigma = 1$, the latter reads
\begin{equation}
\mathcal{L}_{\mathit{comp}} = \frac{1}{4} F^2_W \langle
[\Gamma_{L,\mu} - \Gamma_{R,\mu}]^2\rangle~.
\label{massterm}
\end{equation}
Since at this stage the composite and elementary sectors do not
communicate, there is no link between the mass term (\ref{massterm})
and the elementary gauge bosons $G_{L/R,\mu}$ that couple to
fermions. Furthermore, all fermions remain massless as a consequence
of the symmetry $S_{\mathit{nat}}$.

In order to recover the SM Lagrangian hidden in Eq.~(\ref{lag}), one
has to reduce the linear symmetry $S_{\mathit{nat}}$ back to
$S_{\mathit{ew}}$, imposing suitable $S_{\mathit{nat}}$ invariant {\bf
  constraints} that would eliminate the redundant degrees of freedom
and provide the missing link between the elementary and composite
sectors . We are now going to describe this reduction.

\subsection{The Coset Space $S_{\mathit{nat}}/S_{\mathit{ew}}$  and Spurions}
\label{sec:reduction}
The four $SU(2)$ gauge fields $G_{L/R,\mu}$ and $\Gamma_{L/R,\mu}$
together with the $U(1)$ gauge field $G_{B,\mu}$ that appear in
Eq.~(\ref{lag}) span the linear representation of the local symmetry
group $S_{\mathit{nat}}$. It is conceivable that at ultrahigh energies
such representation is actually realized, with nine among the thirteen involved
fields acquiring a mass $\gg \Lambda_W$. As the energy decreases below
$\Lambda_W$, the linearly realized subgroup is gradually reduced
ending up with $S_{\mathit{ew}} = SU(2)_W \times U(1)_Y$ , i.e., with four
light EW gauge bosons. Since (by definition), all nine gauge fields from
the coset $S_{\mathit{nat}}/S_{\mathit{ew}}$ are very massive, they
can be integrated out and, at low energies, there are no more gauge
fields left in the coset $S_{\mathit{nat}}/S_{\mathit{ew}}$. It
follows that any object that remains in the LEET and carries a local
charge from $S_{\mathit{nat}}/S_{\mathit{ew}}$ must necessarily be a
{\bf non propagating spurion}, since there is no way to write a
corresponding gauge invariant kinetic term. One may anticipate that
the reduction $S_{\mathit{nat}} \to S_{\mathit{ew}}$ will yield three
$SU(2)$-valued scalar spurions, reflecting the structure
\begin{equation}
S_{\mathit{nat}}/S_{\mathit{ew}} = \left [ SU(2) \right]^3
\end{equation}
of the coset space.
Following Ref.~\cite{HS06}, the reduction proceeds by
pairwise identification of $SU(2)$ factors from composite and elementary
sectors. The precise alignment of gauge fields defining this identification 
is uniquely dictated by the requirement that one should end up with        
the couplings of the SM. 
\subsubsection{The left-handed sector}
One first identifies up to a gauge the ``composite''
and ``elementary'' $SU(2)_L$ imposing {\bf the constraint }
\begin{equation}
\Gamma_{L,\mu} =  \mathcal{X} g_L G_{L,\mu} \mathcal{X}^{-1} +
i  \mathcal{X}  \partial_{\mu}  \mathcal{X}^{-1}
\label{xconstraint}
\end{equation}
where $\mathcal{X}$ is a $2 \times 2$ matrix field satisfying the reality
condition
\begin{equation}
\mathcal{X} = \tau_2 \mathcal{X}^{\star} \tau_2  \equiv \mathcal{X}^c .
\end{equation}
This last condition is equivalent to the statement that $\mathcal{X}$ is a
real multiple of an $SU(2)$ matrix
\begin{equation}
\mathcal{X}(x) = \xi \Omega_L(x),\quad     \Omega_L(x) \in SU(2).
\end{equation}
Taking the trace of Eq.~(\ref{xconstraint}), it becomes obvious that {\bf
  $\xi$ must be a constant }. Furthermore, requiring that the
constraint (\ref{xconstraint}) should be invariant under $S_{\mathit{nat}}$,
forces $\mathcal{X}$ to transform as the bifumdamental representation
of the group $\mathcal{G}_L = SU(2)_{\Gamma_L} \times
SU(2)_{G_L}$ , cf.
\begin{equation}
\mathcal{X} \to \Gamma_{L} \mathcal{X} G^{-1}_{L}.
\end{equation}
Given this transformation, the constraint (\ref{xconstraint}) can be
{\bf equivalently rewritten} as
\begin{equation}
D_{\mu}  \mathcal{X} = 0 .
\end{equation}
As anticipated above, the covariant reduction of the product
$\mathcal{G}_L = SU(2)_{\Gamma_L} \times SU(2)_{G_L}$ to the
SM left isospin makes appear one {\bf non propagating spurion
  $\mathcal{X}$}.  The latter is a constant real multiple of a unitary
unimodular matrix transforming as the representation $[1/2,1/2]$ of
the product $\mathcal{G}_L$.  There is a gauge (called standard gauge)
in which the spurion becomes simply the $\xi$ multiple of the unit
matrix. This reduction procedure is represented on the left hand side
of Fig.~\ref{figmoose}b, where the spurion
$\mathcal{X}$ and its transformation properties are shown.
\subsubsection{The right-handed sector}
 
The remaining reduction involves $\mathcal{G}_R =
SU(2)_{\Gamma_R} \times SU(2)_{G_R}$ together with
the factor $U(1)^{B-L}_{G_B}$. It proceeds via a two-step identification
ending up with $U(1)_Y$ of the Standard Model
\begin{equation}
 \mathcal{G}_R \times U(1)^{B-L}_{G_B}  \to  U(1)_Y
\label{rreduction}
\end{equation}
involving two $SU(2)$-valued spurions $\mathcal{Y}$ and $\omega$, see the
right hand side of Fig~\ref{figmoose}b.
 
In the first step one repeats what has been done in the left-handed
sector. One imposes the constraint
\begin{equation}
\Gamma_{R,\mu} = \mathcal{Y}  g_R G_{R,\mu} \mathcal{Y}^{-1}  +
i \mathcal{Y} \partial_{\mu} \mathcal{Y}^{-1}
\label{yconstraint}
\end{equation}
where the spurion $\mathcal{Y} = \mathcal{Y}^c$  is real implying
\begin{equation}
\mathcal{Y} = \eta\, \Omega_R,\quad \Omega_R \in SU(2) .
\end{equation}
$\eta$ is a constant parameter. Requiring the constraint
(\ref{yconstraint}) to be invariant under $S_{\mathit{nat}} $ leads to the
following transformation property of the spurion $\mathcal{Y}$ (see
also Fig.~\ref{figmoose}b)
\begin{equation}
\mathcal{Y}  \to \Gamma_R \, \mathcal{Y}\,  G_R^{-1}.
\end{equation}
This means that the constraint
(\ref{yconstraint}) can equivalently be reexpressed as
\begin{equation}
D_{\mu} \mathcal{Y}  =  0~.
\end{equation}
In the second step it is convenient to represent $U(1)^{B-L}_{G_B}$ as a $SU(2)$
matrix 
\begin{equation}
G_{B} = \exp ( -i \alpha \tau_3 )
\end{equation}
and to identify it with the right-handed isospin defined in the
first step. (The parameter $\alpha$ is the same as in
Eq.~(\ref{trafofermions}).) This amounts to orienting the right-handed
isospin in the third direction, selecting a new (diagonal) $U(1)_{Y}$, where
\begin{equation}
\frac{Y}{2}  =  T^3_R  + \frac{B-L}{2}~.
\label{hypercharge}
\end{equation}
This procedure is equivalent to the constraint
\begin{equation}
\Gamma_{R,\mu} = \omega\, g_B\, G_{B,\mu} \frac{\tau_3}{2} {\omega}^{-1} +           
i \omega \partial_{\mu} {\omega}^{-1},
\label{rconstraint}
\end{equation}
where $G_{B,\mu}$ is the $U(1)^{B-L}_{G_B}$ gauge field and $g_{B}$ stands
for the corresponding gauge coupling. The covariance of the
constraint, Eq.~(\ref{rconstraint}), under $S_{nat}$ is equivalent to
the requirement that the spurion $\omega$ transforms as
\begin{equation}
\omega  \to \Gamma_{R}\, \omega \, G^{-1}_{B},
\end{equation}
as represented in Fig.~\ref{figmoose}b. As a consequence of the reality
condition $\omega ={ \omega}^{c}$, the spurion $\omega$ is a real (constant)
multiple of a $SU(2)$ matrix:
\begin{equation}
\omega = \zeta \,\Omega_{B},\quad    \Omega_{B} \in SU(2) .
\end{equation}

\subsubsection{Lepton number violation}
The reduction (\ref{rreduction}) is necessary to make appear $U(1)_{Y}$ as
required by the Standard Model. It has two immediate consequences which
follow from the existence of the spurion $\omega$ and  reflect the
particular structure of the right-handed sector.
       
First, one can define the projection on up and down components
of right-handed doublets that is covariant under the full symmetry $S_{nat}$.
The real spurion $\mathcal{Y}$ can be decomposed as 
\begin{equation}
\mathcal{Y} = \mathcal{Y}_{\uparrow} + \mathcal{Y}_{\downarrow},\quad
\mathcal{Y}_{\uparrow,\downarrow} = \Pi_{\uparrow,\downarrow}\,
\mathcal{Y} \to \Gamma_{R}\,\mathcal{Y}_{\uparrow,\downarrow}\,
G^{-1}_{R}
\end{equation}
where the covariant projectors $\Pi_{\uparrow,\downarrow}$ are defined as
\begin{equation}
\Pi_{\uparrow,\downarrow} = \omega \frac{1\pm \tau_3}{2} {\omega}^{-1} \to    
 \Gamma_{R} \Pi_{\uparrow,\downarrow} {\Gamma_{R}}^{-1}~.
\end{equation}
Notice that a similar possibility to separate up and down components
respecting the $S_{nat}$ symmetry does not exist for left-handed
doublets.

The second and most important consequence of the existence of the
spurion $\omega$ is the necessary appearance of {\bf Lepton Number violating
operators invariant under ${\bm S_{nat}}$ }. Indeed, from $\omega$ one can
define  the spurion
 $\mathcal{Z}$  carrying two units of the $B-L$ charge
\begin{equation}
\mathcal{Z} = \omega \tau_{+} \omega^{\dagger} \to \exp(i \alpha)
            \Gamma_{R} \mathcal{Z} {\Gamma_{R}}^{-1}
\end{equation}
which in turn allows to construct LNV operators which are invariant
under $S_{\mathit{nat}}$. Such operators will be naturally suppressed
by the parameter $\zeta^2 \ll \xi , \eta$.  Hence, LNV is unavoidable
though its strength cannot be predicted within the LEET alone. It
can be consistently kept small. Consequences for the systematic LEET
description of LNV processes have been discussed elsewhere~\cite{HS06}.

\subsection{The standard gauge}
\label{standardgauge}
In order to make the emergence of the well known
SM interaction vertices from the $ \mathcal{O}(p^2)$ Lagrangian (\ref{lag})
explicit, it is convenient to use the `` standard gauge'' in which
\begin{equation}
\Sigma = \Omega_{L} = \Omega_{R} = \Omega_{B} = 1~.
\end{equation}
The existence of this gauge has been shown in Ref.~\cite{HS06}. The constraint
(\ref{xconstraint}) becomes in the standard gauge
\begin{equation}
\Gamma_{L,\mu}^{i} = g_{L} G_{L.\mu}^{i},\quad i = 1,2,3,
\end{equation}
representing the $SU(2)_{W}$ group of the SM with $ g = g_{L}$.  In
the right-handed sector the constraints (\ref{yconstraint}),
(\ref{rconstraint}) reduce to
\begin{equation}
\Gamma_{R,\mu}^{1,2} = g_{R} G_{R,\mu}^{1,2} = 0
\end{equation}
and to
\begin{equation}
\Gamma_{R,\mu}^{3} = g_{R} G_{R,\mu}^{3} = g_{B} G_{B,\mu}~.
\label{u1}
\end{equation}
Eq.~(\ref{u1}) reflects the relation (\ref{hypercharge}) between
hypercharge, $T_{R}^{3}$ and $B-L$, defining $U(1)_Y$ of the SM. From
the normalization of gauge field kinetic terms , one identifies the SM
coupling $g'$:
\begin{equation}
g'^{-2} = g_{R}^{-2} + g_{B}^{-2}.
\end{equation}
The three $SU(2)$-valued spurions reduce in the standard gauge to three
constants $\xi$, $\eta$ and $\zeta$:
\begin{equation}
\mathcal{X} = \xi \times \underline{1},\quad \mathcal{Y}_{\uparrow,\downarrow} = \eta \,\frac{1 \pm \tau_{3}}{2}
\label{spurionsa}
\end{equation}
and the LNV spurion $\mathcal{Z}$ reduces to
\begin{equation}
\mathcal{Z} = \zeta^2 \tau_{+}.
\label{spurionsb}
\end{equation}

Inserting into the Lagrangian (\ref{lag}) the standard gauge
expression of the constraints (\ref{xconstraint}),
(\ref{yconstraint}), (\ref{rconstraint}), one recovers the Higgsless
part of the SM Lagrangian~\cite{HS04a,HS04b,HS06}. In particular
\begin{itemize}
\item $W$ and $Z$ get standard masses and mixing through the GBs kinetic term
          (\ref{massterm}).
\item  There are no physical scalars left: The three GBs $\Sigma$ are
          absorbed by the longitudinal components of $W$ and $Z$.
\item  Couplings of fermions to $W, Z$ and photon are standard. There are no
          Yukawa couplings. At LO, the right-handed neutrino $\nu_{R}$     
          decouples.
\item  Fermions stay massless as a consequence of the symmetry $S_{nat}$.
\item There exists a huge accidental flavor symmetry acting in the
family space.
\end{itemize}

\subsection{Fermion Masses}

Fermion masses are suppressed with respect to the LEET scale
$\Lambda_W = 4 \pi F_{W} \sim 3$ TeV by powers of spurions: Indeed, in
order to write a {\bf Dirac mass-term} invariant under the whole
symmetry $S_{nat}$, one needs to insert at least one spurion
$\mathcal{X}$ and one spurion $\mathcal{Y}_{a}$ . This can be
seen from Fig.~\ref{figmoose}b: The shortest way from $\Psi_{R}$
to $\Psi_{L}$ necessarily meets both spurions. The resulting mass
operator reads
\begin{equation}
\mathcal{M}_{a} = \bar\Psi_{L} \mathcal{X}^{\dagger} \Sigma             
                     \mathcal{Y}_{a} \Psi_{R}~,\quad a \in \{\uparrow,\downarrow\}~,
\end{equation}
and it is is of
the order $\mathcal{O}(p\, \xi \eta)$.  The natural size of the
low-energy constant multiplying such operator is $\sim \Lambda_{W}$.
The fact that the {\bf highest} fermion mass (i.e. top mass) must be
suppressed as $\mathcal{O}(p)$, suggests the following power counting
of spurion factors $\xi$ and $\eta$:
\begin{equation}
\xi \eta \sim m_{top} / \Lambda_{W} = \mathcal{O}(p)~.
\end{equation}
Adopting this power counting rule for spurions we define the total IR
dimension
\begin{equation}
d^*  =  d  + \frac{1}{2} ( n_{\xi}  +  n_{\eta} ) 
\label{irdim}
\end{equation}
where $n_{\xi}$, $n_{\eta}$ stand for the number of insertions of
spurions $\mathcal{X}$ and $\mathcal{Y}$ respectively. The leading
mass term in the Lagrangian has $d^{\star} = 2 $ as well as the
leading spurion-free Lagrangian (2.25). Together they constitute the
{\bf leading order of the LEET.}
Notice that the formula (\ref{powercounting}) counting the chiral
dimension of a Feynman graph holds if $d$ is replaced by $d^*$.

Further suppression of Dirac fermion masses by additional powers of
spurions is conceivable corresponding to running around the diagram (b)
in Fig~\ref{figmoose} several times. A similar description of fermion mass
hierarchy has been proposed by Froggatt and Nielsen some time
ago~\cite{FN}.

\subsubsection{Majorana mass terms and the unbearable lightness of Neutrinos}

Majorana masses necessarily involve the spurion
$\mathcal{Z}$ and are further suppressed by the corresponding factor
$\zeta^2$ reflecting the scale of LNV. The corresponding mass operator
involving the right-handed neutrino reads
\begin{equation}
\mathcal{M}_{R} = \bar\Psi_{R} \mathcal{Y}^{\dagger} \mathcal{Z} \mathcal{Y}
                 \Psi_{R}^{C} = \mathcal{O}( p \,\zeta^2 \eta^2)~,
\label{majoright}
\end{equation}
whereas the left-handed neutrino mass term reads
\begin{equation}
\mathcal{M}_{L} = \bar\Psi_{L} \mathcal{X}^{\dagger}\Sigma \mathcal{Z}
\Sigma^{\dagger} \mathcal{X} \Psi_{L}^{C} = \mathcal{O}(p \,\zeta^2 \xi^2)~.
\label{majleft}
\end{equation}
 
At this stage several conclusions can be drawn concerning the
smallness of neutrino masses within the present LEET framework: First,
there is no fundamental difficulty of keeping Majorana masses
arbitrarily small, though their smallness can hardly be predicted
within the LEET alone. Next, the Majorana masses of left-handed and
right-handed neutrinos should be expected of the same order of
magnitude unless the spurion parameters $\xi$ and $\eta$ are of
essentially different size~\footnote{The extreme such case, where
$\eta$ would be larger than $\xi$ by several orders of magnitude, does
not seem to be favored by the NLO fits discussed later within this
article.}.  Finally, in order to accommodate the LEET with the
smallness of neutrino masses it is necessary and sufficient to find
the reason of {\bf suppression of neutrino Dirac masses} compared to
the observed masses of charged leptons and quarks.

In Ref.~\cite{HS04a,HS06} it has been suggested that this suppression
finds its origin in a discrete symmetry enjoyed by leptons and absent
for quarks. This discrete symmetry is already present at LO as a
part of a huge accidental flavor symmetry: It is the $Z_2$ {\bf
reflection symmetry} which can covariantly be defined as
\begin{equation}
\mathcal{Y}_{\uparrow} \, l_R \rightarrow - \mathcal{Y}_{\uparrow} l_R, \quad
\mathcal{Y}_{\downarrow} l_R \rightarrow \mathcal{Y}_{\downarrow} l_R
\label{z2a}
\end{equation}
where $l_{R}$ stands for any lepton doublet. In the standard gauge this
transformation simply becomes
\begin{equation}
\nu_{R} \longrightarrow - \nu_{R}~.
\label{z2}
\end{equation}
Notice that the up-component of the RH lepton doublet, i.e.,
$\nu_{R}$ is the only fermion which does not carry any gauge charge
and for which one may expect the reflection symmetry (\ref{z2}) to
extend beyond the leading order.

The reflection symmetry does not prevent the right-handed neutrino to
become massive through the Majorana mass term, Eq.~(\ref{majoright}).
It however forbids the Dirac mass term $\bar \nu_{L} \nu_{R} $. The
further important consequence of this reflection symmetry is the {\bf
absence of charged right-handed lepton currents $\bar e_{R} \gamma
\nu_{R}$} to all orders of the LEET. As will be seen shortly, this
fact has its phenomenological relevance all over this article.

\subsection{Beyond the Leading Order}

In conclusion of this Section we summarize the steps and rules
to be followed in constructing order by order the whole effective Lagrangian. 

\begin{itemize}
\item[i)] Construct all local operators invariant under $S_{nat}$ from
the 13 gauge fields $G_{L,\mu} , G_{R,\mu}, G_{B,\mu},
\Gamma_{L,\mu}, \Gamma_{R,\mu}$ from the Goldstone boson matrix
$\Sigma(x)$, from the chiral fermion doublets, from spurions
$\mathcal{X} , \mathcal{Y}_{\uparrow,\downarrow}$, as well as from the spurion
$\mathcal{Z}$ provided we wish to consider the LNV sector of the
theory. Notice that the latter can be consistently omitted.
\item[ii)] Impose the constraints
\begin{equation}
D_{\mu} \mathcal{X} = D_{\mu} \mathcal{Y}_{a} = D_{\mu} {Z} = 0~,\quad a
\in \{\uparrow,\downarrow\} ,
\end{equation}
go to the standard gauge, eliminate all redundant gauge degrees of
freedom $\Gamma_{L/R , \mu} , G_{R ,\mu}^{1,2} , G_{B ,\mu}$, and
trade the spurions for the constant factors $\xi$, $\eta,$ and $\zeta$,
cf Eqs.~(\ref{spurionsa}, \ref{spurionsb}).
\item[iii)] Collect all invariants with the same infrared dimension
$d^{\star}$, Eq. (\ref{irdim}), into $\mathcal{L}_{d^{\star}}$ ,
$d^{\star} = 2 , 3 {\ldots} $. Associate with each independent
invariant a prefactor (Low Energy Constant (LEC)). The bare LECs are
in general infinite, their divergent parts can be computed using
dimensional regularization and they should cancel the divergences
arising from loops at {\bf the same order $d^{\star}$} according to
Eq.~(\ref{powercounting}). The renormalized LECs defined in this way
depend on the renormalization scale $\mu$. The sum of all terms of a
given $d^{\star}$ should be $\mu$ independent.
\item[iv)] At the scale $\mu \sim \Lambda_W$, the renormalized LECs are
expected to be of the order 1 (say, \mbox{$ 0.1 < \mathrm{LEC} < 10$}), 
unless the LEC carries an inverse power of the mass dimension. In the
latter case an additional suppression may occur and additional
physical input is needed to pin it down.
\end{itemize}

As already stated, the LO coincides with the Higgsless vertices of
the SM and the fermion mass term, the latter being of spurionic origin.
Loops, divergences, oblique corrections, corrections to universality, FCNC
etc only start at NNLO ($d^{\star} \ge 4$) in agreement with their observed 
smallness. On the other hand, New Physics is predicted to start at NLO, i.e. 
$d^{\star} = 3$. At this order there are {\bf only two new operators},
describing non standard couplings of fermions to standard gauge bosons
$W$ and $Z$. They are suppressed by spurion factors: they are of order
$\mathcal{O}(p^2\, \xi^2)$ and $\mathcal{O}(p^2\, \eta^2)$, respectively.
The observable effects of these non standard terms do not interfere with
non leading (loop) effects present in the SM.

For years it has been believed that the most important effects 
beyond the SM should be searched among oblique corrections (parameters
S, T, and U{\ldots}) whereas the non standard vertex corrections should be
tiny. The minimal not quite decoupling LEET does not bear out this wisdom
and predicts NLO modifications of fermion couplings. In the sequel of this
paper we discuss a systematic comparison of this prediction with experiment.

\section{Next-to leading order (NLO)}
\label{NLO}
Let us now specify the operators at next-to-leading order (NLO),
i.e. at the order $d^* = 3$.  These operators necessarily involve
spurions.  In contrast to the usual decoupling scenario, where there are
80 operators at NLO (mass dimension $D = 6$)~\cite{BW86}, only
two operators appear at NLO\footnote{In principle, fermion mass
terms counting as $d = 1, n_\xi + n_\eta=4$ exist, too. Since a discussion
of fermion mass hierarchy is beyond the scope of the present paper, we
will not consider them here.}. They count as $\mathcal{O}(p^2\,\xi^2)$
and $\mathcal{O}(p^2\, \eta^2)$, respectively, and represent
non-standard couplings of fermions to gauge bosons. For left-handed
fermions the unique such operator reads
\begin{equation}
\mathcal{O}_L = \bar \psi_L \mathcal{X}^{\dagger}\gamma^\mu \Sigma D_{\mu}\Sigma^{\dagger}\mathcal{X}\psi_L~,
\label{OpLNLO}
\end{equation}
whereas in the right-handed sector the corresponding operator has four
components (\mbox{$a,b \in\{ \uparrow,\downarrow\}$}) that are
separately invariant under $S_{\mathit{nat}}$
\begin{equation}
\mathcal{O}_R^{a,b} = \bar\psi_R \mathcal{Y}^{\dagger}_{a} \gamma^\mu \Sigma^{\dagger}
D_{\mu}\Sigma \mathcal{Y}_{b}\psi_R~.
\label{OpRNLO}
\end{equation}
Following the expansion scheme of the LEET, these two operators
represent the most important effects of physics beyond the Standard
Model. Oblique corrections only appear at NNLO (\mbox{$d^* = 4$}) and
so do loop corrections.

We will assume that, including LO and NLO effects, all flavor symmetry
breaking effects can be transformed from vertices to the fermion mass
matrix. It means that there exists a flavor symmetric basis in which
the couplings of fermions to gauge bosons are proportional to the unit
matrix in flavor space. This property is shared by many models with
minimal flavor violation~\cite{MFV}. Within the LEET it can be motivated
as follows. The LEET exhibits at LO an (accidental) flavor
symmetry. At NNLO loop-induced effects can break this symmetry. It
would appear  rather unnatural to introduce tree-level flavor symmetry
breaking effects via spurions at NLO (cf the discussion on that point
in Ref.~\cite{HS06}).

\subsection{Formulary}
\label{formulas}
The NLO Lagrangian reads:
\begin{equation}
\mathcal{L}_{\text{NLO}} = \rho_L \mathcal{O}_L(l) + \lambda_L \mathcal{O}_L(q)+
\sum_{a,b}  \rho_R^{a,b} \mathcal{O}^{a,b}_R(l) + \sum_{a,b}  \lambda_R^{a,b} \mathcal{O}_R^{a,b}(q)
\end{equation}
with $l,q$ representing leptons and quarks, respectively. As
discussed before, $\rho_{L/R}$, and $\lambda_{L/R}$ are order one
(unless suppressed by a symmetry) LECs. $\rho_R^{e,\nu}=0$
due to the presence of the discrete symmetry $Z_2$ introduced in
Eqs.~(\ref{z2a},\ref{z2}) which forbids the Dirac mass of the neutrinos.  In
the standard gauge ($s.g.$), see section~\ref{standardgauge}, we have (using
the notation of \cite{HS06})
\begin{eqnarray}
\mathrm{i} \Sigma^{\dag} D_{\mu} \Sigma & \overset{\text{s.g.}}{=}& \frac{e}{2cs} 
\left\{ Z_{\mu} \tau^3 
+ \sqrt{2} c \left( W_{\mu}^+ \tau^+ + W_{\mu}^- 
\tau^- \right) \right\},  
\label{Sigma}
\end{eqnarray}
where s and c are the sine and cosine of the Weinberg angle
\begin{equation} 
s =\frac{g'}{\sqrt{g^2 + g'^2}}~, \ \ c = \frac{g}{\sqrt{g^2 + g'^2}}~.
\end{equation}
It is convenient to write the explicit form of the operators appearing
in the Lagrangian in matrix notation with $\mathrm{U}= (u,c,t)^T, 
\mathrm{D}= (d,s,b)^T, 
\mathrm{N}= (\nu_e, \nu_\mu, \nu_\tau)^T, 
\mathrm{L}= (e, \mu, \tau)^T $. 
We then have 
\begin{eqnarray}
 \mathcal{O}_L(q) & \overset{\text{s.g.}}{=} & - \xi^2  \frac{e}{2 cs} 
  \left\{ \bar{U}_L \gamma^{\mu} Z_{\mu} U_L - \bar{D}_L
  \gamma^{\mu} Z_{\mu} D_L
+\sqrt{2} c \left( \bar{U}_L
  \gamma^{\mu} W^+_{\mu} D_L + \text{h.c.} \right) \right\} .  
\label{OLq}\\
 \mathcal{O}_R^{u,u}(q) & \overset{\text{s.g.}}{=} & \eta^2  \frac{e}{2 cs} 
   \bar{U}_R \gamma^{\mu} Z_{\mu} U_R~. 
\label{ORquu}\\
 \mathcal{O}_R^{d,d}(q) & \overset{\text{s.g.}}{=} &- \eta^2  \frac{e}{2 cs} 
  \bar{D}_R \gamma^{\mu} Z_{\mu} D_R~. 
\label{ORqdd}\\
 \mathcal{O}_R^{u,d}(q) & \overset{\text{s.g.}}{=} & \eta^2  \frac{e}{\sqrt{2}s} 
 \left( \bar{U}_R \gamma^{\mu} W^+_{\mu} D_R + \text{h.c.} \right).
\label{ORqud}
\end{eqnarray}
The operators for the leptons can be obtained by substituting
$U \longmapsto N$, $D \longmapsto L$. 

For convenience, we will now rewrite the Lagrangian up to NLO directly
in terms of effective couplings to the photon, to $Z$ and to $W$.  
Since the symmetry $U(1)_Q$ is unbroken, the coupling to the
photon is unchanged with respect to the SM and is given by
\begin{equation}
\mathcal{L}_{\gamma} = eJ^{\mu} A_{\mu}~.
\label{photon}
\end{equation}
The Lagrangian describing neutral current interactions reads
\begin{eqnarray}
\mathcal{L}_{\text{Z}} & = & \frac{e (\gf)}{2 c s} Z_\mu 
  \Big\{  \bar{\mathrm{N}}_L \gamma^{\mu} N_L 
+ \epsilon^\nu \bar{\mathrm{N}}_R \gamma^{\mu} N_R +
( -1 + 2 \tilde{s}^2 )
\bar{\mathrm{L}}_L \gamma^{\mu} L_L +
( - \epsilon^e + 2 \tilde{s}^2 )
\bar{\mathrm{L}}_R \gamma^{\mu} L_R  \nonumber\\
&& +  ( 1+\delta-\frac{4}{3}\tilde{s}^2)
\bar{\mathrm{U}}_L \gamma^{\mu} U_L + ( \epsilon^u -\frac{4}{3}\tilde{s}^2 )
\bar{\mathrm{U}}_R \gamma^{\mu} U_R  \nonumber\\
&& +  (-(1+\delta)
+\frac{2}{3} \tilde{s}^2 )
\bar{\mathrm{D}}_L \gamma^{\mu} D_L+
( -\epsilon^d+\frac{2}{3} \tilde{s}^2 )
\bar{\mathrm{D}}_R \gamma^{\mu} D_R \Big\}~,
\label{LagrZ}
\end{eqnarray}
and for the charged current we have
\begin{eqnarray}
\mathcal{L}_{\text{W}} &=& \frac{e (\gf)}{\sqrt{2} s}   
\left\{ \bar{\mathrm{N}}_L V_{\mathrm{MNS}} \gamma^{\mu} L_L 
+(1+\delta) 
\bar{\mathrm{U}}_L V_L \gamma^{\mu} D_L + 
\epsilon
\bar{\mathrm{U}}_R V_R \gamma^{\mu} D_R\right\}
W_{\mu}^+ + \text{h.c}~.\nonumber \\
\label{LagrW}
\end{eqnarray}
$V_{\mathrm{MNS}}$ is the mixing matrix in the lepton sector, and 
the two matrices $V_L$ and $V_R$ describe chiral quark flavor mixing. They
arise from the diagonalisation of the quark mass matrices:
\begin{eqnarray}
V_L&=& \Omega^U_L \Omega_L^{D\dagger} \\ \nonumber
V_R&=& \Omega^U_R \Omega_R^{D\dagger}~,
\end{eqnarray} 
where $\Omega^U_L$, $\Omega^U_R$, $\Omega^D_L$ and $\Omega^D_R$ denote
$U(3)$ transformations of $U_L$, $U_R$, $D_L$ and $D_R$, respectively,
to the mass eigenstate basis. In this basis the mass matrices are
diagonal and real. The two mixing matrices $V_L$ and $V_R$ are unitary
by construction. Within the present framework, chiral flavor mixing is
universal up to and including NLO.  Note that in
Eqs.~(\ref{LagrZ}), (\ref{LagrW}) we have factorized $\gf$, the factor
describing the universal modification of the coupling of left-handed
leptons. Defining
\begin{equation}
\sweff=\frac{s^2}{\gf}~,
\label{stilde}
\end{equation}
allows to absorb
the factor $\gf$ into the definition of $G_F$, see next section. 
The effective coupling parameters $\epsilon^i,\delta$ defined above
are then 
related to the spurion parameters and the LECs
$\rho_L,\rho_R,\lambda_L,\lambda_R$ in the following way:
\begin{equation}
(1+\delta)=\frac{1-\xi^2 \lambda_L}{1-\xi^2 \rho_L}~,
\label{delta}
\end{equation}
for the couplings of left-handed quarks and 
\begin{equation}
\epsilon =\frac{\eta^2 \lambda_R^{u,d}}{1-\xi^2 \rho_L}~\ \ ,\ \ \epsilon^{\nu}=\frac{\eta^2 \rho_R^{\nu,\nu}}{1-\xi^2 \rho_L}~\ \ ,\ \
\epsilon^{e}=\frac{\eta^2 \rho_R^{e,e}}{1-\xi^2 \rho_L}~\ \ ,\ \ \epsilon^{u}=\frac{\eta^2 \lambda_R^{u,u}}{1-\xi^2 \rho_L}, \ \, \ \
\epsilon^{d}=\frac{\eta^2 \lambda_R^{d,d}}{1-\xi^2 \rho_L}~,
\label{eps}
\end{equation}
for the couplings of right-handed fermions to $W$ (parameter
$\epsilon$) and $Z$ (parameter
$\epsilon^\nu,\epsilon^e,\epsilon^d,\epsilon^u$).

\subsection{Right-handed couplings and chiral flavor mixing}
At NLO, one major effect is the appearance of direct couplings of
right-handed quarks to $W$.  We thus have to generalize flavor CKM
mixing to include the mixing of right-handed quarks, too.  Indeed,
the charged current interaction (see Eq.~(\ref{LagrW})) contains two
mixing matrices, $V_L$ and $V_R$.  For $n$ families, $V_L$ and $V_R$
are $n \times n$ unitary matrices.  Together they contain $n(n-1)$
angles and $n(n+1)$ phases.  By a redefinition of the quark fields we
can eliminate, as in the case of the SM, $2n-1$ phases. The total
number of independent phases is thus $n^2-n+1$. Their assignment to
$V_L$ and $V_R$ is somewhat arbitrary.  A convenient choice is to use
the freedom of redefining the quark fields to eliminate the
maximum number of phases from $V_L$. Then $V_L$ will have the same
structure as the usual CKM mixing matrix for left-handed quarks in the
SM. The number of CP-violating phases $N_{L/R}$ from $V_L$ and $V_R$,
respectively, is then
\begin{eqnarray}
N_L&=&\frac{(n-1)(n-2)}{2}~, \\ \nonumber
N_R&=&\frac{n(n+1)}{2}~.
\end{eqnarray} 
For three generations, we will have six additional phases compared
with the case without direct coupling of right-handed quarks to $W$.
This generates new CP-violation effects as for instance in electric
dipole moments. The determination of the CP-violating phases is a
subject by itself and beyond the scope of this paper.

Even the analysis of CP-conserving charged current processes at NLO
cannot be reduced to the genuine spurionic parameters $\delta$ and
$\epsilon$, see Eqs.~(\ref{delta}) and (\ref{eps}), but involves in
addition unknown mixing angles for left-handed and right-handed
quarks. For the comparison with experiment, it is convenient for the
following analysis to introduce in Eq.~(\ref{LagrW}) effective vector and
axial-vector couplings as:
\begin{eqnarray}
\veff^{ij}&=&(1+\delta) V_L^{ij}+\epsilon V_R^{ij}+\mathrm{NNLO}~,\nonumber\\
\aeff^{ij}&=&-(1+\delta) V_L^{ij}+\epsilon V_R^{ij}+\mathrm{NNLO}~.
\label{effectivecouplings}
\end{eqnarray}
It is obvious that at NLO, due to the direct coupling of right-handed
quarks to $W$, we have $\veff \not = -\aeff$. At this order, there is
no reason for $\veff$ or $\aeff$ to be unitary. We will return to
this point in Sec.~\ref{sec:chimix}. We should stress that $V_L$ and
$V_R$ are completely general here. In particular, we do not assume, as
is often the case in left-right symmetric models, a (pseudo)-manifest
left-right symmetry which would suggest an alignment of $V_L$ and
$V_R$. It has already been pointed out in Ref.~\cite{Langacker:1989xa}
that allowing for a more general form of $V_R$, much of the stringent
constraints on left-right symmetric models can be released.
\subsection{${\bm {G_F}, \bm{M_W^2}}$}
\label{gfmw}
As it is often done in the SM we will relate the fundamental couplings
of the theory $g$ and $g'$ to the fine structure constant $\alpha$
and the life time of the muon, two quantities which are 
measured to a very high degree of accuracy. One has
\begin{equation}
\alpha(0)=e^2/(4 \pi)~.
\end{equation}
At next to leading order the   
Fermi constant as determined by the muon life time is given by:
\begin{equation}
\frac{G_F}{\sqrt 2} = \frac{4 \pi \alpha (0)}{8 m_Z^2 c^2 s^2 (1-\Delta r)}
(1- \xi^2 \rho_L)^2~. \label{gf}
\end{equation}
Two remarks are in order concerning this equation. First,
loop corrections are in principle appearing first at NNLO. However
the LEET is essentially an expansion for the weak part of the theory. 
Consequently we have kept in Eq.~(\ref{gf}) the electromagnetic loop 
corrections, $\Delta r=\Delta \alpha$ which describe the running of the 
electromagnetic coupling $\alpha$. Second, the spurion contribution 
and loop corrections modify the LO result in exactly the same way.  

As already mentioned, see Eq.~(\ref{gf}), writing any observables in terms of
the Fermi constant as it is done in the SM will absorb the factor $1 -
\xi^2 \rho_L$ appearing in Eq.~(\ref{LagrW}). It is thus not possible
to determine this quantity from charged currents.  However this
quantity appears both in the coupling to Z through $\sweff$ and in
the expression for $G_F$. Having fitted $\sweff$ to the available
data as discussed in the next section one can solve a system of two
equations with two unknowns. This leads to the relations
\begin{equation}
\frac{m_W^2}{m_Z^2}= \frac{h}{h+\tilde s^4} \, , \quad h=\frac {\pi \alpha(0)}
{\sqrt 2 G_F m_Z^2 (1-\Delta r)}
\label{mweq}
\end{equation}
and
\begin{equation}
1 - \xi^2 \rho_L= \frac{\tilde s^2}{h+\tilde s^4}
\label{xi2}
\end{equation}
where use has been made of Eq.~(\ref{gf}) and (\ref{stilde}) as well
as of the on-shell relation
\begin{equation}
s^2 =1-m_W^2/m_Z^2 .
\label{s2onshell}
\end{equation}
Note that the expression for $m_W$ is the same at tree level and at
NLO. However its value differs since at LO $\xi^2 \rho_L=0$ which is
not necessarily the case at NLO. In the following we will use $G_F=1.16637
\, 10^{-5}~\mathrm{GeV}^{-2}$, the canonical value
$\alpha(0)=1/137.03599911$, $m_Z=91.1891$GeV as given in the
PDG~\cite{PDG06} and $\Delta r=0.059$. One thus gets $h=0.1776$ and
one has at LO $\sweff =0.2309$ and $m_W=79.97$ GeV. We will come back
to these values in section~\ref{zpole}.
\section{Couplings to ${\bm Z}$}
\label{Zanalysis}
For the neutral current interaction, there are many accurate
measurements available, in particular the huge amount of precise data
from the experiments at LEP, LEP2 and SLC at energies around the $Z$
resonance and even above. The latter data are parametrized in terms of
effective observables (couplings, masses) including QED and QCD
radiative effects, see Ref.~\cite{Zmeasure} for a thorough discussion
of the definition of these effective ``pseudo-observables''. It is
understood that we will use these effective quantities throughout the
following discussion. Following Ref.~\cite{Zmeasure} we denote these
pseudo-observables by a superscript ``0''.

Altogether six unknown parameters, defined in Eq.~(\ref{delta}) and
Eq.~(\ref{eps}), appear in the couplings to $Z$: $\epsilon^e$,
$\epsilon^{\nu}$, $\epsilon^u$, $\epsilon^d$, $\tilde{s}^2$ and
$\delta$.  In contrast to the charged current where the additional
$Z_2$ symmetry of the LEET in the neutrino sector forbids a coupling
of right-handed leptons to $W$, we have to consider here non-standard
leptonic as well as non-standard quark couplings.

The parameter $\epsilon^\nu$, the coupling of right-handed neutrinos
$\nu_R$ to $Z$, cannot be determined at NLO from the asymmetry
measurements. Since $\nu_R$ are expected to be light enough to be
pair-produced in $Z$ decays, $\epsilon^\nu$ enters the invisible width
of $Z$~\cite{HS06}. But, as $\nu_R$ have no $SU(2)_L\times U(1)$
charge, their coupling to $Z$ always contains powers of
spurions. Therefore there is no interference with the SM
contribution. This implies that $\epsilon^\nu$ appears only
quadratically in this invisible width, being thus additionally suppressed.
The experimental constraints are weak: $\epsilon^\nu$ should
be roughly of the order 0.1 or smaller for the contribution of $\nu_R$
to the invisible width to become smaller than the experimental
uncertainty~\cite{HS06}. At present we cannot determine its value more
precisely.

In the next section we will discuss fits of the remaining five
parameters to $Z$ pole observables.  In addition to the data at the
$Z$ resonance, there are several measurements at lower momentum
transfer which are often included in precision tests of the standard
model. We will comment on some of these measurements in section~\ref{complnc}.

\subsection{Fit to ${\bm Z}$ pole observables}
\label{zpole}
In order to determine the unknown NLO parameters from the available
data, we perform two different fits. For the ``restricted fit'' we
take the following observables related to the $Z$~(see
Ref.~\cite{Zmeasure}, table 8.4): the total width of the $Z$,
$\Gamma_Z$, the hadronic pole cross section $\sigma_h^0$, the ratios
$R_l^0 = \Gamma_h/\Gamma_l$, $R_b^0 = \Gamma_b/\Gamma_h$, $R_c^0 =
\Gamma_c/\Gamma_h,$ the three asymmetries $A_{FB}^{0,l}$,
$A_{FB}^{0,b}$, $A_{FB}^{0,c}$, and $\mathcal{A}_l (P_\tau)$. In the
``full fit'' we include in addition the direct measurement of
$\mathcal{A}_b,\mathcal{A}_c,$ and $\mathcal{A}_e$ from SLD. For the
expression of these observables in terms of the effective couplings
that appear at NLO, see appendix~\ref{zexp}. Note that, since we
assume universal non-standard couplings, we have the following
relation up to small radiative corrections $3\, R_b^0 + 2 R_c^0 =
1$. The way this is obeyed in the fit merely tests the universality
and the radiative corrections. As the parameter $\delta$ appears both
in the couplings to $Z$ and to $W$, we included the leptonic branching
fraction of the $W$ into the fits~\cite{LEPEWWG}. It is sensitive to a
correction induced by the parameter $\delta$ to the hadronic width of
$W$, see section~\ref{wwidth}.  Alternatively, we could have
considered the data for the total width of the $W$.  However, the
assigned experimental error induces a very large error on $\delta$ and
we did not consider the total width here.

\TABLE[h]{
\begin{tabular}{|c||c||c c c c c|}
\hline & Parameter&  & & Correlations & &\\ 

& &  $(\epsilon^e)_{\mathrm{NLO}}$ & $(\epsilon^u)_{\mathrm{NLO}}$ &$(\epsilon^d)_{\mathrm{NLO}}$& $(\sweff)_{\mathrm{NLO}}$& $(\delta)_{\mathrm{NLO}}$ \\ \hline \hline
$(\epsilon^e)_{\mathrm{NLO}}$ &-0.0024(5)& 1.00 & & & & \\ 
$(\epsilon^u)_{\mathrm{NLO}}$ & -0.02(1) & -0.07 & 1.00 & & & \\
$(\epsilon^d)_{\mathrm{NLO}}$ & -0.03(2) & -0.10 & 0.60 & 1.00 & &\\
$(\sweff)_{\mathrm{NLO}}$ & 0.2309(3) & 0.25 & 0.44 & 0.63 & 1.00& \\ 
$(\delta)_{\mathrm{NLO}}$ & -0.005(4) & -0.16 & 0.83 & 0.94 & 0.61 & 1.00\\ \hline
$\chi^2/d.o.f$ & 3.1/5 &&&& & \\\hline
\end{tabular}
\caption{\it \label{ncresultsrest}
Result of the restricted fit (see text) to $Z$ pole data at NLO for
  $\alpha_s(m_Z) = 0.1190$.}
}
\TABLE[h]{
\begin{tabular}{|c||c||c c c c c|}
\hline & Parameter&  & & Correlations & &\\ 

& &  $(\epsilon^e)_{\mathrm{NLO}}$ & $(\epsilon^u)_{\mathrm{NLO}}$ &$(\epsilon^d)_{\mathrm{NLO}}$& $(\sweff)_{\mathrm{NLO}}$& $(\delta)_{\mathrm{NLO}}$ \\ \hline \hline
$(\epsilon^e)_{\mathrm{NLO}}$ &-0.0024(5)& 1.00 & & & & \\ 
$(\epsilon^u)_{\mathrm{NLO}}$ & -0.02(1) & 0.0 & 1.00 & & & \\
$(\epsilon^d)_{\mathrm{NLO}}$ & -0.03(1) & -0.02 & 0.35 & 1.00 & &\\
$(\sweff)_{\mathrm{NLO}}$ & 0.2307(2) & 0.49 & 0.19 & 0.37 & 1.00& \\ 
$(\delta)_{\mathrm{NLO}}$ & -0.004(2) & -0.11 & 0.73 & 0.87 & 0.33 & 1.00\\ \hline
$\chi^2/d.o.f$ & 8.5/8 &&&& & \\\hline
\end{tabular}
\caption{\it \label{ncresultsfull}
Result of the full fit (see text) to $Z$ pole data at NLO for
  $\alpha_s(m_Z) = 0.1190$.}
}
We stated above
that the main QED and QCD corrections are already included
into the definition of pseudo-observables, such that the quantities
assigned with a superscript ``0'' correspond to ``bare'' EW
quantities. This is not entirely the case for the partial widths 
$\Gamma_f$
for the Z decay into a
fermion$f\bar f$ pair.
They have been defined in such a way that they add up to the
total width. This means that they contain factors
$R_A^f,R_V^f$ describing the residual QED/QCD effects, see Ref.~\cite{qcdqed}:
\begin{equation}
\Gamma_f = 4 N_c^f \Gamma_0 \Big((g_V^f)^2 R_V^f + (g_A^f)^2 R_A^f\Big)~.
\label{gammapartialz}
\end{equation}
In this equation,
$N_c^f=1$ for leptons and $N_c^f = 3$ for
quarks,  $\Gamma_0 = (G_F m_z^3)/(24 \sqrt{2}\pi)$ and $g_V^f,g_A^f$
are the effective EW vector and axial couplings, respectively.
Note that, since within the expansion scheme of the LEET weak loops
appear first at NNLO, we 
did not include any weak loop corrections in our NLO analysis (see the
discussion on that point in Sec.~\ref{gfmw}, too).

The hadronic pole cross section, $\sigma^0_h$ and the
total width $\Gamma_Z$ of $Z$ are rather sensitive to the value of
$\alpha_s$. This allows to determine within the SM the value of
$\alpha_s(m_Z) = 0.1190(27)$ rather precisely~\cite{Zmeasure}. In our
case it turns out that the sensitivity of the fit to $\alpha_s$ is
considerably reduced due to the non-standard EW parameters. It is thus
not possible to determine simultaneously $\alpha_s$ and the additional
EW parameters in a reliable way. In the following we use $\alpha_s(m_z) =
0.1190$.

Let us first consider the LO results, 
$ \xi^2 \rho_L=\epsilon^i = \delta = 0$.
$\sweff = s^2$ is fixed by the on-shell relation, see
Eq.~(\ref{s2onshell}) and we get
$(m_W)_{LO}= 79.97~{\rm GeV}$, see the discussion in
section~\ref{gfmw}. Figure ~\ref{figpullLO} show the pulls
\begin{equation}
\frac{(|O^{\mathit{meas}} -
  O^{\mathit{fit}}|)}{\sigma^{\mathit{meas}}}~,
\end{equation}
where $O^{\mathit{meas/fit}}$ is the measured/fitted value for a
observable and $\sigma^{\mathit{meas}}$ denotes the corresponding
experimental error. Note that at LO all parameters are fixed and
$O^{\mathit{fit}}$ is determined from the calculation~\footnote{In
  principle, the value of $\alpha_s$ can be varied for the LO fit. In
  order to better compare with the NLO result, we decided to keep
  $\alpha_s(m_Z) = 0.1190$ fixed for the LO result, too. The value of
  $\alpha_s$ has no influence on $A_{FB}^{0,b}$, whereas the agreement
  with $\Gamma_Z$ can be slightly improved allowing $\alpha_s$ to
  vary.}. One obtains a good agreement except for $\Gamma_Z$ and
$A_{FB}^{0,b}$ which are not well reproduced at this order. 

Let us turn to the NLO calculation. Tables \ref{ncresultsrest} and
\ref{ncresultsfull} give the values of our parameters for the
restricted and the the full fit, respectively.  $\sweff$ is, together
with $\epsilon^e$, severely constrained by the electron
pseudo-observables due to the fact that the vector effective coupling
of the electron $g^e_V$ is very small.  This coupling being well
described already at LO, $\sweff$ is nearly unchanged at NLO and
$(\epsilon^e)_{NLO}$ is very small.  All values for our parameters are
nicely of the order expected from the LEET. Note that $\epsilon_u
\approx \epsilon_d$ and $\delta \ll \epsilon_u, \epsilon_d$, see the
definitions, Eqs.~(\ref{delta}, \ref{eps}). At this stage it is,
however, too early to draw any conclusions on the size of the
constants $\eta, \xi$ and of the LECs.
\FIGURE[h]{
\includegraphics*[scale=0.4]{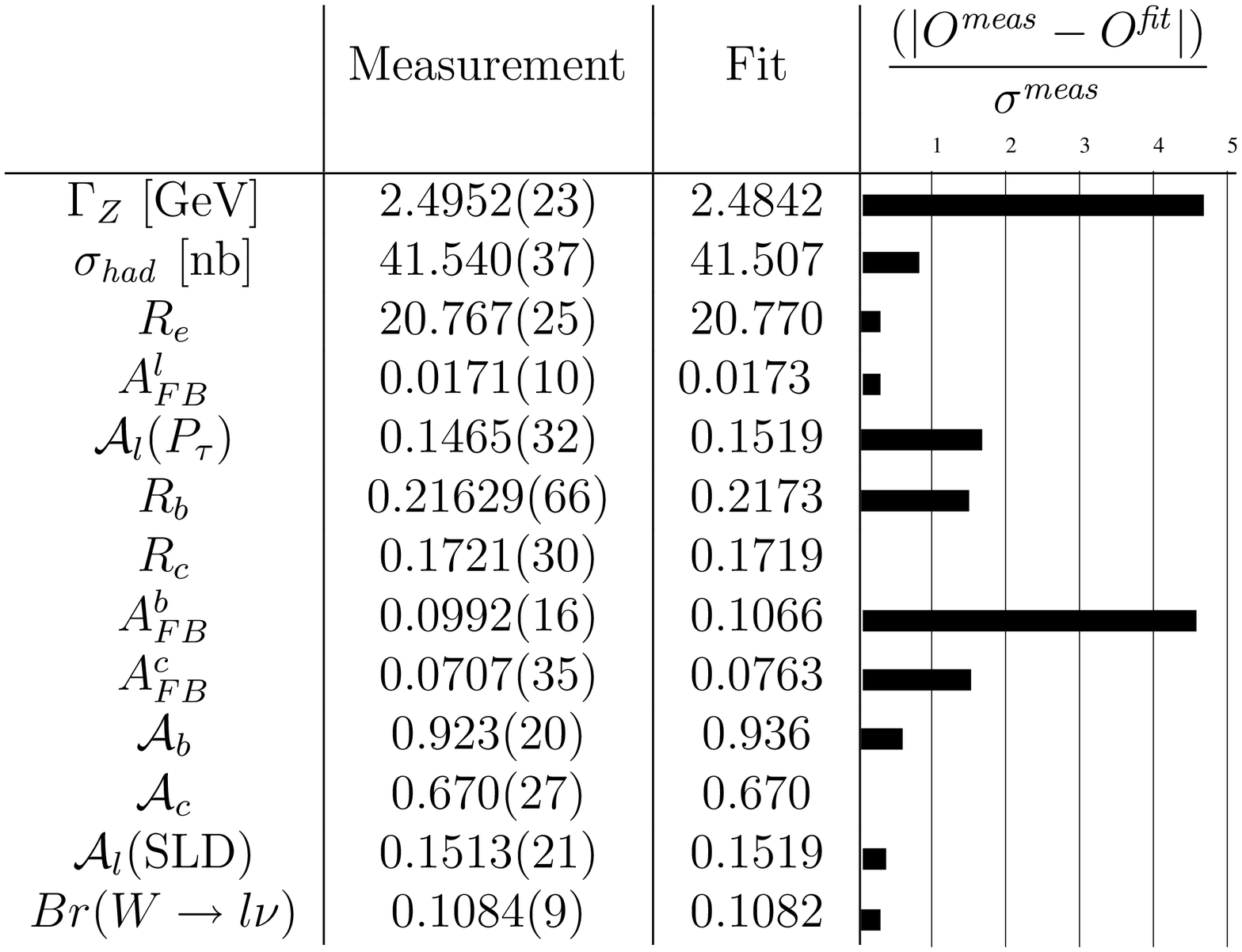}
\caption{\it Pull for
  the $Z$ pole observables at LO.}
\label{figpullLO}
} 
\FIGURE[h]{
\epsfig{width=7.5cm,file=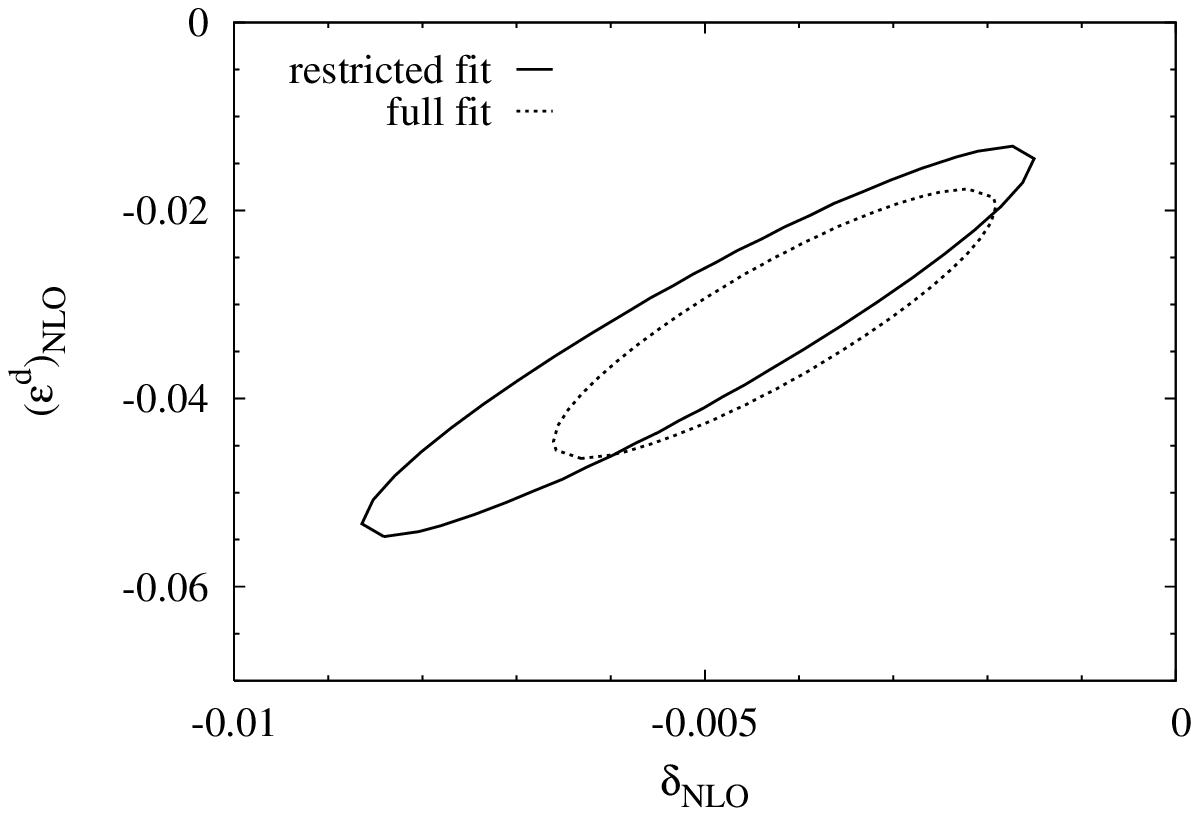}\hfill
\epsfig{width=7.5cm,file=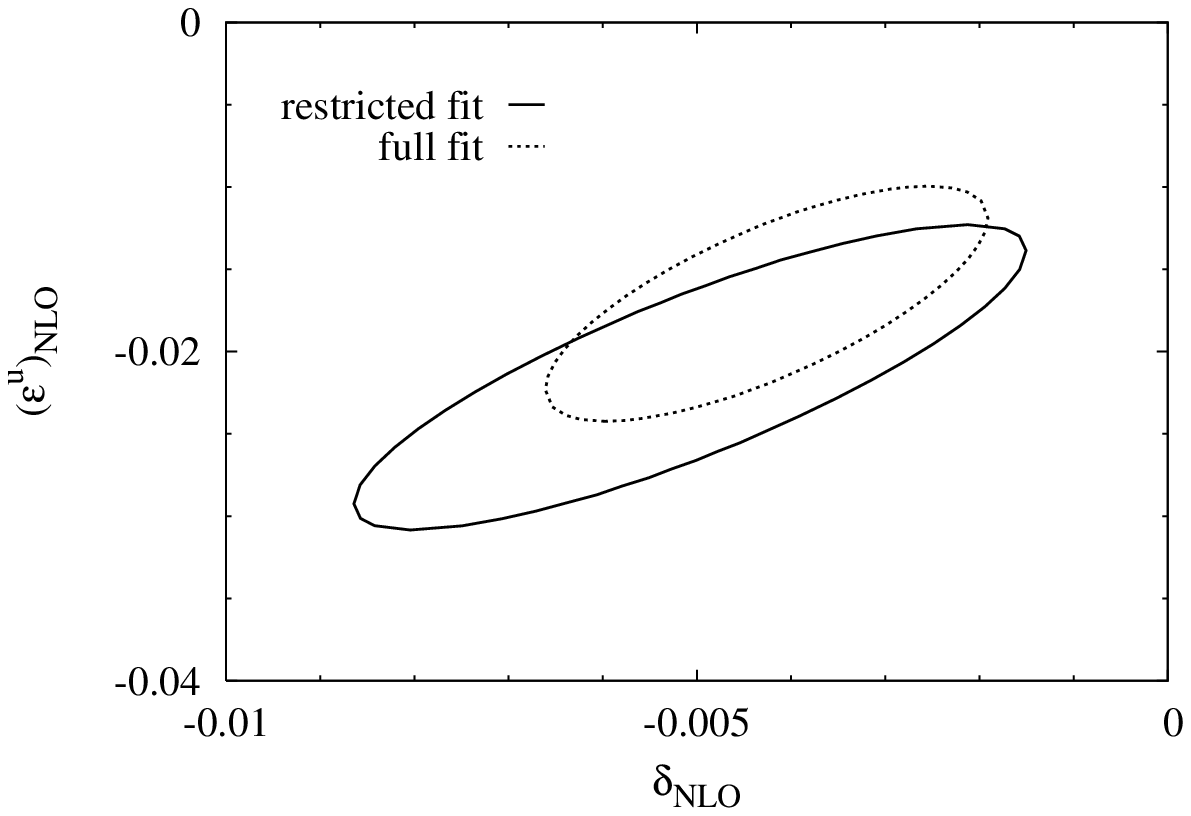}
\caption{\it 1$\sigma$ contours for the parameters $\epsilon^d$ (left
  panel) and $\epsilon^u$ (right panel) versus $\delta$ for the two
  different fits to $Z$ pole data with $\alpha_s(m_Z) = 0.1190$, see
  Table~\ref{ncresultsrest} 
and Table~\ref{ncresultsfull}.  
}
\label{figellipsenc1}
} 
\FIGURE[h]{
\epsfig{width=7.5cm,file=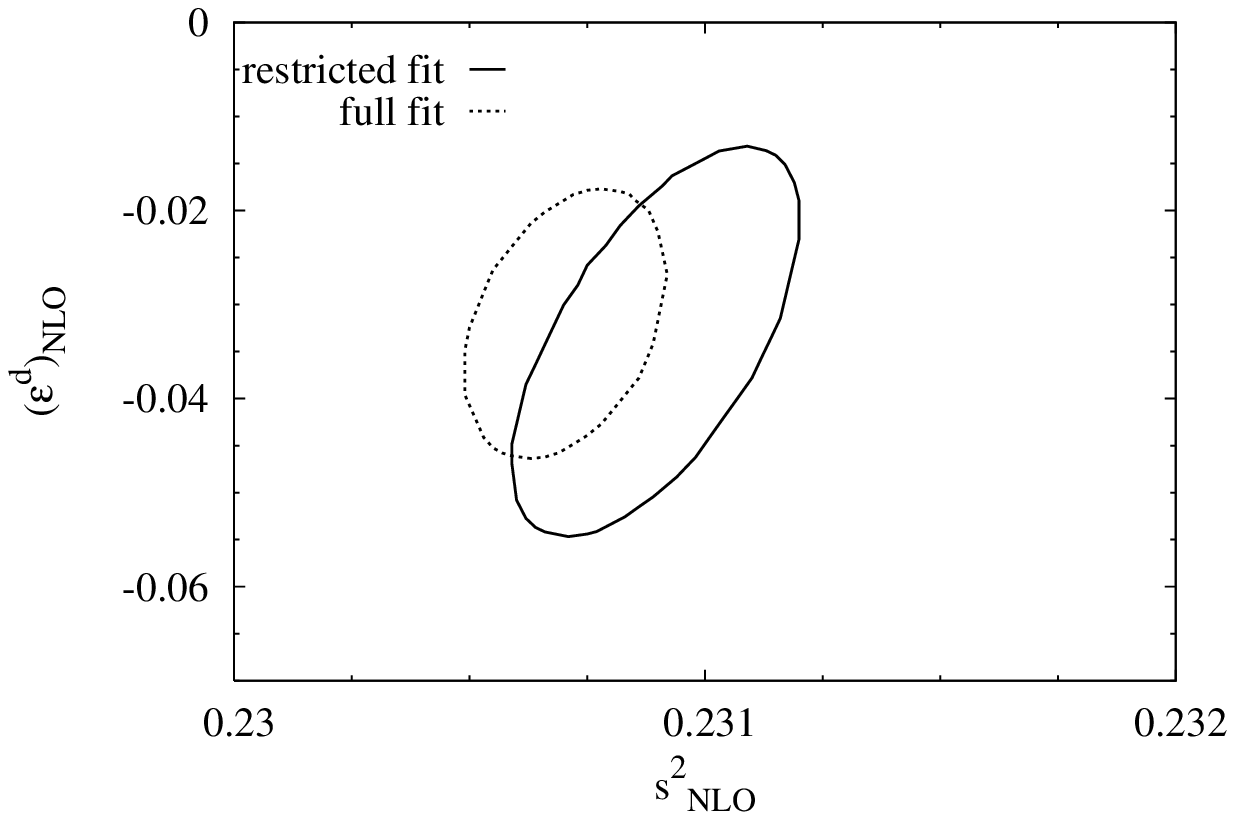}\hfill
\epsfig{width=7.5cm,file=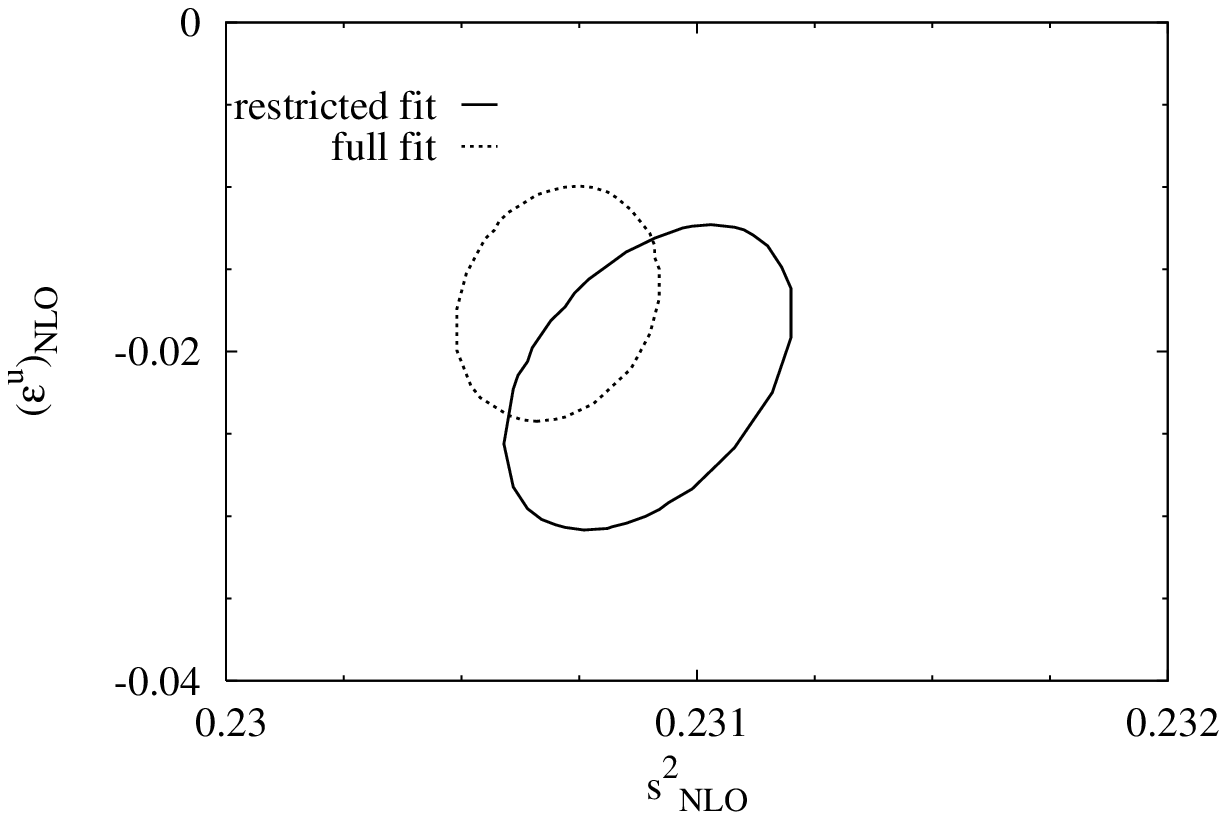}
\caption{\it 1$\sigma$ contours for the parameters $\epsilon^d$ (left
  panel) and $\epsilon^u$ (right panel) versus $\sweff$ for the two
  different fits to $Z$ pole data with $\alpha_s(m_Z) = 0.1190$, see
  Table~\ref{ncresultsrest} and Table~\ref{ncresultsfull}.  }
\label{figellipsenc2}
} 
\FIGURE[h]{
\epsfig{width=7.5cm,file=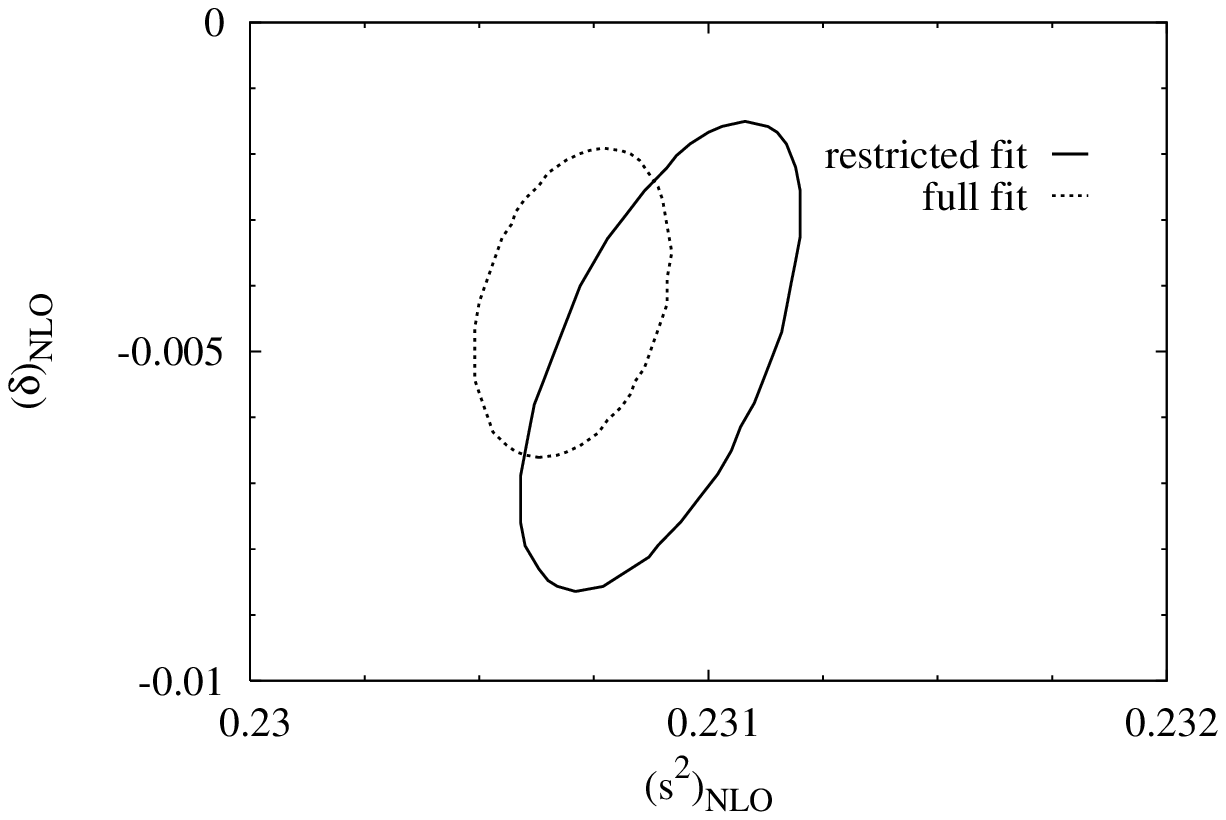}\hfill
\epsfig{width=7.5cm,file=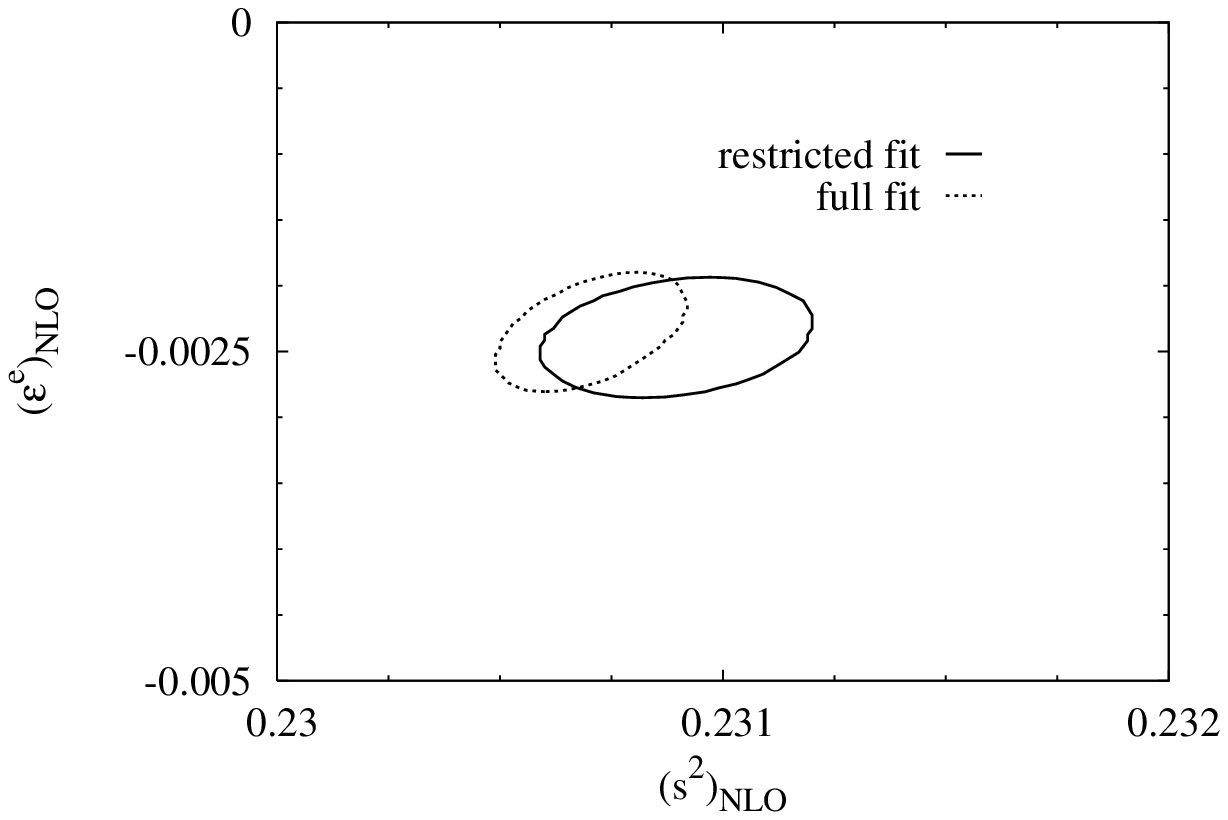}
\caption{\it 1$\sigma$ contours for the parameters $\delta$ (left
  panel) and $\epsilon^e$ (right panel) versus $\sweff$ for the two
  different fits to $Z$ pole data with $\alpha_s(m_Z) = 0.1190$, see
  Table~\ref{ncresultsrest} and Table~\ref{ncresultsfull}.  }
\label{figellipsenc3}
} 
\FIGURE[h]{
\includegraphics*[scale=0.6]{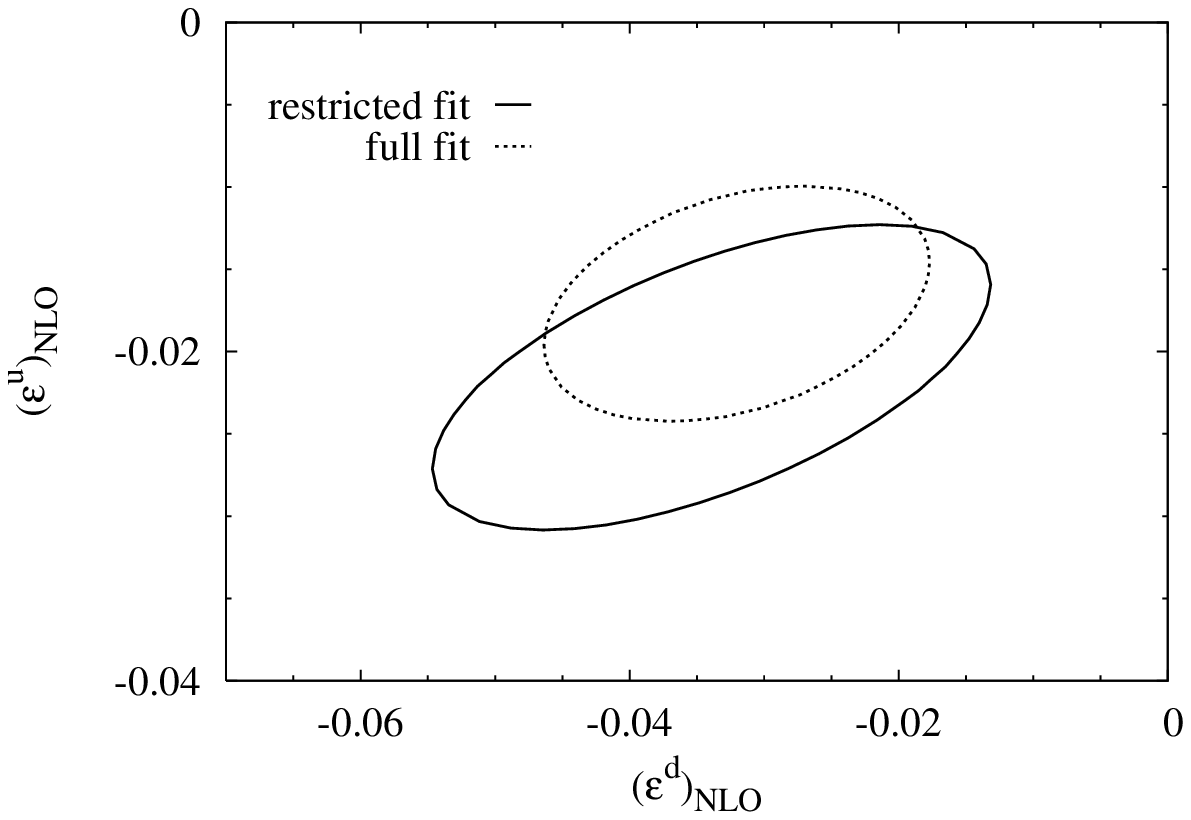}
\caption{\it 1$\sigma$ contours for the parameters $\epsilon^d$ versus 
$\epsilon^u$ for the two
  different fits to $Z$ pole data with $\alpha_s(m_Z) = 0.1190$, see
  Table~\ref{ncresultsrest} and Table~\ref{ncresultsfull}.  }
\label{figellipsenc4}
} 
\FIGURE[h]{
\includegraphics*[scale=0.4]{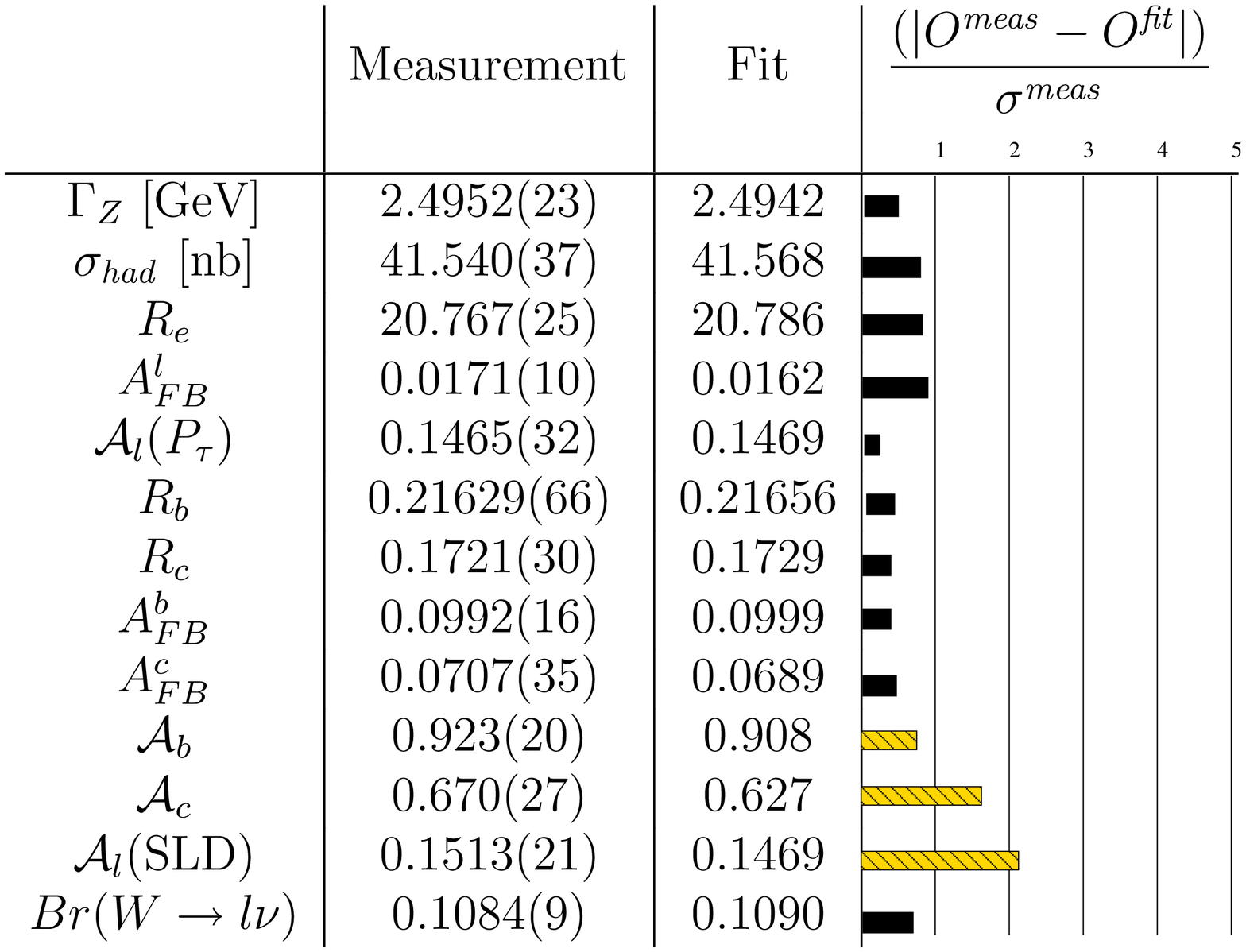}
\caption{\it Pull for the $Z$ pole observables in the restricted fit
  (see Table~\ref{ncresultsrest}). The pull for the quantities
  measured at SLD (which are not included in the fit) is shown in yellow
  (hatched light gray).}
\label{figpullNLOLEP}
} 
\FIGURE[h]{
\includegraphics*[scale=0.4]{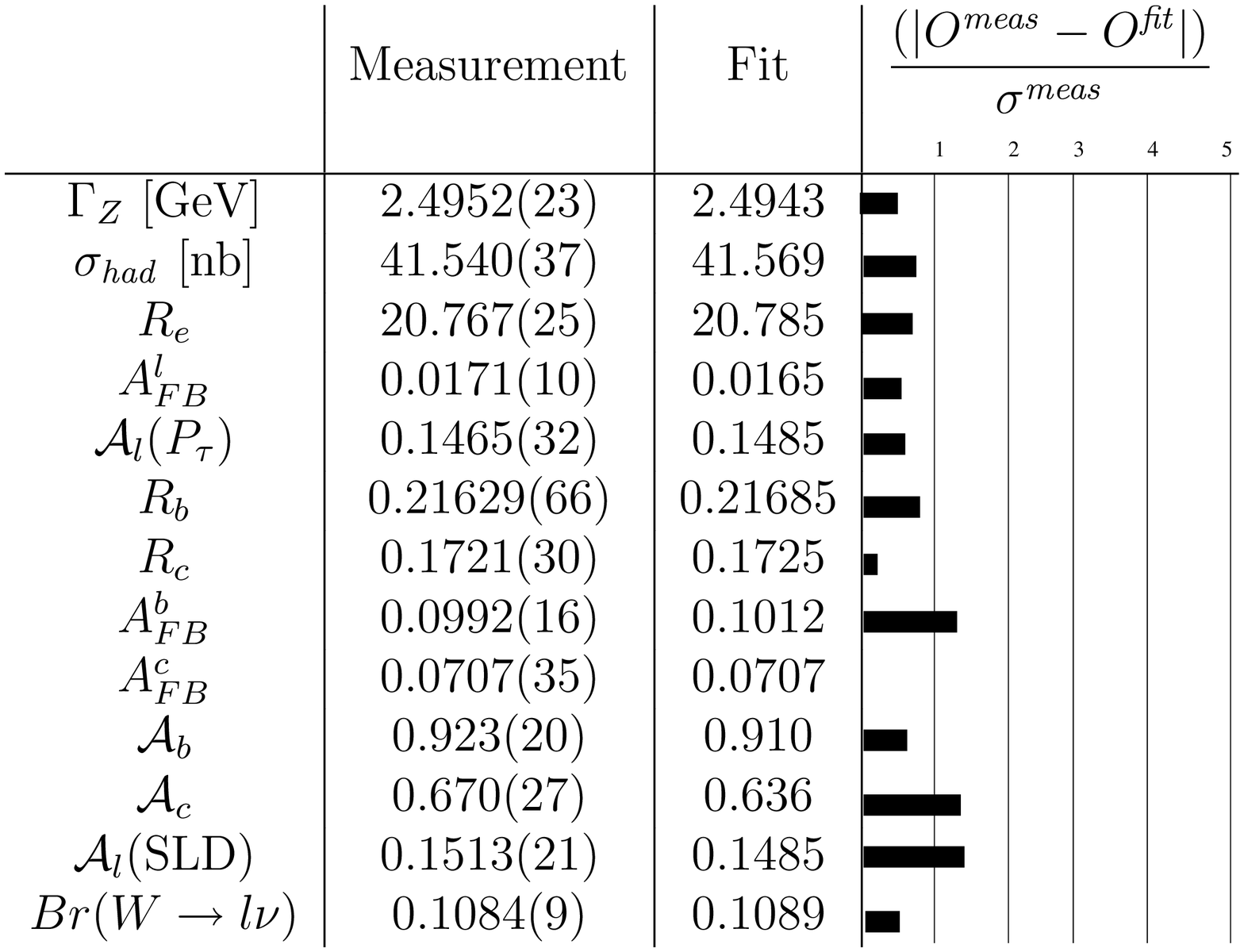}
\caption{\it Pull for
  the $Z$ pole observables in the full fit (see
  Table~\ref{ncresultsfull}). }
\label{figpullNLO}
} 

Varying $\alpha_s(m_Z)$ between $0.115$ and $0.125$, which should be
reasonable values, we observe only very little effect on the fit
result. For the restricted fit, for instance, the central value for
the parameter $\delta$, which is the most sensitive to $\alpha_s(m_Z)$,
varies between $-0.005(4)$ and $-0.006(4)$. The $\chi^2$ is not
considerably modified, neither. It is $3.0/5\, d.o.f$ for
$\alpha_s(m_z) = 0.125$ and $3.2/5\, d.o.f$ for $\alpha_s(m_Z) =
0.115$. It should be noted that the result for $\delta$ is rather
sensitive to the experimental value for the leptonic branching
fraction of $W$. For the present fit we have taken the latter value
from LEP data~\cite{LEPEWWG}.  To illustrate the correlations between
the different couplings we show the $1\sigma$ ellipses for the most
correlated combinations of parameters in
Figs.~\ref{figellipsenc1}, \ref{figellipsenc2}, \ref{figellipsenc3}, 
\ref{figellipsenc4}. The error thereby reflects only the experimental
error. For the presented result $\alpha_s(m_z) = 0.1190$.

The good agreement of our fit with the data can be seen from
Figs.~\ref{figpullNLOLEP} and \ref{figpullNLO}. 
$\Gamma_Z$ and
$A_{FB}^{0,b}$ are  much better reproduced at NLO compared with the LO
result.
Remarkably we can reproduce
simultaneously the data for $A_{FB}^{0,b}$ and $R_b^0$ as well as
$A_{FB}^{0,c}$ and $R_c^0$. This is not straightforward.  In the
recent literature the ``$A_{FB}^b$ puzzle'' has been intensively
discussed and many models have been proposed providing solutions to
this problem (see for instance~\cite{afbpuzzle}). Mostly, the proposed solution
is based on a modification of the couplings of the third generation.
Note that we do not need any non-universal couplings here to well
reproduce the ratio $R_b^0$ and the asymmetry. 
In our theory the effect is
mainly due to (universal) non-standard right-handed couplings. 

Clearly the values of our parameters will be modified at NNLO. Since
we performed a fit they, in fact, implicitly contain at NLO higher
order corrections. How big these are is hard to determine without
doing the calculation.  In our framework the Higgs particle is clearly
absent from the loops and furthermore counter terms have to be added to
the loop contributions. Indeed only the sum of loops plus counter terms
is meaningful within our effective theory. However, if the expansion
is convergent the higher order corrections should be
small. Furthermore, from the fact that loops and our spurion
contributions enter the observables in a similar way, we do not expect
from a NNLO calculation a change in the nice agreement between the
calculated and experimental values of the pseudo-observables.

Having obtained the value of $\sweff$, $m_W$ and $\xi^2 \rho_L$ can be
determined from Eqs.~(\ref{mweq}), (\ref{xi2}). Since $(\sweff)_{NLO}
\sim (\sweff)_{LO}$ the difference between $(m_W)_{\mathrm{NLO}} =
79.97(11)$ GeV (restricted fit), $(m_W)_{\mathrm{NLO}} = 79.99(07)$
GeV (full fit), and $(m_W)_{\mathrm{LO}} = 79.97$ GeV is extremely
small and $(\xi^2 \rho_L)_{\mathrm{NLO}} =0.000(18)$ (restricted fit),
$(\xi^2 \rho_L)_{\mathrm{NLO}} =0.001(12)$ (full fit). This result
contains the electromagnetic loop corrections in the spirit of our
discussion of the way we do the expansion, see section~\ref{gfmw}.  In
this case the NLO corrections are thus extremely small.  Clearly
corrections at NNLO have to be evaluated.  First, $m_W$ does not
receive direct corrections at NLO, it is only modified indirectly via
the factor $1-\xi^2\rho_L$ from the redefinition of $G_F$, whereas at
NNLO direct corrections to $m_W$ will appear, too. Also, higher order
corrections could be important due to numerically large factors of
$(m_t/m_W)^2$.  At NNLO the quantity $\Delta r$ and the value of
$\sweff$ will be modified. One can evaluate the value of $\Delta r_w$,
the quantity which has to be added to $\Delta r$ such that the
physical mass of $W$ is reproduced keeping $\sweff$ fixed.  We obtain
$\Delta r_w =0.046$ (where the subscript $w$ means that they
correspond to the weak contributions) of the size of the
electromagnetic corrections $\Delta \alpha= 0.059$. In the standard
model $\Delta r_w =-0.0242$ and loop corrections to $s_W$ are of the
order of 0.04. Hence the value of $0.046$ is of the expected size. We
will see below another example of a quantity which is accidentally
small at NLO.
\subsection{Low energy observables}
\label{complnc}
Several experiments provide data
at energies below the $Z$ pole. They could  in principle give complementary
information on the couplings. We did not include these observables
into our fit for two reasons. First, in general the energies involved
in these experiments are much smaller than $m_Z$ such that the
couplings involved are probed at a different energy
scale. Second, in some cases the uncertainties are too
large to detect non-standard effects on the percent level. Having fitted our 
parameters we can calculate some of these observables and compare with the experimental data. 
\subsubsection{Atomic parity violation}
Measurements of atomic parity violation probe the coupling of
electrons to the quarks inside the nucleus via the neutral
current. The parity violating part of the amplitude has two
contributions, one from an axial coupling to electrons and a vector
coupling to quarks ($A_e V_q$), and another one from a vector coupling to
electrons and axial coupling to quarks ($V_e A_q$). In order to keep hadronic
uncertainties small, it is preferable to probe vector couplings for
quarks. In this case, due to the conservation of the vector
current, the hadronic matrix elements can be reliably
predicted. The relevant part of the effective four-fermion lepton/quark
interaction Lagrangian is
\begin{equation}
\mathcal{L}_{Z}^{lq} = -\frac{G_F}{\sqrt{2}} \, 4 g_A^e \bar{e} \gamma_\mu
\gamma^5 e \Big(g_V^u \bar{u} \gamma^\mu u + g_V^d \bar{d}\gamma^\mu
d\Big)~. 
\end{equation}
This allows to define the weak charge 
\begin{equation}
Q_W =  - 4 g_A^e \Big( Z \, (2 g_V^u + g_V^d) + N\, (g_V^u + 2 g_V^d) \Big)~,
\label{qw}
\end{equation}
where $Z$ and $N$ denote here the number of protons and neutrons in
the nucleus, respectively. 
We kept here explicitly the dependence on the axial coupling of the
electron since this makes it easier to distinguish non-standard quark
and electron couplings, respectively. The usually defined effective
four-fermion couplings are simply
given by $C_{1q} = 4 g_V^q g_A^e$. At NLO (see
Eqs.~(\ref{gvs}, \ref{gas}) for the couplings) 
 the weak charge is given by:
\begin{equation}
Q_W = (1 - \epsilon^e) \Big( Z (1-4 \sweff + \delta- \epsilon^d+2\,
\epsilon^u) - N (1 + \delta+2\, \epsilon^d- \epsilon^u)\Big)~.
\label{qweak}
\end{equation}
Up to now the most precise measurements are those on $^{133}$Cs
atoms~\cite{PNCexp}.  Spin-dependent measurements allow to eliminate
the small contribution from axial couplings to the nucleus, such that
the result should be relatively reliable. Inserting the values for the
parameters (cf Tables~\ref{ncresultsrest} and \ref{ncresultsfull})
discussed above into Eq.~(\ref{qweak}), we obtain, respectively,
$(Q_W(^{133}\mathrm{Cs}))_{\mathrm{NLO}} = - 70.72 \pm 3.72$
(restricted fit) and $(Q_W(^{133}\mathrm{Cs}))_{\mathrm{NLO}} =
-70.72\pm 4.19$ (full fit)\footnote{These values contain the
correlations listed in tables~\ref{ncresultsrest} and
\ref{ncresultsfull}.}, in agreement with the relatively more precise
experimental value, $Q_W(^{133}\mathrm{Cs}) =
-72.71(49)$~\cite{APVCs}. Here again, we have to keep in mind that the
already relatively large error of our NLO result is presumably subject
to uncertainties related to NNLO corrections.

Interesting results on the weak charge of the proton are to be
expected from the QWEAK experiment at Jefferson Lab. From our fit to
$Z$ pole data we predict $Q_W^p = 0.062(17)$ (restricted fit) and
$Q_W^p = 0.062(22)$ (full fit) for the weak charge of the proton.
Here, we have to stress that the accidental smallness of the NLO
result (keep in mind that $ 1 - 4 \sweff$ is very small), enhances the
sensitivity of the result to sub-leading corrections, including 
loop corrections, too.
\subsubsection{Parity violation in ${\bm e^-e^-}$ (M\o ller) scattering}
Parity violation in $e^-e^-$ scattering at low energies by the SLAC
E158 collaboration provides another determination of the effective
couplings of electrons~\cite{Moller}. The measured asymmetry can be
written in terms of the weak charge for electrons, $Q_W^e$, probing
the $VA$ part of the purely electronic four-fermion interaction.  It
is defined in analogy with the nuclear weak charge defined above, see
Eq.~(\ref{qw}), 
\begin{equation}
Q_W^e = 4\, g_A^e g_V^e~.
\end{equation}
At NLO this reads
\begin{equation}
Q_W^e = 1-4\, \sweff\,(1-\epsilon^e)~.
\end{equation}
Inserting the values for $\sweff$ and $\epsilon^e$ from the two fits
to $Z$ pole data into this equation, we obtain $(Q_W^e)_{\mathrm{NLO}}
= 0.074(1)$\footnote{This value again contains the correlations, see
tables~\ref{ncresultsrest} and \ref{ncresultsfull}.}. This is about
six standard deviations away from the experimental
result~\cite{Moller}, $Q_W^e = 0.041(5)$. This represents another
example where the NLO result is again accidentally small due to the
fact that $4\sweff$ is close to one, such that sub-leading
corrections can play an important role.  We should therefore not be
surprised that there is a discrepancy between our prediction at NLO
and the data in this particular case.

\section{Couplings of light quarks to ${\bm W}$}
\label{Wanalysis}
Let  us now discuss the modified couplings of quarks
to $W$ at NLO.  
We are faced in this case with the problem, how to disentangle QCD 
and non-standard
electroweak effects. It is indeed most acute for the effective couplings to
$W$, $\mathcal{V}_{\mathit{eff}}^{ij}$ and
$\mathcal{A}_{\mathit{eff}}^{ij}$. Their measurement requires an
independent knowledge of the involved QCD parameters like the decay
constants $F_{\pi}$, $F_K$, $F_D$, $F_B$ or the transition form factors
such as $f_{+}^{K^0\pi^{-}}(0) \ldots $.  The unfortunate circumstance
is that the most accurate experimental information on QCD quantities
mentioned above, in turn comes from semi-leptonic transitions of the
type $P \to l\nu$ and $P' \to P l\nu$ where $P={\pi, K, D, B}$ and,
consequently, the result of their measurement depends on (a priori
unknown) EW couplings. 

The chiral generalisation of quark mixing and of CKM unitarity
directly follows from the existence of couplings of right-handed
quarks to $W$. It affects the meaning of the tests of the unitarity of
the CKM matrix: The chiral matrices $V_L$ and $V_R$ have to be
separately unitary, but the effective matrices $\veff$ and $\aeff$,
Eq.~(\ref{effectivecouplings}), which are more directly related to
observables, can exhibit deviations from unitarity. The latter are
expressible in terms of spurion parameters $\delta$ and
$\epsilon$. Even the unitarity triangles (``UT'') representing the
off-diagonal elements of the unitarity condition might be but need not
be affected. This gives a new motivation to the intense studies of UTs
performed during the last years as a possible source of effects beyond
the SM.

We will concentrate here on light quarks $u$, $d$, and $s$. For them
the SM loop effects inducing Right-Handed charged quark Currents
(RHCs) are strongly suppressed by at least two powers of light quark
masses. There are interesting tests in the heavy quark sector, too,
which are certainly worth being studied, but since in the heavy quark
sector additional quark mixing matrix elements become involved, we
will postpone this investigation to future work.

\subsection{Exclusive low--energy tests of couplings of right-handed 
quarks}
\subsubsection{Chiral flavor mixing for light quarks}
\label{sec:chimix}
At NLO, the light quark effective couplings
$\mathcal{V}_{\mathit{eff}}^{ua}$, $\mathcal{A}_{\mathit{eff}}^{ua}$,
$a=d ,s$, see Eq.~(\ref{effectivecouplings}),
can be expressed in terms of three non standard effective EW
parameters: the spurion parameter $\delta$,
Eq.~(\ref{delta}) and two RHCs parameters $\ens$ and $\es$ defined as
(cf Eq.~(\ref{eps}))
\begin{equation}
\ens= \epsilon\ \mathrm{Re}
\Bigl{(}\frac{V_R^{ud}}{V_L^{ud}}\Bigr{)}, \quad
\es =\epsilon\ 
\mathrm{Re} \Bigl{(}\frac{V_R^{us}}{V_L^{us}}\Bigr{)}~.
\label{vrud}
\end{equation}
We obtain
\begin{eqnarray}
|\vfud|^2 &=& |V_L^{ud}|^2 ( 1 +2\, \delta +2\, \ens) \label{vfud}\\
|\afud|^2 &=& |V_L^{ud}|^2 ( 1 +2\, \delta -2\, \ens) \label{afud}\\
|\vfus|^2 &=& |V_L^{us}|^2 ( 1 +2\, \delta +2\, \es) \label{vfus}\\
|\afus|^2 &=& |V_L^{us}|^2 ( 1 +2\, \delta -2\, \es)~,
\label{afus}
\end{eqnarray}
where $V_L^{ud}$ and $V_L^{us}$ are related by the unitarity condition
of the left-handed mixing matrix. Neglecting
$V_L^{ub}$, as suggested by the measurement of 
$|\mathcal{V}_{\mathit{eff}}^{ub}|$ and $|\mathcal{A}_{\mathit{eff}}^{ub}|$, respectively, the unitarity condition can be written as follows, 
\begin{equation}
|V_L^{ud}|^2 + |V_L^{us}|^2 = 1~.
\label{vl}
\end{equation}
Let us discuss these equations.
\begin{itemize}

\item
The only very precisely known quantity in this set of equations
is $|\vfud|$. It is 
determined from nuclear $0^{+} \to 0^{+}$ transitions
relying on the conservation of the vector current (CVC) and its
value is~\cite{MS}~\footnote{Note that our phase convention is
such that $|\vfud| = \vfud$. In this case $V_L$ has not the same
structure as the CKM matrix within the SM, but might have additional phases.}
\begin{equation}
\vfud = 0.97377(26)\equiv cos\hat{\theta}~.  
\label{costheta}
\end{equation}
At LO, one recovers the unitarity of the CKM matrix in the SM, and
$\hat{\theta}$ corresponds to the Cabbibo angle. 
It is useful for the following discussions to rewrite the
effective vector and axial couplings in terms of this quantity,
using the relation between $|\vlud|$ and $|\vfud| = \cos\hat\theta$,
see Eq.~(\ref{vfud}),
\begin{eqnarray}
|\vfud|^2 &=& \cos^2\hat\theta \nonumber\\
|\afud|^2 &=& \cos^2\hat\theta\,( 1 - 4 \, \ens) \nonumber\\
|\vfus|^2 &=& \sin^2\hat\theta\, (1 +
 2\frac{\delta+\ens}{\sin^2\hat\theta}) (1 + 2\, \es - 2\, \ens) \nonumber\\
|\afus|^2 &=& \sin^2\hat\theta\, (1 +
 2\frac{\delta+\ens}{\sin^2\hat\theta}) (1 -2\, \es - 2\, \ens)~,
\label{effcouplings}
\end{eqnarray}
where we used the unitarity of the left-handed mixing matrix, $V_L$,
Eq.~(\ref{vl}), in the last two equations.
We only kept terms up to first order in the spurion parameters
$\delta, \ens,$ and $\es$ except the term proportional to
$1/\sin^2\hat\theta$. For the latter the effect of spurions is 
enhanced due to the smallness of $\sin\hat\theta$.
 
\item
As already pointed out the genuine spurion parameters $\delta$ and
$\epsilon$ are expected to be of the order 0.01.  To obtain bounds on
$\ens$ and $\es$ we can exploit the unitarity of the right-handed
mixing matrix which gives the following condition:
\begin{equation}
|\ens|^2\, |V_L^{ud}|^2 + |\es|^2\, |V_L^{us}|^2 \le \epsilon^2~.
\label{ellipsees}
\end{equation}
Using the unitarity condition of the left-handed mixing matrix and the 
expression of 
$V_L^{ud}$ in terms of $cos \hat \theta$  one obtains:
\begin{equation}
(|\ens|^2\ - |\es|^2)
(1- 2(\delta +\ens)) \,\cos^2 \hat \theta +  |\es|^2
 \le \epsilon^2~.
\label{ellipsef}
\end{equation}
In the following we will see that all the observables can be written
in terms of $\delta +\ens$, $\ens$ and $\es -\ens$. Fixing the value
of $\,\delta +\ens$, the above condition can be visualized as an
ellipse in the plane $\ens/\epsilon,(\es-\ens)/\epsilon$. This is
shown in Fig.~\ref{figellipse} for two typical values of $\delta
+\ens$.  It can be seen from Fig.~\ref{figellipse} that, on the one
hand $|\ens|\lsim\epsilon$ is small.  On the other hand, $\es$ can be
enhanced to a few percent level: $|\es| \lsim 4.5 \epsilon$. This
enhancement of $\es$ is possible for example if the hierarchy in
right-handed flavor mixing is inverted, i.e. $| V_R^{ud}| < |
V_R^{us}|$.
\end{itemize} 

As stressed above, in the presence of non-standard EW couplings, the
effective mixing matrix $\veff$ is not necessarily unitary. Wolfenstein
~\cite{Wolfenstein:1984ay} used this effect to find limits on the
mixing of $W_L$ and $W_R$ in left-right symmetric models.
Here, we can
express the deviation from unitarity at NLO in terms of the EW
parameters $\delta,\es,\ens$:
\begin{equation}
| \vfud|^2 + |\vfus|^2
= 1 + 2 \, (\delta + \ens) +             
2\, (\es-\ens)\, \sin^2\hat{\theta}~.
\label{effunit}
\end{equation}
\FIGURE[t]{
\epsfig{width=8cm,file=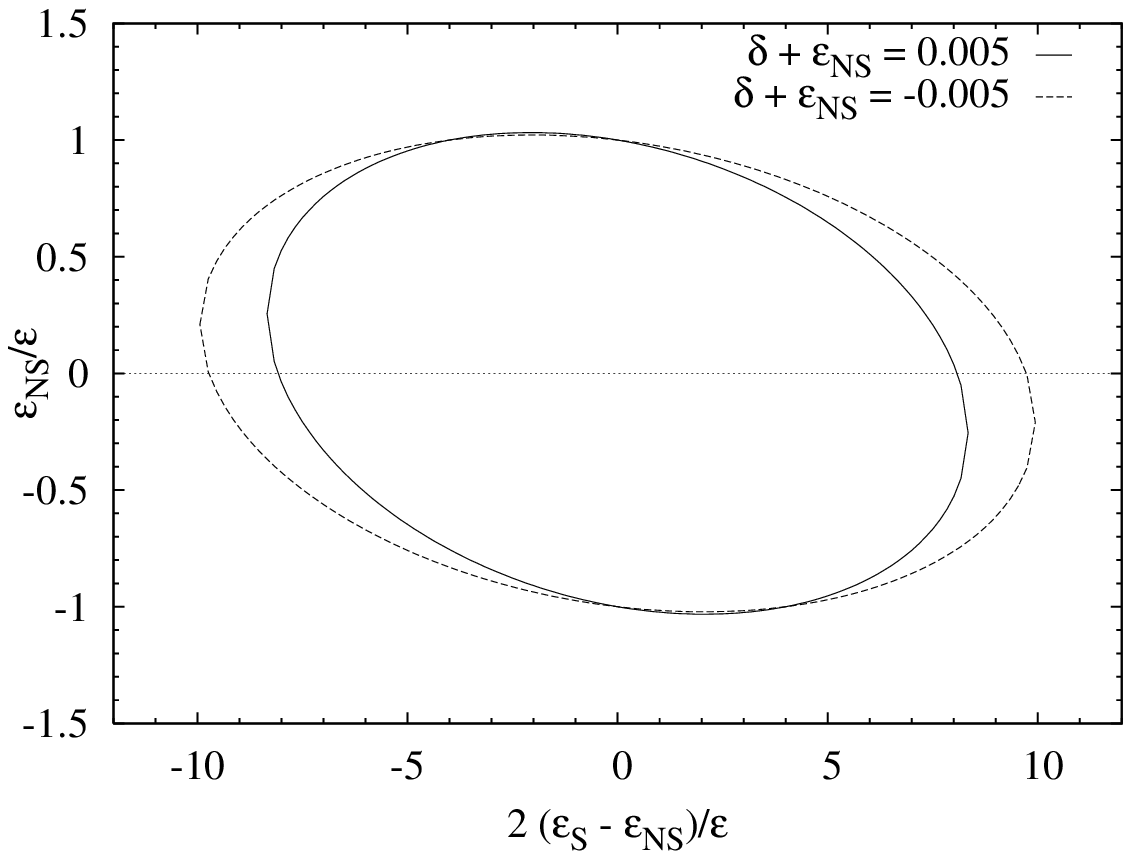}
\caption{\it Maximum values of $\ens/\epsilon$ and $(\es-\ens)/\epsilon$ 
for two different values of $\delta +\ens$ compatible with the unitarity of $V_{L,R}$, cf Eq.~(\ref{ellipsef}). }
\label{figellipse}
}

In this equation the only parameter, $\es$, which can be significantly
larger than $\sim 0.01$ appears multiplied by
$\sin^2\hat\theta$. Since the possible enhancement of $\es$ is due to
the mixing hierarchy of left-handed quarks proportional to
$1/|V_L^{us}|\approx 1/\sin\hat\theta$ the effect should be at most of
the order $\epsilon \sin\hat\theta \sim 0.002$. The deviation from
unitarity due to the spurionic parameters is thus at most of the order
of the genuine spurion parameters $\delta, \epsilon$. It can, however,
be much smaller.  In addition, we cannot a priori say whether the
r.h.s of Eq.~(\ref{effunit}) should be smaller or larger than one.
\subsubsection{Interface of effective electroweak and low-energy QCD couplings}
\label{interface}
As mentioned above, presently the most precise
determinations of couplings of (light) quarks to $W$ arise from
semileptonic decay processes involving QCD parameters. 
\begin{itemize}
\item From neutron life-time
measurements and angular distributions we can extract the value of $|g_A
\afud|/|g_V\vfud|$,
 
\item from the decay rate $\Gamma(\pi_{l2}(\gamma))$ we can infer
$|F_{\pi^+} \afud|$, 

\item from the branching ratio 
Br~$\Big(K^+_{l2}(\gamma)/\pi^+_{l2}(\gamma)\Big)$ we can extract the value of
$|F_{K^+} \afus|/|F_{\pi^+} \afud|$, 

\item and the $K^0_{l3}$ decay rate allows
to determine $|f_+^{K^0\pi}(0) \vfus|$.
\end{itemize}

These decay processes  involve hadronic
matrix elements of vector/axial quark currents as for example the following 
exclusive
matrix elements defining nucleon form factors
\begin{eqnarray}
\langle p(k') | \bar{u}\gamma_\mu d|n(k)\rangle &=& 
g_V(q^2)\,\bar{u}_p(k')\gamma_\mu u_n(k) + \dots \\\ 
\langle p(k')| \bar{u}\gamma_\mu\gamma_5 d|n(k)\rangle &=& g_A(q^2)\,\bar{u}_p(k')
\gamma_\mu\gamma_5
u_n(k) + \dots ~,
\end{eqnarray}
with $q^2 = (k'-k)^2$, or
the decay constants of pseudoscalar mesons 
\begin{equation}
\langle 0|\bar{u} \gamma_\mu \gamma_5 d (0)|\pi^+(p)\rangle = i \sqrt{2} F_{\pi^+} p_\mu~, \quad
\langle 0|\bar{s} \gamma_\mu \gamma_5 u(0)|K^+(p)\rangle = i \sqrt{2} F_{K^+} p_\mu 
\end{equation}
as well as the hadronic matrix element describing
the $K^0_{\mu3}$ decay. It can be written in terms of two form factors
\begin{equation}
\langle \pi^-(p') | \bar{s}\gamma_{\mu}u | K^0(p)\rangle =
(p'+p)_\mu\  f^{K^0\pi^-}_+ (t) + (p-p')_\mu\  f_-^{K^0\pi^-} (t),         
\label{hadronic element}
\end{equation}
where $t = (p'-p)^2$.  Here the form factors and decay
constants stand for radiatively corrected genuine QCD quantities.
It is further understood that all isospin breaking
effects due to $m_d - m_u$ are included.
These QCD quantities are subject
to Chiral Perturbation Theory (ChPT) or lattice studies.
Strictly speaking, without any theoretical input, none of them is 
experimentally accessible.

If the EW effective couplings of quarks to $W$ are given by the SM,
i.e. $\delta=\ens=\es = 0$, the above mentioned branching ratios and
decay rates allow to determine the corresponding QCD quantities rather
precisely since the EW effective couplings are all precisely
determined by the value of $\cos\hat\theta$, see Eq.~(\ref{costheta}).
We will denote these QCD quantities extracted from semileptonic decay
data assuming SM weak interactions with a hat. Their values are~\cite{PDG06}
\begin{eqnarray}
\hat{r}_A &=& \left|\frac{g_{A} \afud}{g_{V}\vfud}\right| = 1.2695(29),\nonumber \\
\hat{F}_{\pi^+} &=& 92.4(3)~\mathrm{MeV} ,\nonumber\\
\hat{F}_{K^+}/\hat{F}_{\pi^+} &=& 1.182(7),\nonumber\\
\hat{f}^{K^0\pi^-}_{+} (0) &=& 0.951(5)~.
\label{hat}
\end{eqnarray}
Here the errors merely reflect the experimental uncertainties in the
measured branching ratios. Note that the value of
$\hat{F}_{K^+}/\hat{F}_{\pi^+} = 1.182(7)$ is significantly lower than
the value largely used in ChPT studies, $F_K/F_\pi = 1.22$ (cf. for
instance~\cite{GL85, Bijnens}).

In the presence of non-standard couplings of quarks to $W$ the values of
these QCD quantities extracted from semileptonic branching ratios
are modified. Using Eq.~(\ref{effcouplings}) the genuine QCD quantities 
can be written in terms of the
corresponding quantities with a hat and the spurion parameters, e.g.:
\begin{equation}
\left(\frac{F_{K^+}}{F_{\pi^+}}\right)^2 
= \left(\frac{\hat{F}_{K^+}}{\hat{F}_{\pi^+}}\right)^2 
\frac{\sin^2\hat\theta}{\cos^2\hat\theta} \frac{|\afud|^2}{|\afus|^2}
 =\left (\frac{\hat{F}_{K^+}}{\hat{F}_{\pi^+}}
\right)^2 \frac{1+2\,(\es-\ens)}{1+\frac{2}{\sin^2\hat{\theta}}
(\delta+\ens)}~.
\label{fkfpi} 
\end{equation}
In a similar manner we can write
\begin{eqnarray}
|r_A|^2  &=& \hat{r}_{A}^2 (1 + 4\, \ens)\nonumber\\
|F_{\pi^+}|^2 &=& \hat{F}^2_{\pi^+} (1 + 4\, \ens)\nonumber \\
|f^{K^0\pi^-}_+(0)|^2 &=& 
\left[ \hat{f}^{K^0\pi^-}_+(0)\right]^2 \,
\frac{1-2(\es-\ens)}{1+\frac{2}{\sin^2\hat{\theta}}
(\delta+\ens)}~.
\label{qcd}
\end{eqnarray}

To constrain the three NLO EW parameters $\delta$ , $\ens$ and $\es$
from the above relations, we need information on the QCD quantities
like $F_{\pi^+}$, $F_{K^+}$, $f_+^{K^0\pi}(0)$ which is independent of
their extraction from semi-leptonic transitions.  Such information
could in principle originate from lattice simulations, from ChPT, or
from short-distance constraints on QCD observables combined with
purely strong/electromagnetic processes.

\subsubsection{Neutron $\beta$-decay and Adler-Weisberger Sum Rule}
Let us illustrate the problem with two examples which historically
played an important role in establishing the $V$-$A$ character of the
weak interaction. The first one is neutron $\beta$-decay. There exist
precise measurements of various angular and spin correlations in the
(polarized) neutron $\beta$-decay and more experimental results are
expected (see e.g.~\cite{neutron}). These are often presented as
accurate tests of the chirality of fermion couplings to $W$. Given the
NLO minimal LEET expression of these couplings, Eq.~(\ref{LagrW}),
one may ask what is the impact of these measurements on the RHCs
parameter $\ens$ defined in Eq.~(\ref{vrud}). Since at NLO the
standard $V$-$A$ couplings of {\bf leptons} to $W$ are not modified, any
observable in neutron $\beta$-decay can be expressed in terms of the
Fermi constant and two EW parameters concerning more particularly $u$
and $d$ quarks: $|g_{V} \vfud|$ which normalizes the
decay rate via the neutron lifetime and the relative
parameter $\hat{r}_A$, see Eq.~(\ref{hat}).
The compatibility of various extractions of $\hat{r}_{A}$ from
different measurements of independent correlations does, indeed,
represent a valuable test of the {\bf $V$-$A$ character of the coupling of
leptons to $W$}. These tests are so far compatible with a pure $V$-$A$
leptonic coupling.

However, as precise they could possibly be, the neutron $\beta$-decay
experiments alone say nothing about the quark RHCs unless one specifies the
a priori unknown QCD quantity
\begin{equation}
r_{A} = g_{A}/g_{V}
\end{equation}
which does not coincide with the experimentally known $\hat{r}_{A}$
provided there exist right-handed $\bar ud$ currents, i.e., 
$\ens \neq 0$ . Actually, one has (cf Eq.~(\ref{qcd}))
\begin{equation}
r_{A} = \hat{r}_{A} ( 1 + 2 \ens) .
\end{equation}
Hence, $\ens$ could be determined if $r_{A}$ was known.

Soon after the idea of universal $V$-$A$ weak interactions has
appeared~\cite{FGM}, the issue of its tests for hadrons has been
considered in the light of the current algebra charge relation $ [
Q_{5} , Q_{5}^{\dagger} ] = 2 I_{3} $ which provides an absolute
normalization of the axial current and gives a precise meaning to the
ratio $r_{A}$.  Combined with chiral symmetry, this relation yields
the Adler-Weisberger sum rule~\cite{AW}, which may be written as
\begin{equation}
  1   =   r_{A}^2 + F_{\pi}^2 \,\frac{2}{\pi}\, \int dk \frac{k^2}{\omega^3(k)}
     [\sigma^{\pi^{-}p} (k)  -  \sigma^{\pi^{+} p} (k) ] +          
       \mathcal{O}(m_{\pi}^2)~,
\label{AW}
\end{equation}
where $k$ and $\omega$ are pion laboratory momentum and energy,
respectively.  Since the charge current algebra, as well as chiral
symmetry, are today integral parts of QCD and can be proven from first
principles, the above relation is an exact QCD low-energy theorem
which holds independently of the EW effective couplings $\veff$ and
$\aeff$.  Using the expressions (cf Eq.~(\ref{qcd})) of $r_{A}$ and
$F_{\pi}$ in terms of experimentally known quantities $\hat{r}_{A}$
and $\hat{F}_{\pi}$, the Adler-Weisberger relation may be written as
a sum rule for the RHCs parameter $\ens$:
\begin{equation}
1 - 4 \ens = \hat{r}_{A}^2 + \hat{F}_{\pi}^2\, \frac{2}{\pi}
\int dk \frac{k^2}{\omega^3 (k)} [\sigma^{\pi^{-} p} (k) -              
\sigma^{\pi^{+} p} (k) ] + \mathcal{O}(M_{\pi}^2).
\end{equation}
Hence, $\ens$ can, in principle, be inferred from observable
quantities under two conditions: i) The chiral symmetry breaking
corrections to the low-energy theorem, Eq.~(\ref{AW}), can be reliably
estimated to high precision and ii) the sum rule integral can be
evaluated with a sufficient precision out of measured and radiatively
corrected $\pi N$ total cross sections (see for exemple
Ref.~\cite{piN} and references therein).

Since $\ens$ is expected to reach at most the percent level, it does
not appear realistic to control the sum rule, Eq.~(\ref{AW}), to this
degree of precision. The previous discussion illustrates the typical
problems one has to face extracting the spurion parameters $\delta,
\ens,$ and $\es$ on the basis of chiral low-energy theorems. 

An additional remark is in order: The Goldberger-Treiman low-energy
theorem is insensitive to the modification of EW effective couplings
considered here. The reason is that $\ens$ cancels in the ratio
$r_A/F_\pi = \hat{r}_A/\hat{F}_\pi$ reflecting the fact that QCD does
not know about EW couplings.
\subsubsection{How to measure $F_\pi$ in non-EW processes?}
\label{pi02gamma}

i)  $\pi^0 \to 2 \gamma$ 

One possible determination of $F_{\pi^+}$ comes from
the $\pi^0 \to 2 \gamma$ partial width. This
process has no interface with the EW couplings and it is 
is independent of the standard determination based on
the $\pi_{l2}$ decay rate. It could thus provide a measurement of $\ens$
through Eqs.~(\ref{hat}) and (\ref{qcd}). The process $\pi^0\to
2\gamma$ is governed by the anomaly which exactly predicts the value
of the amplitude in the chiral limit. Corrections up to 
$\mathcal{O}(p^6,e^2 p^4)$ and to first order in $m_d-m_u$ have
recently been calculated~\cite{AM02,pi02y}. They are dominated by isospin
breaking corrections and are of the order $10^{-2}$. 
For the moment we will not pursue this possibility further for an
experimental and a theoretical reason.
\begin{itemize}
\item The experimental situation for
the $\pi^0$-lifetime does not allow to determine the partial width
$\pi^0 \to 2\gamma$ better than with an error of $7.1\%$ (current
world average~\cite{PDG06}). This induces an error of at least
several percent on the determination of $F_\pi$. The upcoming result
of the Primex experiment at Jefferson Lab will certainly improve on
this situation aiming at a precision~\cite{primex} of 1.5\% for the
partial width. Then in principle it becomes conceivable to look for
effects of the order of percent relating Eqs.~(\ref{hat}) and
(\ref{qcd}).
 
\item Upon relating the
unknown $\mathcal{O}(p^6)$ low-energy constants of the
Wess-Zumino-Witten Lagrangian to the $\eta$ decay width employing a
three-flavor framework~\cite{AM02,pi02y}, the dominant corrections to the
chiral limit involve the isospin breaking quark mass ratio $R =
(m_s-\hat{m})/(m_d - m_u)$. Thus from the theoretical side a
sufficiently precise
determination of $F_\pi$ from the $\pi^0\to 2\gamma$ partial width
requires a good knowledge of the isospin breaking parameter
$\epsilon^{(2)} = (\sqrt{3}/4) \ (1/R)$. This can probably be achieved in the
near future comparing the high statistics measurements of charged and
neutral $K_{l3}$ decays. 
\end{itemize}

\noindent ii) $\pi\pi$ scattering.

In principle, another possibility to extract a value of $F_\pi$ independently
arises from $\pi\pi$ scattering. In the low-energy domain the
$\pi\pi$ scattering amplitude is strongly constrained by chiral
symmetry because of the Goldstone-boson character of the pions. Asking
in addition for the amplitude to satisfy crossing symmetry and
unitarity, it can be written in terms of six sub-threshold
parameters~\cite{pipi}, the pion mass and $F_\pi$. Matching the
phenomenological description of the amplitude from the solution of Roy
equations~\cite{DFGS,ACGL} with the chiral representation, it should
be possible to extract the value of $F_\pi$. Presently, however, the
errors on the extracted sub-threshold parameters, assuming $F_\pi
=\hat{F}_\pi$ are at least of the order of one percent such that it
seems difficult to reliably determine $F_\pi$ with a precision of less
than a percent. 
\subsection{The gold plated test: $\bm{K^L_{\mu 3}}$ decays}
\label{kaons}
In Ref.~\cite{BOPS} it has been shown that a stringent test involving
the EW coupling $\es - \ens$ can be devised in $K^L_{\mu 3}$
decay. Indeed,  combining the measurement of the scalar
$K\pi$ form factor in $K^L_{\mu 3}$ decays with the Callan-Treiman
low-energy theorem, it is possible to measure the ratio
$F_{K^+}/F_{\pi^+} f^{K^0\pi^+}_+(0)$ independently from the above mentioned
semileptonic branching ratios and decay rates. 

Let us briefly resume here the results of Ref.~\cite{BOPS}. 
We will concentrate
on the normalized scalar form factor, see Eq.~(\ref{hadronic element})    
\begin{equation}
f(t)=\frac{f^{K^0\pi^-}_S(t)}{f^{K^0\pi^-}_+(0)} = \frac{1}{f^{K^0\pi^-}_+(0)}
\left(f^{K^0\pi^-}_+(t) + \frac{t}{\Delta_{K\pi} } f^{K^0\pi^-}_-(t)\right)
\ \ ,\ \ f(0)= 1 .
\label{defnffactor}
\end{equation}
The Callan-Treiman low-energy theorem (CT)~\cite{Dashen:1969bh} fixes
the value of $f(t)$ at the point
$t=\Delta_{K\pi}=m_{K^{0}}^2-m_{\pi^+}^2$ in the $\mathrm{SU}(2)\times
\mathrm{SU}(2)$ chiral limit:
\begin{equation}
C \equiv f(\Delta_{K\pi})= \frac{F_{K^+}}{F_{\pi^+}}\frac{1}{f_{+}^{K^0\pi^-}(0)}+  
\Delta_{CT},
\label{C}
\end{equation}
where the CT discrepancy $\Delta_{CT}$ defined by Eq.~($\ref{C}$) is
expected to be small and calculable in ChPT.
It is proportional to $m_u$ and/or $m_d$. In the
limit $m_d=m_u$ at NLO in ChPT one has 
$\Delta_{CT}^{\mathrm{NLO}}= - 3.5 \times
10^{-3}$~\cite{Gasser:1984ux}. We will focus the discussion on the
neutral kaon mode since the analysis of the charged mode is subject to
larger uncertainties related, in particular, to $\pi^0\eta$
mixing~\cite{BOPS} which could easily enhance the CT discrepancy by one
order of magnitude.

In the physical region the form factor can be parameterized accurately in
terms of only one parameter, $\ln C$, in a model independent
way using the dispersive representation proposed in
Ref.~\cite{BOPS}. This allows for a direct measurement of $\ln C$
in $K^L_{\mu 3}$ decays recently performed by the NA48
collaboration~\cite{NA48}. They obtain
\begin{equation}
\ln C_{\mathit{exp}} = 0.1438 \pm 0.0138~.
\label{lncna48}
\end{equation}
This
value can be combined with the determination from the branching ratios
Br~$K^+_{l2(\gamma)}/\pi^+_{l2(\gamma)}$~\cite{PDG06a}, the inclusive
decay rate $K^L_{e3 (\gamma)}$ ~\cite{fplus}, and the value of
$|\vfud|$ known from superallowed $0^+\to 0^+$ nuclear
$\beta$-decays~\cite{MS} as (see Eqs.~(\ref{fkfpi}, \ref{qcd}))
\begin{equation}
C = B_{\mathit{exp}} \, r + \Delta_{CT}\label{lnc}~,
\end{equation}
with 
\begin{equation}
B_{\mathit{exp}} = \Bigl{|}\frac{F_{K^+} \afus}
{F_{\pi^+} \afud}\Bigr{|}
\frac{1}{|f_+^{K^0\pi^-}(0)
\vfus|}|\vfud|
\end{equation}
 and 
\begin{equation}
r = \Bigl{|}\frac{\afud\vfus}{\vfud
\afus}\Bigr{|}~.
\end{equation} 
This gives \footnote{There is a small difference between the value used in Ref.~\cite{NA48} and the value given here because we use only the $K^L_{e3}$ data to evaluate $|f_+^{K^0\pi^-}(0) \vfus|$.} to first
order in $\epsilon$:
\begin{equation}
\ln C  = 0.2182 \pm 0.0035 + \tilde\Delta_{CT} + 2 (\es -
\ens) = 0.2182 \pm 0.0035 + \Delta\epsilon
\label{lncexp}
\end{equation}
where $\tilde \Delta_{CT} = \Delta_{CT}/B_{\mathit{exp}}$. From the 
experimental result Eqs.~(\ref{lncna48}) one gets:
\begin{equation}
\Delta\epsilon = -0.074\pm0.014~.
\label{epsmeasure}
\end{equation}
This is an interesting result. Within our framework it suggests that $\es$ could be
enhanced as a consequence of an inverted mixing hierarchy in
the right-handed sector, see Section~\ref{sec:chimix}. 
\FIGURE[t]{
\epsfig{file=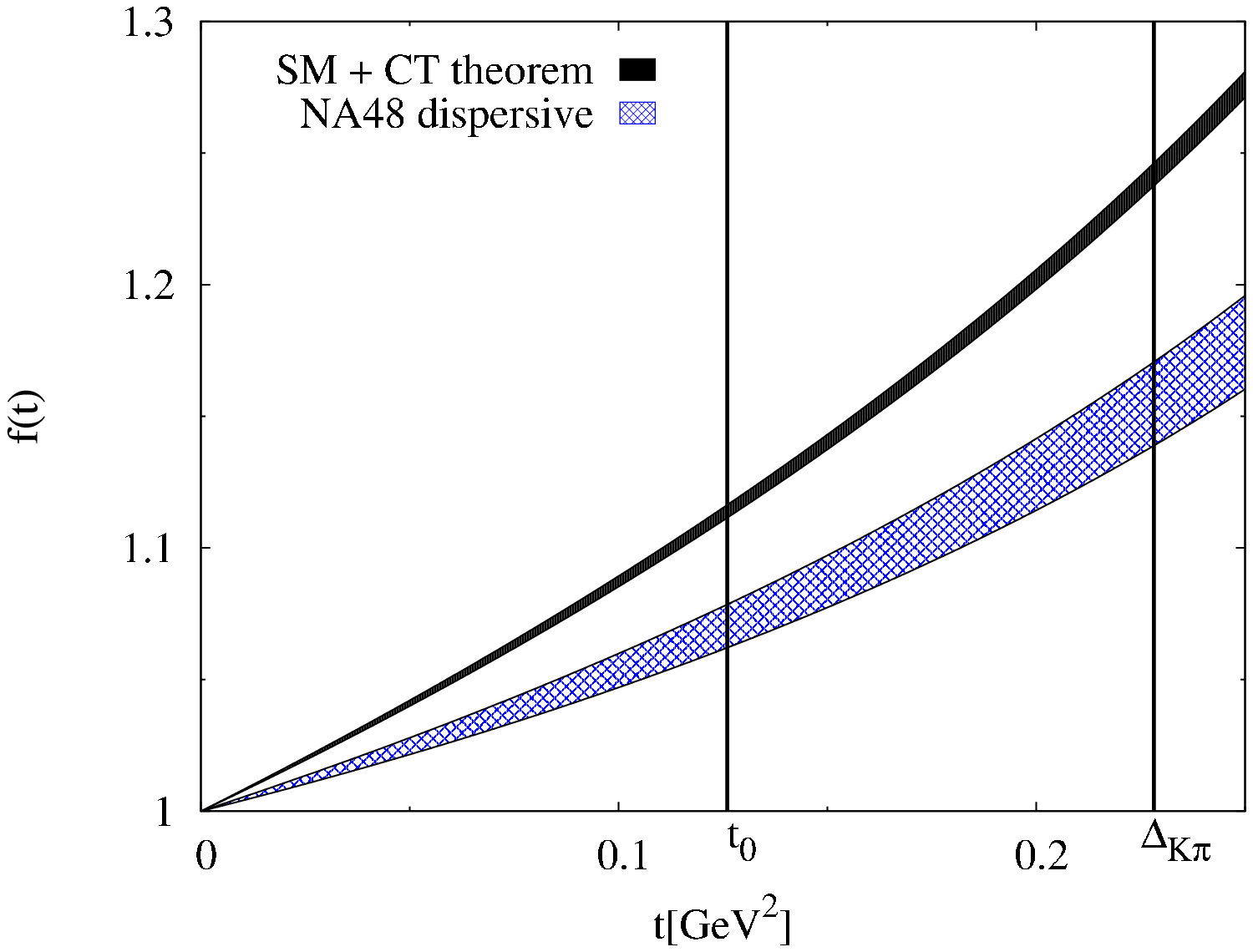,width=7cm}
\caption{\it The normalized scalar $K\pi$ form factor as a function of $t$. }
\label{figkmu3}
}
The strong deviation between the SM prediction and the measurement is
clearly shown in Fig.~\ref{figkmu3} which display $f(t)$. The upper 
black curve corresponds to the SM prediction, i.e. $\es =\ens = 0$. The
assigned error is purely experimental. The full blue curve correspond
to the NA48 value using the dispersive representation. Of course, this
result should be confirmed by other independent measurements of the
scalar $K\pi$ form factor based on the dispersive representation which are underway. For more discussion see
Refs.~\cite{BOPS,BOPS07}.

\subsection{Inclusive OPE based tests}
In addition to the processes, where chiral dynamics controls the QCD
part, we can investigate (semi-)inclusive processes where the QCD part
is dominated by short-distance dynamics and can be evaluated applying
operator product expansion techniques. We will
discuss here three types of such processes: inelastic neutrino
scattering, $W$ boson (semi-) inclusive decays and hadronic tau
decays.
\subsubsection{Inelastic neutrino scattering}
The hadronic tensor contributing to the cross section for inelastic
(anti-) neutrino-nucleon scattering is given by
\begin{equation}
W^{\mu\nu}(p,q) = \frac{1}{4\pi} \int d^4 x \sum_\sigma \mathrm{e}^{iqx} \langle p,\sigma|[J^\nu (x), (J^\mu(0))^\dagger]|p,\sigma\rangle~, 
\end{equation}
where $J^\mu$ is the EW hadronic current (see Eq.~\ref{LagrW})
\begin{equation}
J^\mu = (1+\delta) 
\bar{\mathrm{U}}_L V_L \gamma^{\mu} D_L + 
\epsilon
\bar{\mathrm{U}}_R V_R \gamma^{\mu} D_R~
\label{hadcc}
\end{equation}
and $\sigma$ indicates the sum over spins.
The hadronic tensor can be splitted into contributions from different
chiralities: 
\begin{equation}
W^{\mu\nu} = (1+\delta)^2\, W^{\mu\nu}_{LL} + \epsilon\,(1+\delta)
W^{\mu\nu}_{LR} + \epsilon^2\, W^{\mu\nu}_{RR}~.
\label{wchiral}
\end{equation}
This decomposition makes obvious that we do not expect a sensitive
test of the coupling of right-handed quarks to $W$ from inelastic
neutrino scattering. The only term linear in the spurion parameter
$\epsilon$, the second term on the right-hand side of
Eq.~(\ref{wchiral}), is proportional to $W_{LR}^{\mu\nu}$ which, due
to its chiral structure, does not contribute to the leading twist.
Sizeable contributions could arise only from heavy quarks.  These are,
however, suppressed by the mixing hierarchy of left-handed quarks. The
third term on the r.h.s of Eq.~(\ref{wchiral}), which contributes to
the leading twist, is in turn suppressed by two powers of the spurion
parameter $\epsilon$. According to the order of magnitude estimates
from the LEET, this contribution should not be much larger than about
$10^{-4}$ and can therefore hardly be disentangled from higher twist
left-handed contributions. The bound on the RHCs contribution in deep
inelastic neutrino scattering obtained by the CDHS collaboration,
later confirmed by CCFR~\cite{Abramowicz:1981iq} from the
$y$-dependence of neutrino and antineutrino scattering cross sections
is thus not very restrictive in the present case.

In conclusion, the parameter $\delta$ is the only one which in
principle can sensibly be extracted from inelastic neutrino
scattering. To first order in the spurionic parameters, we can write
\begin{equation}
W^{\mu\nu} = (1+2\delta)\,\hat{W}^{\mu\nu}~,
\end{equation}
where $\hat{W}^{\mu\nu}$ denotes the hadronic tensor assuming SM weak
interactions. Hence the determination of the parameter $\delta$
amounts to the precise knowledge of the absolute normalisation of the
neutrino scattering cross section. As we discussed in
Section~\ref{zpole}, $\delta \lsim 0.01$ and we would have to control
the cross sections to a better accuracy. Note that currently the
uncertainties in parton distribution functions (PDFs) (normalisation)
do not reach this precision.

We would like to point out that there is a 
possibility to overcome this difficulty based on Adler's neutrino sum
rule~\cite{Adler:1965ty} which can be written as follows for the
scattering on a proton target:
\begin{equation}
\int_0^\infty \Big(W_2^{(\bar{\nu} p)}(q^2,\nu)  - W_2^{(\nu p)}(q^2,\nu)\Big)
d\nu = 2\,(1+2\delta)~,
\label{adler}
\end{equation}
where the structure function $W_2$ is defined via
\begin{equation}
W_{\mu\nu}^{(\nu)}(p,q) = \frac{p_\mu p_\nu}{m_N^2}\Big(\theta(\nu) W_2^{(\bar\nu p)}(q^2,\nu) 
-\theta(-\nu) W_2^{(\nu p)}(q^2,-\nu)\Big) 
+ \mathrm{independent\ tensor\ structures}.
\end{equation} 
$m_N$ is the nucleon mass and the integration variable is $\nu = p
q/m_N$.  Let us recall that the sum rule, Eq.~(\ref{adler}), is an
exact QCD statement (not receiving any $\alpha_s$ corrections), valid
for fixed $q^2 < 0$.  Hence, in principle, the Adler sum rule provides a
test of EW couplings which is independent of high $q^2$ approximations
and/or precise knowledge of PDFs.
In practice, it is however not easy to evaluate this sum rule
precisely from existing data because not all values of $\nu$ are
equally well accessible and there are complications due to heavy quark
thresholds. Notice further that the spurion parameter $\delta$ enters
the overall normalisation of the cross section, together with the
factor $(m_W^2/(-q^2 + m_W^2))^2$ from the propagator of the exchanged
$W$-boson.  This fact could be exploited to sharpen the combined EW and QCD
analysis~\cite{H1}.
\subsubsection{$W$ boson (semi-) inclusive decays}
\label{wwidth}
The non-standard charged currents couplings ($\veff$ and $\aeff$)
affect, among other things, the decay of $W$ into hadrons.
Consequently, these new couplings will appear in the description of
the decay ratio and we can try to extract them from the corresponding
data.  The total hadronic decay width of $W$ bosons can be obtained
from the absorbative part of the corresponding two-point correlation
function
\begin{equation}
\Pi^{\mu\nu}(q) = i \int d^4 x \mathrm{e}^{iqx} \langle
0|T(J^\mu (x) (J^\nu(0))^\dagger|0\rangle~. 
\end{equation}
The expression up to NLO within
the LEET can be deduced from Eq.~(\ref{hadcc}).
Considering only the effects of first order in $\epsilon$ and $\delta$,
we obtain:
\begin{equation}
\Gamma (W \mapsto h)= (1+2\delta)~\hat{\Gamma} (W \mapsto h)~,
\end{equation}
where $\hat{\Gamma} (W \mapsto h)$ is the hadronic W width extracted
assuming SM interactions. Again, as in the case of inelastic neutrino
scattering, only the parameter $\delta$ appears. Perturbative QCD
predicts the value of $\hat\Gamma (W\to h)$ as a series in
$\alpha_s(m_{W})$~\cite{PDG06}:
\begin{equation}
\hat{\Gamma} (W\to h)=\frac{G_FM_W^3}{6\sqrt{2}\pi}~6~ R_W~,
\end{equation}
where the factor 
\begin{equation}
R_W = 
\Big{[}1+\frac{\alpha_S(m_W)}{\pi}+
1.409\Big{(}\frac{\alpha_S(m_W)}{\pi}\Big{)}^2-12.77\Big{(}\frac{\alpha_S(m_W)}{\pi}\Big{)}^3\Big{]}
\end{equation}
arises from QCD corrections. For the 
total $W$ decay width we obtain:
\begin{equation}
\Gamma_{W}=\frac{G_FM_W^3}{6\sqrt{2}\pi}\Big{[}3+6~(1+2\delta)~R_W\Big{]}~.
\end{equation}
Let us recall that the leptonic contribution is not changed with
respect to the SM because at NLO within the LEET, the universal
modification of the couplings of left-handed leptons is absorbed into
the definition of $G_F$. Right-handed charged leptonic currents are
forbidden due to the additional $Z_2$ symmetry for the neutrinos.

Hence, the measurements sensitive to the hadronic decay width of $W$
should allow us to extract the parameter $\delta$ that modifies the
coupling of $W$ to the left-handed fermions.  This is a common feature
to all inclusive charged current processes at high energies. This can
be understood as follows. We are looking at a correction to the SM
result where the charged current interaction is purely
left-handed. The parameter $\epsilon$ related to right-handed charged
quark currents can therefore only appear to first order in connection
with quark masses or other non-perturbative quantities inducing a $LR$
structure, i.e. they are typically suppressed by a factor $m_i
m_j/m_W^2$.
 
Let us now discuss the different measurements.
\begin{itemize}
\item
There are direct and indirect experimental measurements of the total
decay width of W available from LEP and
Tevatron.  At the
moment, the direct measurements
\footnote{We can only use the direct measurements because the indirect
  ones use some input from measurements at the Z pole, as for instance
  the value of the branching ratio $Z\mapsto e^+e^-$, which are
  extracted assuming the SM and which can get modified within the
  present framework.}  are not precise enough to be sensitive to a
  value for $\delta$ on the percent level.  With the different data
  from~\cite{LEPEWWG,unknown:2005ij} we obtain roughly $-0.03 <
  \delta < 0.03$.  In view of the experimental effort undertaken to
  improve on the precision for the $W$ decay width, it will probably
  become possible to test the value of $\delta$ more precisely from
  the decay width of $W$ in the near future.
\item Another possibility is to take the measured leptonic branching
fraction $\Gamma(W\to l\nu)/\Gamma_W$ which is known with high
precision. We have included this quantity into the fit determining the
spurionic parameters at NLO in the couplings to $Z$, since the same
parameter $\delta$ enters the couplings of left-handed quarks to
$Z$. This fit has been discussed in detail in Section~\ref{zpole}. The
resulting value is $\delta = -0.006(4)$ at $\alpha_s(m_Z) = 0.1190$,
taking the value for the leptonic branching fraction from LEP~\cite{LEPEWWG},
$\Gamma(W\to l\nu)/\Gamma_W = 0.1084(9)$.
\item
If it was possible to measure precisely partial decay widths into
hadrons, it would be conceivable to determine $\delta$ and the
corresponding matrix element of $V_L$ simultaneously. Here again, the
contributions from couplings of right-handed quarks to $W$ are
strongly suppressed since they appear only with powers of masses or
other quantities with $LR$ structure, divided by the $W$
mass. For example, the partial width into $c\bar{s}$ is given by
\begin{equation}
\Gamma(W\to c\bar{s}) = 
\frac{G_FM_W^3}{6\sqrt{2}\pi}(1+2\delta)|V_L^{cs}|^2~R_W~.
\end{equation}
Currently, although recently there has been some effort in order to
measure partial widths, the assigned experimental errors~\cite{PDG06,OPALvcs}
are much too large to determine reliable $\delta$ and/or the
corresponding mixing matrix elements.
\end{itemize}
\subsubsection{Hadronic tau decays}
\label{tau}
The hadronic tau decays are semileptonic decays involving the charged
current.  Even though the different analysis of these decays done so
far \cite{Davier:2005xq} have not yet reported any evidence of physics
beyond the standard model, it seems interesting to reconsider
them in the light of our generalization of the electroweak charged
current. 

For our analysis
we will consider the normalized total hadronic width given by the ratio 
\begin{equation}
R_{\tau,c} = \frac{\Gamma(\tau^-\to \nu_\tau \mathrm{hadrons}
(\gamma))}{\Gamma(\tau^- \to \nu_\tau e^-\bar\nu_e)}~,
\label{rtau}
\end{equation}
where $c$ can be $V,A$ or $S$ depending whether one considers the
vector, axial or strange channel, respectively. Additional information
is provided by the moments $R_{\tau,c}^{(kl)}$ which explore the invariant 
mass distribution of final state hadrons,
\begin{equation}
R_{\tau,c}^{(kl)} = \int_0^{m_\tau^2} ds\,
\left( 1-\frac{s}{m_\tau^2} \right)^k\, \left( \frac{s}{m_\tau^2} \right)^l\,
\frac{dR_{\tau,c}}{ds}~.
\label{moments}
\end{equation}
The two experimental collaborations ALEPH~\cite{Schael:2005am} and
OPAL~\cite{Ackerstaff:1998yj,Abbiendi:2004xa} have presented precise results
for the total ratio $R_{\tau} = R_{\tau,V} + R_{\tau,A} + R_{\tau,S}$ as well
as separate results for the vector, the axial, and the strange channel,
respectively. In addition they give results for different measured moments. 
More precise data on
tau decays are to be expected from the $B$-factories.

The inclusive character of these quantities allows for a theoretical
description of the hadronic part in terms of the operator product expansion
(OPE). This has triggered much work testing QCD -- in particular quark masses
and the value of $\alpha_s$ -- at the tau mass scale (for a recent review
see~\cite{Davier:2005xq}) assuming standard model weak interactions.  We shall
focus our discussion on the differences arising with respect to this standard
analysis of inclusive hadronic decay rates due to non-standard charged current
interactions. In particular, we will not discuss in detail the description of
the hadronic part, but refer the reader to the comprehensive literature on
that subject (for a recent review see~\cite{Davier:2005xq}).

The ratios, Eqs.~(\ref{rtau}) and (\ref{moments}), can be written
as follows~\cite{Braaten:1991qm,LeDiberder:1992fr}:
\begin{eqnarray}
R_{\tau,V}^{(kl)} &=& \frac{3}{2}\ S_{EW}\ 
|\vfud|^2 \Big( r_{kl} + \delta^{(0),kl} + \delta^{'}_{EW} +
 \sum_{D = 2,4,...} \delta^{(D),kl}_{ud,V}\Big) \nonumber \\
R_{\tau,A}^{(kl)} &=& \frac{3}{2}\ S_{EW}\ 
|\afud|^2 \Big( r_{kl} + \delta^{(0),kl} +\delta^{'}_{EW} +
 \sum_{D = 2,4,...} \delta^{(D),kl}_{ud,A}\Big) \nonumber \\
R_{\tau,S}^{(kl)} &=& \frac{3}{2}\ S_{EW}\ \Big( 
|\vfus|^2 [r_{kl} + \delta^{(0),kl} + \delta^{'}_{EW} +
 \sum_{D = 2,4,...} \delta^{(D),kl}_{us,V}] \nonumber \\ && \quad 
+|\afus|^2 [ r_{kl} + \delta^{(0),kl} + \delta^{'}_{EW} +
 \sum_{D = 2,4,...} \delta^{(D),kl}_{us,A}]\Big)~. 
\end{eqnarray}
$S_{EW} = 1.0198$~\cite{Davier:2005xq,Davier:2002dy} denotes a small
electroweak radiative correction. The residual electroweak correction,
$\delta^{'}_{EW} = 0.0010$~\cite{Braaten:1990ef} will be neglected in
what follows. $r_{kl}$ is a normalization coefficient for the purely
perturbative part. It determines the parton level prediction for the
decay rates and moments. For $k = l =0, r_{00} =
1$. The other values are listed in appendix~\ref{apptau}.
The quantities $\delta^{(D),kl}_i$ are QCD corrections. They are
functions of several QCD parameters: $\alpha_s$, quark masses and
non-perturbative condensates.  $\delta^{(0),kl}$ describes the
massless perturbative contribution, and $\delta^{(2),kl}_i$ are
corrections due to non-zero quark masses. The terms $\delta^{(D),kl}$
for $D \ge 4$ comprise non-perturbative contributions within the OPE
expansion.  In the following discussion we will be interested in the four quantities:
\begin{eqnarray}
\Delta^{+,kl}_{ui} &=& \frac{1}{r_{kl} + \delta^{(0),kl}}\,\frac{1}{2}\sum_{D = 2,4,6,...}
(\delta^{(D),kl}_{ui,V} + \delta^{(D),kl}_{ui,A}) \nonumber \\
\Delta^{-,kl}_{ui} &=& \frac{1}{r_{kl} + \delta^{(0),kl}}\,\frac{1}{2}\sum_{D = 2,4,6,...}
(\delta^{(D),kl}_{ui,V} - \delta^{(D),kl}_{ui,A})~.
\end{eqnarray}
with $i = d$ or $s$. 
\paragraph{Spectral functions}
Much effort has been devoted to
obtain values for the non-perturbative condensates in the $VV-AA$
channel employing different weighted sum rules based on the vector and
axial hadronic spectral functions extracted from hadronic tau decay
data (see e.g.~\cite{Cirigliano:2003kc,Narison:2004vz, Almasy}). In our case,
this analysis would involve the parameter $\ens$, too.  
In analogy with Eq.~(\ref{qcd}) we can write the spectral functions
in the $VV-AA$ channel as: 
\begin{eqnarray}
v(s) &=& \hat{v}(s)\\
a(s) &=& (1 + 4\, \ens) \, \hat{a}(s)\\
v(s) -
a(s) &=& (1+ 2\, \ens) (\hat{v}(s) - \hat{a}(s)) - 2\, \ens
(\hat{v}(s) + \hat{a}(s))~,
\label{vaeq}
\end{eqnarray}
where the quantities with a hat represent again the quantities
extracted from experiment
assuming SM electroweak interactions, i.e., under the assumption $\veff
= -\aeff$.  It is clear from Eq.~(\ref{vaeq}) that, although
the difference between the vector and the axial current is measured
rather precisely, this is only
of limited usefulness in our case since we cannot easily disentangle
electroweak ($\ens$) and QCD quantities ($v(s),a(s)$). We shall
mention in particular one point. The function $v(s)-a(s)$ should
vanish for sufficiently large values of $s$, when the perturbative
regime is reached. This does not necessarily imply that
$\hat{v}(s) - \hat{a}(s)$ should vanish: for a nonzero value of
$\ens$, this difference is proportional to $\ens (v(s) + a(s))$. The
expected values for $\ens$ are, however, much too small for this
remark to be relevant for the discussion of quark-hadron duality
violations at the tau mass scale from the ALEPH and OPAL data.

\paragraph{Non-strange sector}

Let us begin the discussion with the non-strange sector.  The only
parameter involved in that case is $\ens$. Indeed since we are only
considering tree-level charged current processes, the non-strange
sector will not furnish us any information on the parameter $\es$
describing RHCs involving strange quarks. Furthermore, since
$|\vfud|^2$ can be determined rather precisely from superallowed beta
decays, the parameter $\delta$ does not appear, either, see
Eq.~(\ref{delta}).

\vspace{0.3cm}
\noindent i) VV+ AA
\vspace{0.1cm}

We will first discuss the $VV+AA$ channel because the total non-strange
rate $R_{\tau,V+A}$ and the corresponding moments are more easily
accessible experimentally than the separate quantities in the vector
and the axial channel and it is generally assumed that in the $VV+AA$
channel the non-perturbative contributions are extremely small, of the
order of $10^{-3}$. This is based on the following theoretical
consideration.  Since in the non-strange sector the contributions
proportional to light quark masses are negligible, the dominant
non-perturbative contribution arises from $D=6$ condensates. In the
large-$N_c$ limit, these condensates factorize\footnote{Employing the
large $N_c$ argument, the factorized expression in
Ref.~\cite{Braaten:1991qm} should be multiplied by a factor $(N_c^2 -1
)/N_c^2 = 8/9$.}. In the
$VV+AA$ channel the large-$N_c$ result should be rather reliable,
because it does not involve any order parameter of chiral symmetry
breaking. In this case the
contribution of the $D=6$ condensates in the $VV+AA$ channel remains
small due to a partial cancellation between the vector and axial
channel. A fit to the data presented by the ALEPH collaboration, seems
to confirm the small value~\cite{Schael:2005am}. It should, however,
be stressed that the theoretical argument is essentially based on
prejudices and that the fit to the data has been done assuming SM weak
interactions. We do not have any firm knowledge about the exact value
of the non-perturbative contributions.

We can, of course, perform a combined fit to the data, trying to
determine the QCD part and $\ens$ at the same time. Before discussing
this option in more detail, it is instructive to have a closer
look at the expression for the decay rate in
the $VV+AA$ channel.  To first order in $\ens$ we obtain:
\begin{eqnarray}
R_{\tau,V+A} &=& 
3\ S_{EW}\,|\vfud|^2  \Big(1 +
    \delta^{(0)}\Big)\Big(1 - 2\,
    \ens\Big)\Big(1+\Delta^{+}_{ud}\Big)\nonumber \\
&\approx& 
3\ S_{EW}\,|\vfud|^2  \Big(1 +
    \delta^{(0)}\Big)\Big(1 - 2\,
    \ens +\Delta^{+}_{ud}\Big)~,
\label{rns}
\end{eqnarray}
where we have neglected in the last line products of $\ens$ with the
non-perturbative contributions $\dpns$ which should be at most of the
order of $10^{-4}$. In the absence of non-standard electroweak
interactions, the above relation determines the value of $\atau$
rather precisely since the non-perturbative contribution,
$\Delta^+_{ud}$, is presumably very small. On the contrary, allowing
for a coupling of right-handed quarks to $W$, we clearly see the
strong correlation between the extracted value of $\atau$ and $\ens$
(assuming that the non-perturbative part is indeed small). Similar
conclusions can be drawn for the corresponding moments in the $VV+AA$
channel, which can be written in essentially the same way as
Eq.~(\ref{rns}).

\vspace{0.3cm}
\noindent ii) Ratio of axial and vector channel
\vspace{0.1cm}

Another possible way to extract the value of $\ens$ is to combine
measured quantities in the vector and the axial channels. This is,
however, less favorable than in the $VV+AA$ channel for a theoretical
and an experimental reason.

\begin{itemize}
\item 
Let us look at the ratio of
the decay rate in the axial and in the vector channel which reads to first
order in $\ens$:
\begin{equation}
\frac{R_{\tau,A}}{R_{\tau,V}} =
 \frac{1 + \Delta^+_{ud} - \Delta^-_{ud}} { 1 + \Delta^+_{ud} + \Delta^-_{ud}}  
(1-4\
 \ens) \approx (1 - 2\,\Delta^-_{ud} - 4\, \ens)~.
\label{rav}
\end{equation}
We have again neglected products of $\ens$ with the non-perturbative
contributions $\Delta^{+/-}_{ud}$ which should be of the order of
$10^{-4}$.  Note that the purely perturbative contribution cancels in
this ratio. It measures the combination $4\ \ens + 2 \Delta^-_{ud}$.  The
latter describes non-perturbative contributions in the $VV-AA$ channel.
Since the corresponding correlator represents an order parameter of
chiral symmetry breaking, corrections to the large-$N_c$ limit are
expected to be large~\cite{Descotes-Genon:1999uh,Stern:1998dy}.
Present estimates give values of the order of $\Delta^-_{ud} \approx
10^{-2}$, thus of the same order as the expected effect of $\ens$. 
$\Delta^-_{ud}$ should therefore be known  to a very high precision in order to
extract the value of $\ens$.  The same argument applies to all
quantities involving separately the vector and the axial channel: the
determination of $\ens$ from these quantities is strongly correlated
with the determination of non-perturbative contributions in the $VV-AA$
channel, in particular the $D=6$ condensates.  
\item
Experimentally for some decay channels, in particular those involving
kaons, it is not easy to disentangle vector and axial channel, such
that the data are
correlated~\cite{Ackerstaff:1998yj,Abbiendi:2004xa,Schael:2005am} and
subject to larger uncertainties.
\end{itemize}

\vspace{0.3cm}
\noindent iii) Fits
\vspace{0.1cm}

We can certainly try to infer the value of the non-perturbative
contributions from a combined fit of these QCD quantities together
with $\ens$ to tau decay data in the non-strange sector including
$R_{\tau,V}, R_{\tau,A}$ and the different measured spectral moments,
see Eq.~(\ref{moments}).  Of course, we cannot pretend to determine
within these fits the QCD parameters reliably enough to be able to
disentangle really quantitatively non-standard EW couplings of quarks
to $W$ of the order of percent.  In addition, as already mentioned,
the extracted value of $\ens$ will be strongly correlated with the QCD
parameters.  The analysis is nevertheless worth doing in order to show
that no inconsistencies appear.  We use the data provided by the OPAL
collaboration for the total rate $R_{\tau,V+A}$ and the spectral
moments, see Eq.~(\ref{moments}), $R_{\tau,V+A}^{ij}$ with ${(i,j)}
\in \{(1,0),(1,1),(1,2),(1,3),(2,0),(2,1),(3,0),(4,0)\}$
\cite{Ackerstaff:1998yj,Abbiendi:2004xa} as well as the rates
$R_{\tau,V/A}$ and the spectral moments for axial and vector channel,
$R_{\tau,V/A}^{ij}$ with ${(i,j)} \in \{(1,0),(1,1),(1,2),(1,3)\}$
\cite{Ackerstaff:1998yj}.  A crosscheck with the ALEPH data is at the
moment not possible because the experimental correlations between the
vector and the axial channel necessary to perform the combined fit are
at present not available.  Our fits should thus primarily be seen as
an illustration of the possibilities offered by the analysis of tau
decay data, awaiting some details about the ALEPH data and more data
from the B-factories.

Our assumptions for the theoretical description of the QCD part are as
follows.  
\begin{itemize}

\item
  
  To evaluate the purely perturbative part, we employ two different
  prescriptions, fixed order perturbation theory (FOPT) and contour
  improved perturbation theory (CIPT). We go up
  to the order $\alpha_s^6$~\cite{LeDiberder:1992fr, Davier:2005xq}.
  In the former case the series does not converge fast and it involves
  unknown coefficients already at order $\alpha_s^4$. We follow
  Ref.~\cite{Davier:2005xq} for the unknown coefficients of
  $\mathcal{O}(\alpha_s^4),\mathcal{O}(\alpha_s^5),$ and
  $\mathcal{O}(\alpha_s^6)$ assuming a geometric growth for
  $\mathcal{O}(\alpha_s^5,\alpha_s^6)$, see appendix~\ref{apptau}. The
  latter method allows to partially resum higher order logarithmic
  contributions and it improves the convergence of the perturbative
  series. It should thus in principle be more reliable (see also the
  discussion in Ref.~\cite{ciptjamin}).

\item In the non-strange channel, the $D = 2$ contribution is
  completely negligible, since it is suppressed as
  $m_{u/d}^2/M_\tau^2$.  For simplicity we therefore take $m_u = m_d =
  0$. In this limit the $D=4$ contribution contains only terms
  proportional to $m_s\langle\bar s s\rangle$ and to the gluon
  condensate, $\langle \alpha_s/\pi GG\rangle$.  The contribution of
  the latter to the total rate is very small, since it is suppressed
  as $\atau^2$, but for different moments, it enters with an
  $\mathcal{O}(1)$ coefficient.  The value of the quark condensate is
  fixed using the Gell-Mann--Oakes--Renner (GMOR) relation namely 
  $-m_s\langle\bar s s\rangle = F_K^2 m_K^2$, whereas our theoretical
  knowledge about the value of the gluon condensate is rather poor. On
  the classical level it can be shown that it should be positive, but
  quantum corrections could obstruct this result. Since it is not
  protected by any symmetry, it is even not really well defined
  because of ambiguities arising from additive corrections.  Present
  determinations from tau decay data give values of the order
  $|\langle \alpha_s/\pi GG\rangle| \sim 0.01 \mathrm{GeV}^4$ or
  smaller with large errors~\cite{Davier:2005xq,Ackerstaff:1998yj}.
  In particular, the determinations from the different channels do not
  always agree. Other phenomenological determinations, e.g. from
  different QCD sum rules (see for instance
  \cite{Yndurain:1999pb,Ioffe:2002be}) give values of the same order,
  subject to large errors, too. Most determinations give $\asgg > 0$.
  The most natural thing to do in this case is to include the gluon
  condensate as free parameter into the fit. But, this will induce
  large errors because due to the large correlations discussed above,
  the fits are not sensitive to many different non-perturbative
  parameters.

\item
Concerning the $D = 6$ contribution, we have employed different
prescriptions in the $VV+AA$ and the $VV-AA$ channel.  For the former we
take the factorized expression with $\langle\bar q q\rangle = -(270
\mathrm{MeV})^3$. The latter value can be motivated taking the GMOR
relation and a reasonable value for the strange quark mass. As is
clear from the above discussion, due to a partial cancellation between
the vector and the axial part, the $D=6$ contribution remains small if
we use the factorized expression, such that the precise value of the
quark condensate is not important for our analysis in this case.  In
the $VV-AA$ channel we parametrize the $D=6$ contribution with two
parameters $a_{6,V-A}$ and $b_{6,V-A}$ (see for example
Ref.~\cite{Cirigliano:2003kc} for an explanation of the notation as
well as for the relation between these parameters and the expectation
values of the corresponding operators which are known for $D=6$). 
For $D = 8,10$ we include only two
parameters $a_{8,V+A},a_{8,V-A},$ and $a_{10,V+A},a_{10,V-A}$.  In the
$VV-AA$ channel different analysis employing sum rules based on the
vector and axial spectral functions extracted from tau decay data show
that non-perturbative contributions for $D > 10$ can in principle be
important (see e.g. Ref.~\cite{Cirigliano:2003kc,Narison:2004vz}). We
refrained from including higher order condensates, since, in
particular in the $VV+AA$ channel, with the given data, the sensitivity
of the fits to these contributions is very low.
\end{itemize}
\FIGURE[t]{
\epsfig{file=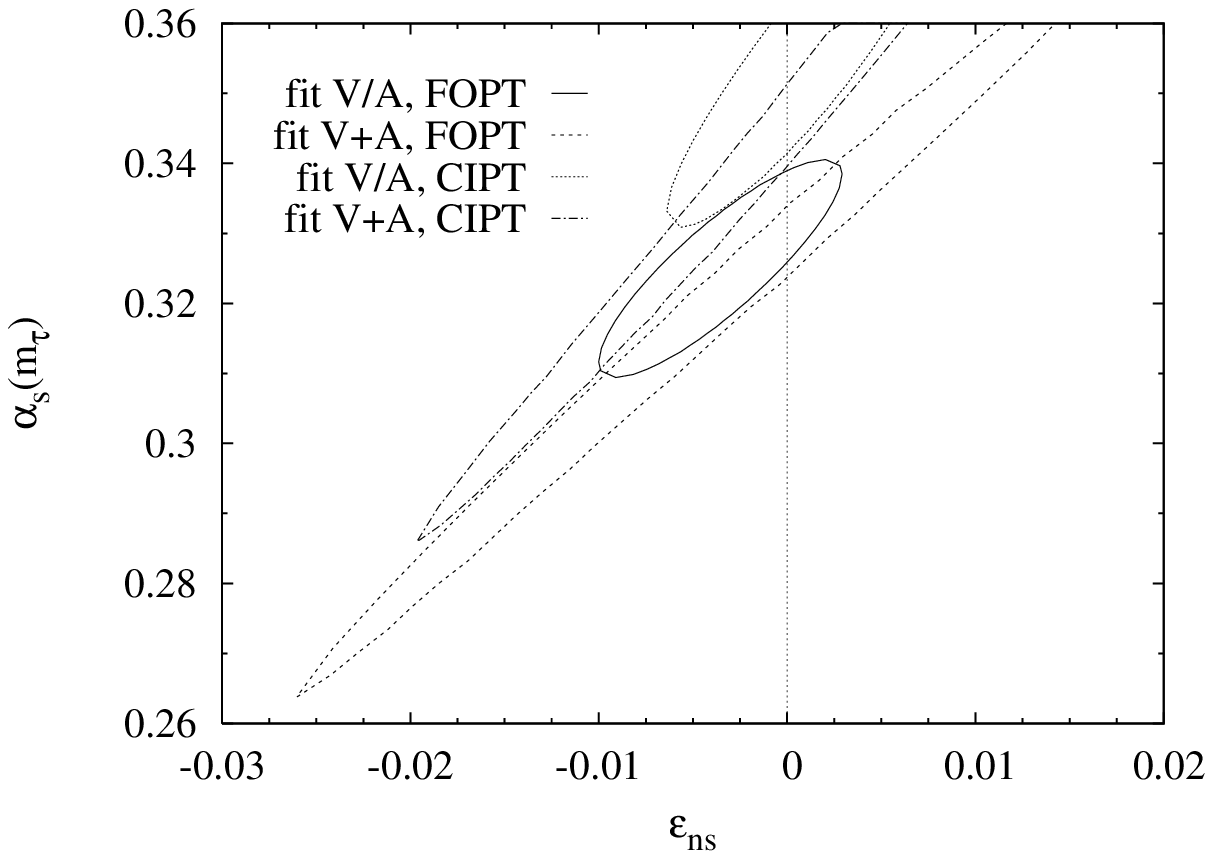,width=7.5cm}\hfill
\epsfig{file=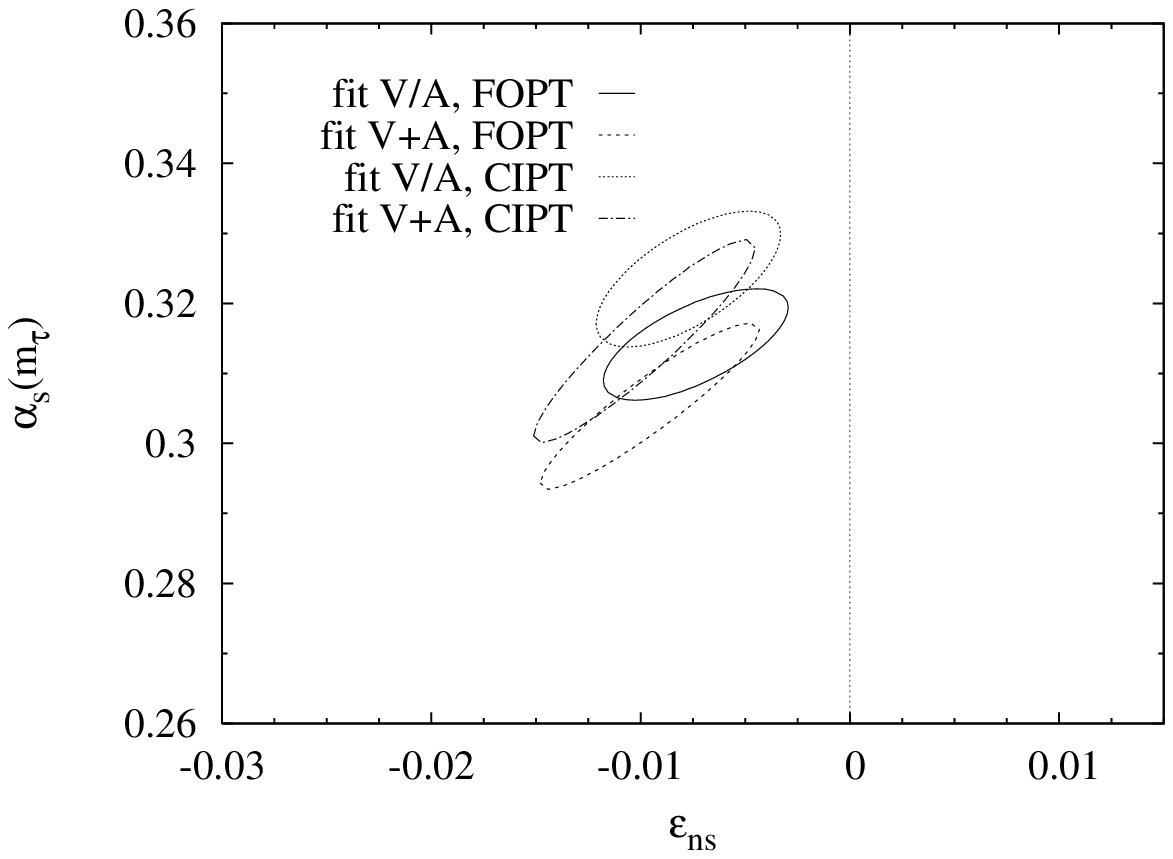,width=7.5cm}
\caption{\it One-sigma contours for the extraction of $\atau$
  and $\ens$ from the $VV+AA$ and $V/A$ channel. Data are from
  the OPAL
  collaboration~\cite{Ackerstaff:1998yj,Abbiendi:2004xa}. Left: Result
  including the gluon condensate as fit parameter, right: fixing the
  value of the gluon condensate at $\asgg = 0$.}
\label{figrns}
}

We obtain good fits to the data, and the results of all fits are
compatible. In particular, the values for the non-perturbative
parameters are perfectly compatible within errors between the $V/A$
fits and the $VV+AA$ fits.  This is not the case for the fits assuming
SM weak interactions presented by the experimental
collaborations~\cite{Schael:2005am,Ackerstaff:1998yj}. But, as already
anticipated, the fits are not sensitive to the many unknown
non-perturbative coefficients. Especially the value of the gluon
condensate induces a large error, which makes a really quantitative
determination of $\ens$ difficult. This is reflected in the elongated
$1\sigma$ contours which we show as a function of $\ens$ and $\atau$
in Fig.~\ref{figrns}, left panel. Fixing the gluon condensate at some
value, the result is much better determined, see Fig.~\ref{figrns},
right panel, where we show as an example the result with $\asgg = 0$.
Increasing the value for $\asgg$ shifts the ellipses in direction of
smaller values for $\ens$ and $\atau$, whereas upon decreasing the
value of $\asgg$ shifts the ellipses towards higher values of $\ens$ and
$\atau$. 
Another important effect already anticipated from the discussion of
Eq.~(\ref{rns}) is obvious: the extracted value of $\ens$ is strongly
correlated with that of $\atau$, see Fig.~\ref{figrns}.  Vice versa,
this implies that the determination of $\atau$ from hadronic $\tau$
decay data depends on the assumptions made for the electroweak charged
current interaction.  It can be observed that the CIPT fits give
slightly larger values of $\atau$ than the FOPT fits. The same effect
has already been seen in the determinations of $\atau$ from hadronic
tau decay data assuming SM weak
interactions~\cite{Davier:2005xq,Ackerstaff:1998yj}. In general, our
values for $\atau$ are smaller than those obtained assuming SM
couplings of quarks to $W$, if $\ens < 0$ and larger if $\ens > 0$ in
agreement with Eq.~(\ref{rns}).
 
Another source of uncertainty concerning these results should be
mentioned. The quantitative outcome of our fits is rather
sensitive to the treatment of the perturbative part.  Assigning an
error of 100\% to the unknown perturbative coefficients of
$\mathcal{O}(\alpha_s^4),\mathcal{O}(\alpha_s^5),\mathcal{O}(\alpha_s^6)$
the central value of $\atau$ changes by $\sim 0.1$. For the fits
employing FOPT the change is even more pronounced if we cut the
perturbative expansion at lower orders in $\alpha_s$. The ellipses are
then shifted towards larger values for $\ens$ and $\atau$. 
 
We should of course ask, can we improve on the precision, using input
on the QCD parameters, for example $\atau$, from other sources ?  There
are many different ways to determine the value of $\alpha_s$ (cf, for
example, Ref.~\cite{alphas}). The most precise
determinations, at the $Z$-pole and from tau decay data, depend on
electroweak physics, see also Section~\ref{zpole}.  There are
determinations of $\alpha_s$, for instance from jet and event-shape
observables, which do not suffer from this drawback.  Unfortunately,
the precision reached, in particular for the determinations which do
not depend on electroweak physics, is not high enough to further limit
the range and increase the precision on $\ens$.

Our conclusion is that presently the uncertainties are such that the
analysis of tau decay data in the non-strange channel does not allow
us to determine $\ens$ quantitatively. All that can be said is good
fits are obtained for values of roughly $-0.02 \lsim \ens \lsim 0.02$.
These values are perfectly in agreement with the order of magnitude
estimates from the LEET.

\paragraph{Strange sector}
Recently a lot of work (see for instance
Refs.~\cite{Maltman:2006mr,Gamiz:2006nj}), assuming
the absence of RHCs, has been devoted to the extraction of $V_{us}$
and $m_s$ from Cabbibo suppressed tau decays. The main quantity in
this context is~\cite{Barate:1999hj,Pich:1999hc}
\begin{equation}
\delta R_{\tau} = \frac{R_{\tau,V+A}}{|V_{ud}|^2} - 
\frac{R_{\tau,S}}{|V_{us}|^2}~.
\end{equation}
This quantity vanishes in the $SU(3)$ flavor limit such that
theoretical uncertainties are reduced. In the presence of RHCs,
$\delta R_\tau$ cannot be properly normalized since the couplings in
the axial and vector channel are no longer the same and in the strange
sector, the vector and axial parts cannot be separated experimentally.
Instead, we will look at the ratio of the strange and the non-strange
contribution.  It can be written to first order in the spurionic
parameters as:
\begin{equation}
\frac{R_{\tau,S}}{R_{\tau,V+A}} 
= \frac{\sin^2\hat\theta}{\cos^2\hat\theta}\, \Big(1+2\ \frac{\ens+ \delta}{\sin^2\hat\theta}\Big)\, (1 + \Delta^+_{us}-\Delta^+_{ud})
\label{rsrns}
\end{equation}
where again we have neglected terms of the form $\ens \Delta^-_{ud}$ and 
$\es \Delta^-_{us}$, which are of the order $10^{-4}$.
Contrary to what one would naively expect, this ratio is in fact independent
of $\es$ precisely due to the fact that this quantity only enters together
with $\Delta^-_{us}$. Note that the hadronic part of this ratio contains the
same $SU(3)$ breaking quantity as $\delta R_\tau$ within the SM.

The QCD corrections in this case are dominated by the mass corrections
to $\Delta^+_{us}$ which are proportional to $m_s^2(m_\tau)/m_\tau^2$.
Unfortunately, the perturbative series for the Wilson coefficient of
the $D=2$ term proportional to $m_s^2(m_\tau)/m_\tau^2$ converges
badly~\cite{Pich:1998yn}, such that there are large theoretical
uncertainties concerning this coefficient.  In the literature one can
find different attempts to cure this problem. One possibility is to
try to improve on the convergence of the series, for instance by
employing contour improved perturbation theory instead of fixed order
perturbation theory.  More phenomenologically, one can replace the OPE
by a direct integration of a (model dependent) parametrization of the
hadronic spectral function~\cite{Gamiz:2002nu,Gamiz:2004ar} to better
determine the non-perturbative contributions in the strange sector
(cf. the discussion in Ref.~\cite{Davier:2005xq}).  Including
sub-leading $D=4$ corrections~\cite{Braaten:1991qm} involving in
particular terms proportional to $m_s^4(m_\tau)/m^4_\tau$ and
$(m_s\langle \bar s s \rangle - m_d\langle\bar d d\rangle)/m_\tau^4$
values for $\Delta^+_{us}$ in the literature are in the range $-0.06$
to $-0.15$ depending on the value of the strange quark mass. The range
given here corresponds to varying $m_s$ from $80$ to $200$ MeV.

Since $r_s = R_{\tau,S}/R_{\tau,V+A} \
\cos^2\hat\theta/\sin^2\hat\theta$ is of the order one and the
hadronic correction $\Delta^+_{us} -\Delta^+_{ud}$ is at most on the
10 percent level, the above equation is rather sensitive to $\ens +
\delta$ (remember that $1/\sin^2\hat\theta$ is about 20).
Let us now be more specific: what does Eq.~(\ref{rsrns}) tell us about
the value of $\ens + \delta$ ? With the ALEPH
data~\cite{Schael:2005am} we obtain $r_s = 0.84 \pm 0.03$ and with the
OPAL data~\cite{Abbiendi:2004xa} $r_s = 0.89 \pm 0.01$.  This value
has to be compared with the possible values of $\ens + \delta$ and the
QCD corrections. We have seen previously that the latter are not well
determined and that in addition, the strange quark mass is also not
well known. Thus, a precise determination of $\ens+ \delta$ is at
present not possible. 
However we can have an estimation of its order of magnitude by
comparing different prescriptions for the calculation of the $D=2$
Wilson coefficient and varying $m_s(m_\tau)$ between 60 and 250
MeV. This should be a conservative estimate for possible values of the
strange quark mass.
\begin{itemize}

\item
The first observation is that with the present estimates of
$\Delta^-_{us}$ we need rather large values for the strange quark mass
($m_s(m_\tau) \sim 150-200$ MeV) if we impose $\delta+\ens = 0$,
i.e. in the SM case. This is consistent with the finding that $V_{us}$
should be somewhat smaller than the value obtained from unitarity with
values of $m_s$ of the order of 95 MeV (cf for
instance~\cite{Gamiz:2004ar}).

\item
The second observation is that, for not too large values of the
strange quark mass, the extracted value of $\delta + \ens$ lies
between roughly -0.005 and 0.005 whatever method used. This indicates
that $\delta+\ens$ should be below 1\%. 
\end{itemize}
In principle it is possible to exploit experimental information on the
moments $R^{kl}_{\tau,S}/R^{kl}_{\tau,V+A}$. But without any further
independent information on the QCD corrections, this will not
determine $\delta+\ens$ more precisely. 
\section{Discussion}
\label{sec:discussion}
Let us now compare the different sources of information we have on the
different parameters. One of our main results is that, apart from
$K^L_{\mu 3}$ decays, discussed in Section~\ref{kaons}, it is
difficult to find any other clean manifestation of RHCs due to the
parameter $\es$, the only parameter which can be larger than the
genuine spurion parameter $\epsilon$. In particular, the data on
hadronic tau decays in the strange sector are not, in contrast to
naive expectations, sensitive to $\es$. The data by the NA48
collaboration~\cite{NA48} on $K^L_{\mu 3}$ decays indicate indeed an
enhancement of $\es$, i.e. an inverted hierarchy for the flavor mixing
of right-handed quarks.

One remark of caution is in order here: much of the numerology
discussed within this section depends on the value of $|\vfud| =
\cos\hat\theta$. We took the value (see Eq.~\ref{costheta}) from
$0^+\to 0^+$
nuclear beta decays. This is by far the most precise one: the error is
about one order of magnitude smaller than the error on the alternative
determinations from pion and neutron beta decays (cf for instance
Ref.~\cite{Hardy07}). We should nevertheless keep in mind that a
change in the value of $|\vfud|$ affects the numbers in particular for
the hatted quantities.

The result for the value of $\es-\ens$ discussed in
Section~\ref{kaons} has another interesting application. From
Eqs.~(\ref{fkfpi},\ref{qcd}) we see that the two quantities
$f_+^{K^0\pi^-}(0)$ and $F_{K^+}/F_{\pi^+}$ depend on the same two
combinations of spurion parameters: $\delta+\ens$ and $\es-\ens$.
Their values are therefore related.
\FIGURE[t]{
\epsfig{width=10cm,file=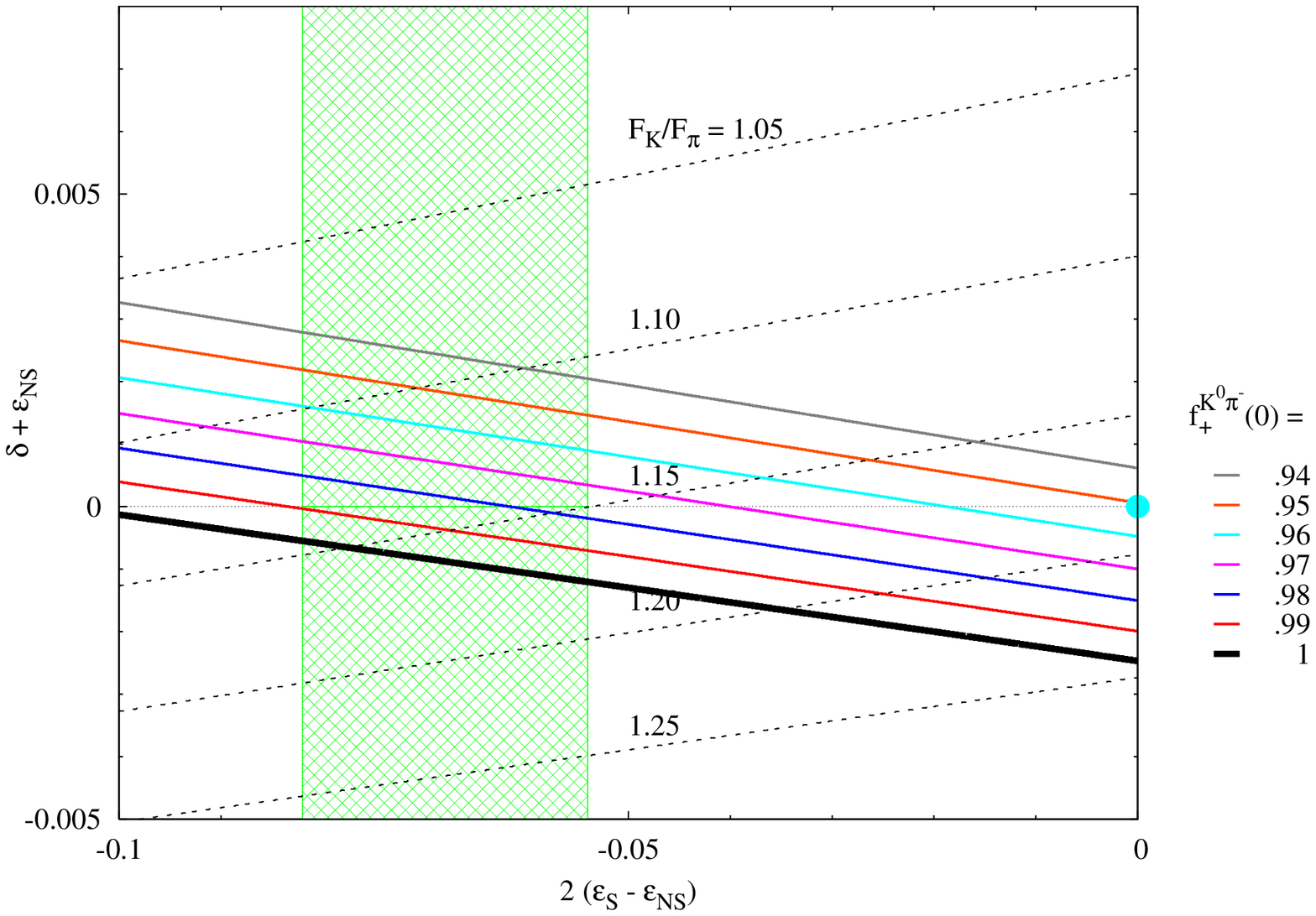}
\caption{\it Lines of constant values for $F_{K+}/F_{\pi+}$ and $f_+^{K^0\pi}(0)$ in the plane
  $\delta+\ens$ and $2 (\es-\ens)$ as resulting from
  Eqs.~(\ref{fkfpi},\ref{qcd}).  The vertical band indicates the range
  for $\Delta\epsilon - \Delta_{\mathrm{CT}}^{\mathrm{NLO}}$ from the
  NA48 data~\cite{NA48}.}
\label{figfkfpi}
}
In Fig.~\ref{figfkfpi} we display
lines of constant values for $f_+^{K^0\pi}(0)$ and $F_{K^+}/F_{\pi^+}$
in the plane $\delta+\ens$ and $2(\es-\ens)$. Both,
$f_+^{K^0\pi^-}(0)$ and $F_{K^+}/F_{\pi^+}$ decrease with increasing
$\delta+\ens$, i.e. from bottom to top. We have indicated
$f_+^{K^0\pi^-}(0) = 1$ with a thick line since we expect this value
to represent an upper bound~\cite{fp0}. In the large $N_c$-limit this
bound becomes exact. The shaded region indicates the determination of
$2 (\es-\ens)$ from the NA48 data~\cite{NA48} assuming
$\Delta_{\mathrm{CT}} = \Delta_{\mathrm{CT}}^{\mathrm{NLO}}$. Taking
the upper bound for $f_+^{K^0\pi^-}(0)$ seriously, this imposes in turn
an upper bound on $F_{K^+}/F_{\pi^+} < 1.19$. If we assume the recent
determination of $f_+(0) = 0.9680(16)$ from lattice
simulations~\cite{fplattice} we obtain an even smaller value,
$F_{K^+}/F_{\pi^+}  = 1.12(2)$. The isospin breaking corrections
relating $F_{K^+}/F_{\pi^+}$ to the value of $F_{K}/F_{\pi}$
conventionally used in ChPT are presumably very small.

One interesting point should be mentioned. The ChPT
prediction~\cite{GL85} for the slope of the scalar $K\pi$ form factor
depends strongly on the value of $F_K/F_\pi$. Taking the values for
$F_K/F_\pi$ discussed above, the ChPT prediction is in agreement
with the values for the slope from NA48~\cite{NA48} and KTeV~\cite{Ktevl0},
i.e., the ChPT prediction is in complete agreement with our
findings.

Note that the effect of $\delta+\ens$ is enhanced in the relations,
Eqs.~(\ref{fkfpi}),(\ref{qcd}), because of the factor
$1/\sin^2\hat{\theta}$. We therefore expect $\delta+\ens$ to remain
well below 1\%.  This is consistent with the finding from the hadronic
tau decays in the strange sector: for reasonable strange quark masses,
$\delta+\ens$ does not exceed half a percent in this case, see
section~\ref{tau}.

We have another indication that $\delta + \ens$ should be small. To
that end let us look at the sum of the elements of the effective
mixing matrix, $\vfud$ and $\vfus$, squared, see
Eq.~(\ref{effunit}). Inserting the NA48 value for $\es - \ens$ with
$\Delta_{\mathrm{CT}} = \Delta_{\mathrm{CT}}^{\mathrm{NLO}}$ into
the second term of Eq.~(\ref{epsmeasure}), the latter equation can be
rewritten as
\begin{equation}
|\vfud|^2 + |\vfus|^2 = 1 - 0.0036(7) + 2 (\delta + \ens)~. 
\label{veffnum}
\end{equation}
We can now look at different instructive examples. If $\veff$ was
unitary, the left hand side of Eq.~(\ref{veffnum}) would be equal to
one and $(\delta + \ens) = 0.0018(4)$. Considering the recent
theoretical evaluations of $f_+^{K^0\pi^-}(0)$, the left-hand side of
Eq.~(\ref{veffnum}) is indeed very close to one. To be more precise,
with the lattice result~\cite{fplattice} $f_+(0) = 0.9680(16)$ we have
$\delta + \ens = 0.0008(5)$ and with the two-loop ChPT 
evaluation~\cite{Bijnens:2003uy, Cirigliano:2005xn},
$f_+^{K^0\pi^-}(0) = 0.984(12)$, we obtain $\delta + \ens = 0. \pm
0.0007$. This indeed indicates that $\delta + \ens$ is probably very
small. 

\FIGURE[t]{
\epsfig{width=10cm,file=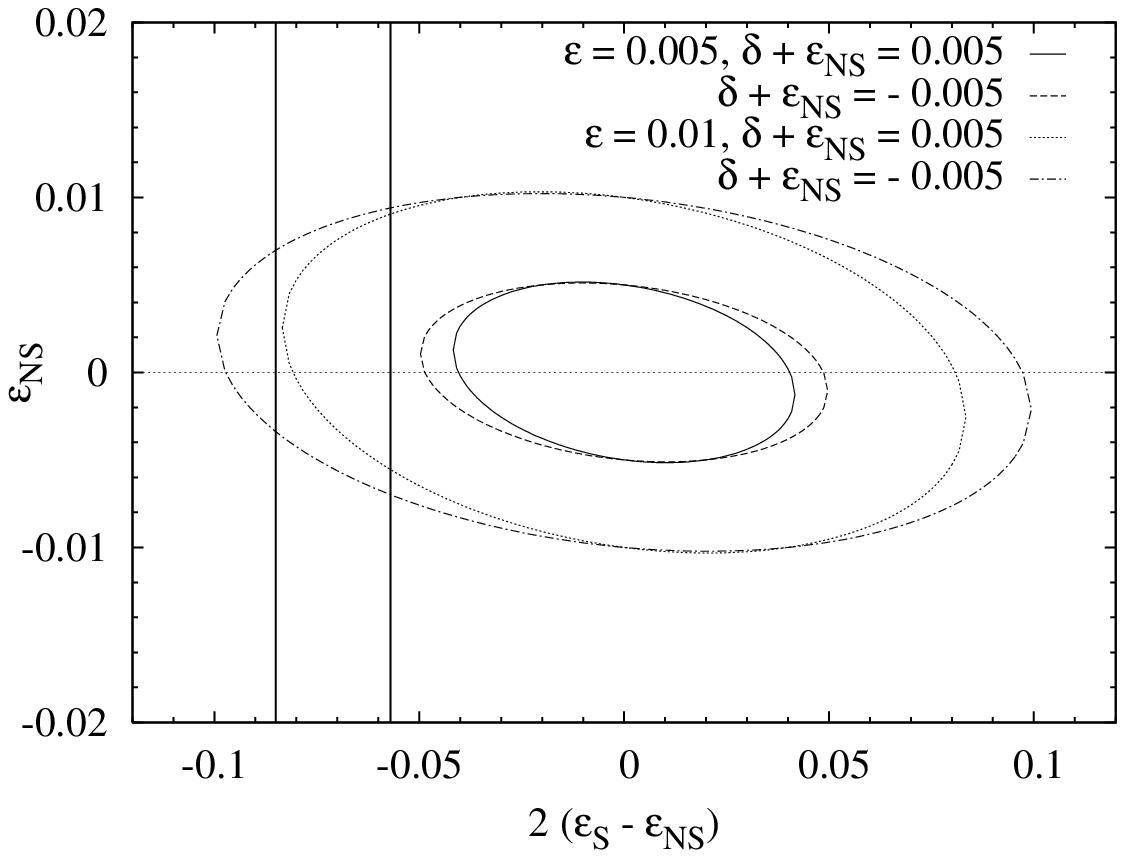}
\caption{\it Maximum values of $\ens$ and $\es-\ens$ compatible with
  the unitarity of $V_{L,R}$, cf
  Eq.~(\ref{ellipsef}), for two different
  values of $\epsilon$ and $\delta + \ens$. 
The vertical lines
  indicate the range for $2 (\es - \ens)$ from the NA48
  data~\cite{NA48} with $ \Delta_{\mathrm{CT}} =
  \Delta_{\mathrm{CT}}^{\mathrm{NLO}}$. }
\label{ellipsens}
}

The determination of $2(\es - \ens)$ from $K_{\mu 3}^L$ decays, see
section~\ref{kaons}, has an impact on the bounds on the spurion
parameters we obtain from the unitarity condition of the mixing
matrices $V_{L,R}$.  In Fig.~\ref{ellipsens} we show the ellipses
giving the maximum values for $\ens$ and $\es-\ens$ compatible with
unitarity of the mixing matrices, Eq.~(\ref{ellipsef}) for two
different values of $\epsilon$ and $\delta + \ens$.  The vertical
lines thereby indicate the range for $2 (\es - \ens)$ from the NA48
data~\cite{NA48} with $ \Delta_{\mathrm{CT}} =
\Delta_{\mathrm{CT}}^{\mathrm{NLO}}$. Note, that in contrast to
Fig.~\ref{figellipse}, we show here absolute values choosing two
values of $\epsilon$. Given a determination of $\es-\ens$ we can thus
obtain two informations. First, for too small values of $\epsilon$,
there are no parameter values compatible with the unitarity of the
mixing matrices and the range of allowed values for $\es-\ens$ at the
same time. Thus we obtain a lower bound on $|\epsilon|$ which is for
the NA48 data, $\epsilon \gsim 0.006$. Second, once $\epsilon$ is
fixed, $\ens$ is constrained. As an example, for $\epsilon = 0.01$ the
NA48 data indicate $-0.007 \lsim \ens\lsim 0.009$ for
$\delta+\ens = -0.005$. These values are perfectly compatible
with the results from the analysis of tau decay data.
\section{Other possible tests}
\label{sec:otp}
Within this section we would like to discuss some other possible
tests. A quantitative exploitation of these tests is beyond the scope
of the present paper but should be considered as prospect for future
work.

\subsection{NLO analysis: hyperon decays and heavy quark sector}
In the right-handed sector, in addition to the genuine spurion
parameters, the quark mixing matrix elements have to be determined. For the
moment, we have only considered the light quark sector. There, the
parameter $\es$ can be enhanced, if $\vrus$ is enhanced with respect
to $\vlus$. We should therefore look for processes where $\es$
enters.  In principle, we can have sizable effects due to $\es$ in
hyperon decays. Recently an experimental effort has been undertaken to
improve the precision on the data (cf e.g.~\cite{Batley:2006fc}).
From the theoretical side, however, the $SU(3)$ breaking effects are
not yet under quantitative control. This constitutes a
severe limitation to the precision attainable in the analysis of
hyperon decays~\cite{Mateu:2005wi}.

Evidently, we will have to explore the heavy quark sector looking for
tests of $V+A$ interactions. There exists already some work in this
direction (cf e.g. \cite{Dassinger:2007pj}), but for the moment the
effective theories used in the heavy quark sector have not reached
sufficient accuracy to
determine the electroweak parameters precisely.
\FIGURE[t]{
\epsfig{width=6cm,file=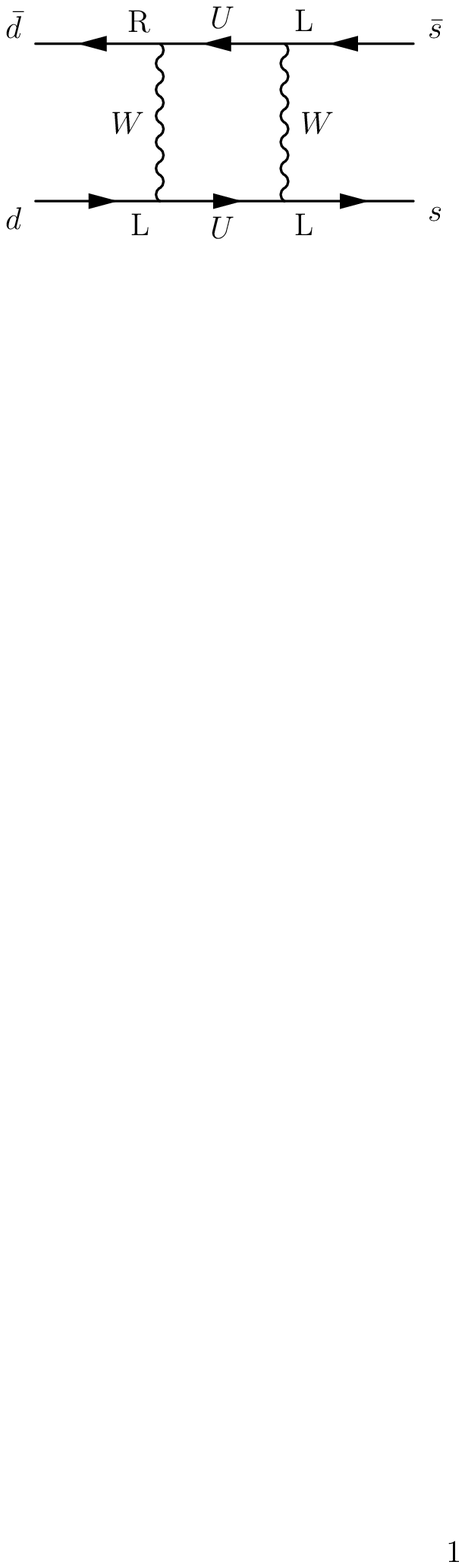}
\caption{\it Box diagram contributing to the $\Delta F = 2$ effective
interaction with one insertion of a right-handed vertex. }
\label{boxgraph}
}

We should therefore look for processes where, as in $K^L_{\mu 3}$
decays, the hadronic part is well controlled (by symmetry
considerations).  Since the Callan-Treiman theorem is based on
$SU(2)\times SU(2)$ chiral symmetry, it can be applied not only to the
$K\pi$ scalar form factor but to scalar form factors involving heavy
quarks, too, for instance the $D\pi$ scalar form factor. It could
thus be measured in $D_{\mu 3}$ decays. In this case the
kinematics is very different: because $m_D \gg m_\pi$, the endpoint of
the physical region, $(m_D- m_\pi)^2= 2.98~\mathrm{GeV}^2$, the
Callan-Treiman point, $m_D^2 - m_\pi^2 = 3.46~\mathrm{GeV}^2$, and the
threshold for $D\pi$ scattering, $(m_D + m_\pi)^2 = 4.02~\mathrm{GeV}^2$ 
are very close together. At present, it does not seem
very promising to exploit the Callan-Treiman theorem in $D$ decays in
the same way as we did in $K$ decays because the actually available
experimental data on $D_{l 3}$ decays are not as precise as for the
$K_{l 3}$ decays. In addition, it is still not possible at the moment
to extract $D\pi$ phase shifts which are crucial to establish the
dispersive representation of the form factor allowing to extrapolate
the data in the physical region of $D_{\mu 3}$ decays to the
Callan-Treiman point.

The decay $t \to Wb$ is a very promising process to directly access
the chirality of the $\bar t W$ coupling measuring the $W$
polarisation.  There has been a first pioneering attempt from
Tevatron~\cite{Abazov:2005fk}. At present the uncertainties are too large to
disentangle a small admixture of RHCs, but hopefully there will be
more data from LHC, which will start to operate soon.

It has been further observed that a coupling of right-handed quarks to
$W$ would alter the chiral structure of the $\mathcal{O}(p^2)$
tree-level effective weak Hamiltonian, and, in combination with
soft-pion theorems, bounds on right-handed couplings can be derived,
for example, from $K\to \pi\pi$, $K \to \pi\pi\pi$
decays~\cite{Donoghue:1982mx}.  Unfortunately, $\mathcal{O}(p^4)$
chiral corrections, in particular long distance loop corrections and
final state interaction can be rather important and can upset small
$\mathcal{O}(p^2)$ effects.

Note that at the present order no significant modifications of the muon
$(g-2)$ arise. First, no additional contributions arise within the
LEET at this order since charged right-handed currents in the lepton
sector are absent.  Second, the Higgs contribution within the SM is
several orders of magnitude smaller than the dominant contributions,
such that the absence of this contribution in the LEET does not change
the conclusions.

\subsection{Loop effects: flavor changing neutral current processes and CP violation}
Flavor changing neutral current (FCNC) processes can give stringent
limits on new physics contributions because the SM contributions are
generally very small due to the GIM mechanism. Constraints arising on
couplings of right-handed quarks to $W$ have been considered in this
context already for a long time within left-right symmetric models.
The strongest constraint in this case comes from $K^0-\Bar{K}^0$
mixing~\cite{Beall:1981ze}. 
Interesting constraints can come from the
(rare) $B$ meson decay processes, too, for which recently a large amount of new
data has become availible from the B-factories. A very prominent
example is here the radiative decay $b\to s\gamma$~\cite{Cho:1993zb}. 

It is clear that a comprehensive analysis of FCNC processes within the
LEET merits to be performed, but is beyond the scope of the present
paper. Here we only want to stress what the systematic power counting
of the LEET tells us about the different contributions to FCNC
processes and their respective suppression. Let us consider the box
diagram contribution to $\bar s d\to \bar d s$ shown in
Fig.~\ref{boxgraph}. Following the generalized Weinberg power
counting, see Eqs.~(\ref{powercounting},\ref{irdim}), the dominant SM
contribution counts as $d^* = 4$, because there is one loop with only
$\mathcal{O}(p^2)$ vertices without any spurions.  The diagrams with
one insertion of a vertex $\mathcal{O}(p^2\,\xi^2)$ or
$\mathcal{O}(p^2\,\eta^2)$ consequently count as $d^* = 5$ and with
two vertices $\mathcal{O}(p^2\,\xi^2)$ or $\mathcal{O}(p^2\,\eta^2)$
as $d^* = 6$. That means that contributions with one right-handed
vertex, as shown in Fig.~\ref{boxgraph}, have dimension $d^* = 5$ and
with two right-handed vertices they have dimension $d^* = 6$. At $d^*
= 6$ there are in addition contributions from diagrams with one vertex
of dimension $d_v = 4$ which is clearly beyond NLO, see
Eqs.~(\ref{LagrZ},\ref{LagrW}). In contrast to left-right symmetric
models, where the dominant non-standard contribution arises from a
$W_R$-boson exchange, we expect the dominant non-standard effect in
the LEET to show up at order $d^* = 5$, i.e., with only one insertion
of a ``non-standard'' vertex. In addition, we have to keep in mind,
that at each order new counter terms will arise\footnote{In principle,
four-fermion operators contributing to FCNC processes can be written
at $\mathcal{O}(p^2)$.  They are, however, dimensionally suppressed by
a factor $\Lambda^{-2}$, where $\Lambda \gg \Lambda_W$ is an energy
scale exterior to the LEET, see the discussion in v),
section~\ref{LEET}.  Here we will limit the discussion to the
operators generated within the LEET.} which have to be included in a
quantitative analysis.

The non-standard contribution of order $d^* = 5$, which should a
priori be the most important one, merits a further comment.  In
general, we expect that unusual operators, not considered in the SM
and its extensions so far, appear in the effective four-fermion
interaction. Let us consider as an example the box diagram shown in
Fig.~\ref{boxgraph} which contributes to $H_W^{\Delta S = 2}$, i.e.,
to $K^0$-$\bar{K}^0$ mixing. Inserting the vertices from the
Lagrangian at NLO, Eq.~(\ref{LagrW}), the leading contribution from
this diagram is clearly of dimension $d^* = 5$ since we have three
(left-handed) $\mathcal{O}(p^2)$ vertices and one (right-handed)
$\mathcal{O}(p^2\,\eta^2)$ vertex.  We find, for instance, the
following type of four-fermion operator
\begin{equation}
i \epsilon\, \frac{m_{u_i}}{m_W^2}\, \bar{s}_R \sigma^{\alpha\tau} d_L
\bar{s_L}\gamma_\tau \partial_\alpha d_L
\end{equation}
and the corresponding permutations.  The factor $\epsilon$ reflects
the spurion suppression. Note that there is an additional suppression
from the derivative which appears because of Lorentz invariance. Apart
from the power counting arguments, inherent to the LEET, we can at
present give no more quantitative estimate of the actual numerical
value of the non-standard contributions to FCNC processes.

Our analysis concerns only the real part of the mixing matrix elements
in the light quark sector. The constraints on the phases, for example
from the electric dipole moment of the neutron, have to be considered
separately.
\section{Summary and conclusion}
\label{sec:summary}
An effective theory framework is a very elegant and efficient way to
treat effects beyond the SM without relying on a specific model. The
basis of the present work is a ``not-quite decoupling'' alternative to
the usually applied decoupling effective theory framework. Within the
not-quite decoupling LEET, the heavy particles adherent to an extended
symmetry $S_{\mathit{nat}} \supset S_{\mathit{ew}} = SU(2)_L\times
U(1)_Y$ at high energies decouple, but the symmetry becomes partly
non-linearly realised at low energies and constrains the effective
interactions at low energies. The classification of the different
operators is thereby based on infrared power counting. The symmetry
$S_{\mathit{nat}}$ can be inferred within this ``bottom-up'' approach
from the requirement that at lowest order we want to recover the
(higgsless) vertices of the SM and nothing else. We have worked with
the minimal version of $S_{\mathit{nat}} = [SU(2)]^4\times U(1)^{B-L}$
fulfilling this requirement, which has been constructed in
Ref.~\cite{HS04a,HS04b,HS06}.

This LEET predicts that a priori the most important effects beyond the
SM are (universal) non-standard couplings of fermions to the gauge
bosons $W$ and $Z$ appearing at NLO. From general arguments (see the
discussion on that point in Ref.~\cite{HS06}) these non-standard
couplings are expected to be of the order of percent. The aim of this
paper was to perform a phenomenological analysis of these non-standard
couplings. Let us summarize our results
\begin{itemize}
\item In the neutral current sector we obtain a good agreement with
  the $Z$-pole data. In particular, we can solve the long-standing
  $A_{FB}^b$ puzzle without introducing non-universal effects in the
  couplings. The important point here is the NLO modification of the
  right-handed couplings. Also low-energy data such as atomic parity
  violation are well reproduced. There are only exceptional cases, when
  the LO + NLO contribution is accidentally small, where we fail to
  reproduce the data. One such example is the $e^-e^-$ M\o ller
  scattering. For these cases it is particularly important to extend
  the analysis to higher orders.
\item The most striking NLO effect is probably the direct coupling of
  right-handed quarks to $W$. We should emphasize here that due to an
  additional (discrete) symmetry in the lepton sector, intended to
  suppress the neutrino Dirac mass, there are no lepton right-handed charged
  currents. That means that many of the stringent tests on a
  right-handed $W_R$ boson present in left-right symmetric extensions
  of the SM do not apply in our case. Due to quark confinement it is
  very difficult to establish a stringent test of quark couplings. We
  always face the problem that it is not easy to disentangle
  QCD from electroweak effects. This is most obvious in semileptonic
  decays. To determine the axial and vector effective EW couplings,
  one has to know the QCD parameters like decay constants and form
  factors and vice versa. The presence of right-handed charged quark
  currents implies that quark mixing is modified: we have to consider
  two (a priori independent) unitary mixing matrices, $V_L$ and
  $V_R$. This means in particular that axial and vector couplings have
  to be considered independently.  
  
  We have focused on the light quark sector. There, one stringent test
  is conceivable. The Callan-Treiman low energy theorem allows for a
  very precise prediction for the value of the scalar $K\pi$ form
  factor at the Callan-Treiman point with only a small hadronic
  correction, $\Delta_{CT}$. The first direct measurement of the form
  factor in $K^L_{\mu 3}$ decays by the NA48 collaboration~\cite{NA48}
  indicates a 5$\sigma$ deviation with the SM prediction. It is hard
  to imagine that this deviation could be explained entirely as the
  deviation from the Callan-Treiman theorem. In this case the
  $\mathcal{O}(p^4)$ ChPT calculation of $\Delta_{CT}$ would have to
  be wrong by a factor of 20, i.e., there should be anomalously large
  higher order corrections. This deviation, can however, be explained
  by a direct coupling of right-handed quarks to $W$ with a
  (partially) inverted mixing hierarchy in the right-handed
  sector. Within the LEET considered here, this explanation is
  unique. Before drawing any firm conclusion, this effect needs
  additional experimental verification.
  
  The extracted values for the decay constants $F_\pi$ and $F_K/F_\pi$
  from the decay rate $\Gamma(\pi_{l2} (\gamma))$ and the branching
  ratio Br$(K_{l2}(\gamma)/\pi_{l2}(\gamma))$, which are needed as
  input for many ChPT calculations, are modified in the presence of
  RHCs. For instance, the value for $F_K/F_\pi$ could be as low as
  $\approx 1.12$ if we take the RHCs parameters from the scalar $K\pi$
  form factor. We should mention in this context, that the ChPT
  prediction for the slope of the scalar $K\pi$ form factor is in
  perfect agreement with the value from the NA48 measurement if the
  value of $F_K/F_\pi$ is changed accordingly.
  
  At the present order there are hardly any other sensitive tests. The
  analysis of hadronic tau decay data in the non-strange channel, for
  example, only indicates that the non-standard couplings stay on the
  percent level. A more precise determination seems not possible for
  the moment. In the strange channel the contribution from
  non-standard couplings is enforced due to the mixing hierarchy in
  the left-handed sector. Here, new data from the B-factories and an
  improved treatment of the QCD part in the near future can give
  interesting new results. 
\item Another consequence of the correlation between the values for the
  QCD and the EW parameters extracted from experiment, is that the
  sensitivity to $\alpha_s$ is lost in the two processes furnishing up
  to now the most precise determination of the strong coupling
  constant: In the analysis of $Z$-pole data as well as in the
  analysis of hadronic tau decays, the additional EW parameters make a
  sensible extraction of the value of $\alpha_s$ difficult. 
\end{itemize}   
To conclude, for the moment our analysis at NLO does not show any sign
for an inconsistency of experimental data with the order by order LEET
estimates. The values we obtained for the parameters are of the order
of magnitude expected from the LEET. Of course, the higher order
contributions should be investigated with care, especially in the
charged current sector where stringent constraints on the coupling of
right-handed quarks can arise from processes at NNLO.
\section*{Acknowledgements}
We would like to thank M.~Antonelli, A.~Ceccucci, M.~Davier,
S.~Descotes-Genon, A.~Djouadi, P.~Franzini, J.~Hirn, M. Knecht,
H.~Leutwyler, U.-G.~Mei{\ss}ner, B.~Moussallam, A.~Pich, J~.J.~Sanz
Cillero, M.~Veltri, and R.~Wanke for their interest, suggestions and
help. We are grateful to D.~Plane and S.~Menke (OPAL collaboration)
for providing us with the precise correlation matrix concerning the
OPAL hadronic tau decay data. This work has been partially supported
by EU contract MRTN-CT-2006-035482 (``Flavianet'') and the EU
Integrated Infrastructure Initiative Hadron Physics
(RH3-CT-2004-506078).
\begin{appendix}
\section{Expressions for the $Z$-pole observables}
\label{zexp}
We will list here the expressions for the $Z$-pole observables in
terms of the effective couplings up to NLO.
The couplings discussed in section~\ref{formulas} can be rewritten in
terms of effective couplings of left-handed fermions to $Z$,
\begin{equation}
g_L^u = \frac{1 + \delta}{2} - \frac{2}{3}\tilde{s}^2 \qquad 
g_L^d = - \frac{1 + \delta}{2} + \frac{1}{3} \tilde{s}^2 \qquad 
g_L^e = -\frac{1}{2} +  \tilde{s}^2\qquad 
g_L^\nu = \frac{1}{2}~,
\label{glz}
\end{equation}
and effective couplings of right-handed fermions to $Z$
\begin{equation}
g_R^u = -\frac{2}{3} \tilde{s}^2 + \frac{\epsilon^u}{2} \qquad 
g_R^d = \frac{1}{3} \tilde{s}^2 - \frac{\epsilon^d}{2}\qquad 
g_R^e = \sweff - \frac{\epsilon^e}{2}\qquad 
g_R^\nu = \frac{\epsilon^\nu}{2}~.
\label{grz}
\end{equation}
The corresponding effective couplings for vector and axial channel are
obtained from $g_A^f = g_L^f - g_R^f, g_V^f = g_L^f + g_R^f$. This gives
\begin{equation}
g_V^u = \frac{1 + \delta}{2}
-\frac{4}{3}\tilde{s}^2+\frac{\epsilon^u}{2} 
\qquad 
g_V^d = - \frac{1 + \delta}{2} 
+ \frac{2}{3} \tilde{s}^2 - \frac{\epsilon^d}{2}\qquad 
g_V^e = -\frac{1}{2} + 2 \tilde{s}^2- \frac{\epsilon^e}{2}\qquad 
g_V^\nu = \frac{1}{2} + \frac{\epsilon^\nu}{2}~,
\label{gvs}
\end{equation}
and 
\begin{equation}
g_A^u = \frac{1 + \delta}{2} - \frac{\epsilon^u}{2} \qquad 
g_A^d = - \frac{1 + \delta}{2} +  \frac{\epsilon^d}{2}\qquad 
g_A^e =  -\frac{1}{2} + \frac{\epsilon^e}{2}\qquad 
g_A^\nu = \frac{1}{2} - \frac{\epsilon^\nu}{2}~.
\label{gas}
\end{equation}

The asymmetries can be written as
\begin{equation}
A_{FB}^f = \frac{3}{4} \frac{ (g_L^e)^2 - (g_R^e)^2}{ (g_L^e)^2 +
  (g_R^e)^2} \frac{ (g_L^f)^2 - (g_R^f)^2}{ (g_L^f)^2 + (g_R^f)^2}~,
\end{equation}
and
\begin{equation}
\mathcal{A}_f = \frac{ (g_L^f)^2 - (g_R^f)^2}{ (g_L^f)^2 + (g_R^f)^2}~,
\end{equation}
where we did not explicitly write the expressions up to first order
in the spurionic parameters although this is a straight forward
manipulation because the expression then becomes
somewhat cumbersome.

The ratios $R_q^0$ are defined as 
\begin{equation}
R_q^0 = \frac{\Gamma_q}{\Gamma_h}~,
\end{equation}
where we have calculated the corresponding partial widths according to
Eq.~(\ref{gammapartialz}). For the ratio $R_l^0$ we have
\begin{equation}
R_q^0 = \frac{\Gamma_h}{\Gamma_l}~.
\end{equation}
The total width is obtained from the sum
over all partial widths
\begin{equation}
\Gamma_Z = \sum_f \Gamma_f~.
\end{equation}  
The hadronic pole cross section is defined as
\begin{equation}
\sigma^0_h = \frac{12\pi}{m_Z^2} \frac{\Gamma_e \Gamma_h}{\Gamma_Z^2}~.
\end{equation}
\section{Hadronic tau decays: description of the perturbative part}
\label{apptau}
In this appendix we will mention some details concerning the different
prescriptions employed to describe the purely perturbative part for
the total tau hadronic decay rate and the related moments.
It can be written as~\cite{Davier:2005xq,Braaten:1991qm,LeDiberder:1992fr}
\begin{equation}
  \delta^{(0),kl} = \sum_{n=1}^\infty \tilde{K}_n(\zeta)
  A^{(n,kl)}(a_s)~,
\end{equation} 
with $a_s = \alpha_s/\pi$. The functions $\tilde{K}_n(\zeta)$ contain
the perturbative coefficients $K_n$. The dependence on the
renormalization scale parameter $\zeta$ is determined by the condition
that physical quantities are independent of $\zeta$. The values of the
perturbative coefficients can be inferred from the calculation of the
$e^+e^-$ inclusive cross section. $K_0 = K_1 = 1$ are universal,
whereas the remaining coefficients depend on the renormalization
scheme used. They have been calculated up to $n = 3$. In the
$\bar{\mathrm{MS}}$ scheme they are given by (for three flavors)
$K_2 = 1.640$ and $K_3 = 6.317$.
The functions $A^{(n,kl)}$ are defined as
\begin{eqnarray}
A^{(n,kl)} &=& \frac{1}{2\pi i} \oint\limits_{|s|  = m_\tau^2} \frac{ds}{s}
\Big( 2 \Gamma(3+k) \left(\frac{\Gamma(1+l)}{\Gamma(4+k+l) } + 2
  \frac{\Gamma(2+l)} {\Gamma(5+k+l)} \right) \nonumber \\ && - 2
\,I(\frac{s}{s_0},1+l,3+k) - 4\, I(\frac{s}{s_0},2+l,3+k)\Big)
a_s^n(-\zeta s)~,
\label{ankl}
\end{eqnarray} 
with $I(x,a,b) = \int_0^x t^{a-1} (1-t)^{b-1} dt$. $\alpha_s(s)$ is
  given by the solution of the renormalization group equation,
\begin{equation}
\frac{d a_s}{ds} = -a_s^2 \sum_n \beta_n a_s^n~,
\end{equation}
\TABLE[h]{
\begin{tabular}{cc|ccccc}
$k$&$l$ & $g_2$& $g_3$&$g_4$&$g_5$&$g_6$ \\ \hline
0&0&3.56&19.995&78.00&14.25 $K_4$ -391.54&17.81 $K_5$ + 45.11
$K_4$ + 1.58 $\beta_4$ - 8062.\\
1&0&4.17&28.35&161.06& 225.58 + 16.69 $K_4$&-5424.+
1.85 $\beta_4$ + 102.1 $K_4$ + 20.87 $K_5$\\
1&1&2.59&5.16&-90.59& -1932. +10.35 $K_4$&
-18683. + 1.15 $\beta_4$-60.88 $K_4$+12.94 $K_5$ \\
1&2&1.94&-2.03&-135.98&-1875. + 7.76 $K_4$&-12337. + 0.862 $\beta_4$-104.6 $K_4$+9.70 $K_5$\\
1&3&1.57&-5.60&-151.4&
-1730.+6.27 $K_4$&-8394. + 0.697 $\beta_4$-124.1 $K_4$+7.84 $K_5$\\
2&0&4.67&35.59&239.7&899.7 + 18.68 $K_4$&-1281. + 2.10 $\beta_4$ + 153.0 $K_4$ + 23.34 $K_5$\\
2&1&2.97&9.40 &-63.77& -1965. +
11.88 $K_4$&-22433. + 1.32 $\beta_4$ - 35.06 $K_4$+14.85 $K_5$\\
3&0&5.08&41.99&313.9&1600. + 20.34 $K_4$&3889. + 2.26 $\beta_4$ + 199.0 $K_4$ + 25.42 $K_5$\\
4&0&5.44&47.73 &383.9& 2310. + 21.76 $K_4$&9781. + 2.42 $\beta_4$ + 240.9 $K_4$ + 27.20 $K_5$
\end{tabular}
\caption{\it Coefficients for the FOPT expansion \label{tablepert}}
}
where the coefficients are known up to $n =
3$~\cite{vanRitbergen:1997va}: $\beta_0 = 9/4, \beta_1 =
4,\beta_2^{\bar{\mathrm{MS}}} = 10.0599,\beta_3^{\bar{\mathrm{MS}}} =
47.2306$. For FOPT one integrates Eq.~(\ref{ankl}) inserting a Taylor
expansion for the running of $\alpha_s(s)$ around some reference point
$s_0$ which we chose $s_0 = m_\tau^2$. This gives
\begin{equation}
\delta^{(0),kl} = r_{kl} \sum_{n=1}^6 (K_n + g_n^{kl})
a_s^n(m_\tau)~. 
\end{equation} 
We have $g_1^{kl} = 0$. The numerical values of the other functions
$g_n^{kl}$ are listed in Table~\ref{tablepert} up to $g_6^{kl}$.
\TABLE[h]{
\begin{tabular}{cc|c}
$k$&$l$& $r_{kl}$ \\ \hline
0&0&1\\
1&0&7/10 \\
1&1& 1/6\\
1&2&13/210 \\
1&3&1/35 \\
2&0&8/15 \\
2&1& 11/105\\
3&0& 3/7\\
4&0& 5/14\\
\end{tabular}
\caption{\it Normalization coefficients determining the parton level predictions.\label{tableparton}}
} 
In Table~\ref{tableparton} we list the parton level predictions for
the different moments.
For CIPT, one numerically integrates the RGE equation for the
running of $\alpha_s$ on the contour to evaluate the function
$A^{(n,kl)}$.  Following the discussion in Ref.~\cite{Davier:2005xq}
we took $K_4 = 25$, $K_5 = 98 $ and $K_6 = 384,$ and $\beta_4 = 222$
for our calculations.
\end{appendix}

\end{document}